\documentclass[a4paper,12pt]{amsart}
\usepackage[left=1in, right=1in, top = 1.5in, bottom = 1.5in]{geometry}
\usepackage[round]{natbib}
\usepackage[utf8]{inputenc}
\usepackage[all]{xy}
\usepackage{amssymb}
\usepackage{amsmath}
\usepackage{csquotes}
\usepackage{subfig}
\usepackage{enumerate}
\usepackage{graphicx}
\usepackage{amsaddr}
\usepackage{color}
\usepackage{mwe}
\usepackage{appendix}
\usepackage{centernot}
\usepackage{pgfplots}
\pgfplotsset{compat=1.18}
\usepackage{mathtools}
\usepackage{comment}
\usepackage{makecell}
\usepackage[ruled,vlined]{algorithm2e}
\usepackage{enumitem}

\usepackage{soul}
\def\@centernot#1#2{%
  \mathrel{%
    \rlap{%
      \settowidth\dimen@{$\m@th#1{#2}$}%
      \kern.5\dimen@
      \settowidth\dimen@{$\m@th#1=$}%
      \kern-.5\dimen@
      $\m@th#1\not$%
    }%
    {#2}%
  }%
}
\makeatother

\newcommand{\independent}{\perp\mkern-9.5mu\perp}

\DeclareRobustCommand{\frac}[3][0pt]{%
  {\begingroup\hspace{#1}#2\hspace{#1}\endgroup\over\hspace{#1}#3\hspace{#1}}}

\usepackage{tikz}
\usetikzlibrary{arrows.meta,shapes}
\usetikzlibrary{arrows,shapes.arrows,shapes.geometric,shapes.multipart,
decorations.pathmorphing,positioning,shapes.swigs,}
\newcommand{\bbP}{\mathbb{P}}

\usepackage{setspace}
\newtheorem{lemma}{Lemma}
\newtheorem{proposition}{Proposition}
\newtheorem{corollary}{Corollary}
\theoremstyle{definition}
\newtheorem{assumption}{Assumption}

\newtheorem{definition}{Definition}
\theoremstyle{remark}
\newtheorem{remark}{Remark}

\title{Variant Specific Treatment Effects with Applications in Vaccine Studies}
\author{Gellért Perényi \& Mats J. Stensrud}
\address{
Institute of Mathematics, Ecole Polytechnique Fédérale de Lausanne, Switzerland 
}

\begin{document}

\begin{abstract} 
Pathogens usually exist in heterogeneous variants, like subtypes and strains. Quantifying treatment effects on the different variants is important for guiding prevention policies and treatment development. Here we ground analyses of variant-specific effects on a formal framework for causal inference. This allows us to clarify the interpretation of existing methods and define new estimands. Unlike most of the existing literature, we explicitly consider the (realistic) setting with interference in the target population: even if individuals can be sensibly perceived as iid in randomized trial data, there will often be interference in the target population where treatments, like vaccines, are rolled out. Thus, one of our contributions is to derive explicit conditions guaranteeing that commonly reported vaccine efficacy parameters quantify well-defined causal effects, also in the presence of interference. Furthermore, our results give alternative justifications for reporting estimands on the relative, rather than absolute, scale. We illustrate the findings with an analysis of a large HIV1 vaccine trial, where there is interest in distinguishing vaccine effects on viruses with different genome sequences. 
\end{abstract}

\maketitle

\noindent%
\vfill

\newpage
\clearpage

\section{Introduction}
Infectious diseases are severe threats to human health, and vaccination is one of the most successful strategies for preventing them. However, a characteristic of infectious agents is their heterogeneity and rapid change. One example is HIV \citep{gaschen_diversity_2002, barouch_challenges_2008, johnston_hiv_2008}, which exists in two main types, both of which have different variants. The heterogeneity is a challenge for the development of treatments, such as vaccines, because the treatment effect often depends on the characteristics of the circulating strain of a pathogen, and the presence of strains varies over time. For example, many vaccines are designed to target particular genetic sequences. These vaccines might offer less protection towards, say, evolving pathogens with different sequences in the target regions.

To describe existing strategies for quantifying treatment effects on heterogeneous variants, consider first a randomized controlled trial (RCT) where participants are assigned to vaccine or placebo treatment. Suppose that pathogens from infected individuals in each arm were recorded \citep{rolland_genetic_2011, hertz_study_2016, ouattara_epitope-based_2020}, and the recordings showed that the genetic sequences of the infected individuals in the vaccine arm differed from those in the placebo arm. Such differences have been attributed to heterogeneity in the effects of the vaccine on different variants, called "sieve effects" (not to be confused with sieve estimators). This heuristic approach might be used as a test of a null hypothesis of equality of effects across variants, see Appendix \ref{APP: connect two sieves}. Nevertheless, this approach does not adequately quantify the protective effect of the treatment on the different variants, which arguably is of primary interest for decision-makers.

Alternatively, there exist statistical "sieve" methods for differentiating treatment effects against different variants of a pathogen \citep{gilbert_statistical_1998,gilbert_interpretability_2001, gilbert_2-sample_2008, sun_proportional_2009, juraska_mark-specific_2013, benkeser_estimating_2019, yang_causal_2022}. This literature builds on results from competing events in survival analysis \citep{gilbert_comparison_2000}: individuals are considered to be at risk of experiencing an infection with different "competing" variants over time. 
These sieve analysis methods have, for example, been applied to study effects of vaccination against HIV \citep{rolland_increased_2012, rolland_genetic_2011, zolla-pazner_vaccine-induced_2014, hertz_study_2016}, malaria \citep{neafsey_genetic_2015, ouattara_epitope-based_2020} and SARS-CoV-2 \citep{rolland_sieve_2021}.

The sieve methods that consider parameters on the cumulative incidence scale \citep{gilbert_interpretability_2001, gilbert_sieve_2001, gilbert_2-sample_2008, sun_proportional_2009,rolland_increased_2012, zolla-pazner_vaccine-induced_2014, neafsey_genetic_2015,benkeser_estimating_2019, benkeser_assessing_2020, yang_causal_2022},  can be endowed with a causal interpretation as total effects in iid settings \citep{robins_identifiability_1992, young_causal_2020}. Nevertheless, the relevance of these parameters requires more justification in infectious disease settings, even in a blinded randomized trial with perfect adherence. Unlike the study of non-communicable diseases, the cumulative incidence of each strain often varies due to differences in the number of infected and immune individuals across the strains \citep{garnett_sexual_1996}. Furthermore, the prevalence of the strains changes rapidly over time. To guide practice, e.g., large scale vaccination programs, estimands should sensibly reflect, or be insensitive to, such changes. Finally, interference between units might be a small problem in a perfectly executed trial examining a vaccine that is not yet available for public use; the interactions between the trial participants will often be negligible, which seems to be the justification for the use of iid assumptions in major vaccine trials. Nevertheless, interference will most likely be a concern if a vaccine is rolled out in a larger target population. Thus, to make the trial results relevant to the future decision setting, we need to argue that the parameters estimated under iid assumptions from the trial quantify the effects of interest in the relevant target population.

In this article, we develop causal methodology for quantifying vaccine efficacy against different disease-causing variants. Using explicit causal theory and assumptions, we clarify when we can make meaning to statements such as \textit{the treatment has the same effect on different variants}. 
Our results also resolve concerns regarding interference, which threatens the generalization of the trial results to the target population. We achieve this by considering parameters that are defined by conditioning or intervening on (a possibly unmeasured) exposure status. We further elaborate on the implications of interference in Appendix \ref{APP: interference}.

In practice, results of vaccine trials are usually presented on the relative scale. Yet, absolute effects are arguably more relevant to practical decision-making in many settings \citep{vanderweele_tutorial_2014}. Indeed, claims favoring the relative scale, such as the stronger heterogeneity of the risk difference, have been suggested to have inadequate evidence \citep{poole_is_2015}. 
Our results can serve as justification for the relative measures, supporting claims made in the literature \citep{tsiatis_estimating_2022, huang_improved_2023}; in infectious disease settings, we show that measures on the relative scale can be identified under assumptions that do not allow identification of measures on the additive scale. The relative measures will also have certain stability properties. Furthermore, we show that identification formulas of different causal effects on the relative scale are equal under explicit assumptions, whereas the corresponding effects on the additive scale are generally different. To fix ideas, consider a running example on HIV vaccination. 

\subsection{The ALVAC/AIDSVAX vaccine against HIV}\label{subsec: running example hiv}
The efficacy of the ALVAC/\\
AIDSVAX vaccine was assessed in the RV144 RCT (NCT00223080, \cite{rerks-ngarm_vaccination_2009}). Data were collected from 16,395 healthy men and women, aged between 18 and 30 years in Thailand, initiated in October 2003. Treatment recipients were tested for HIV1 infection and viremia at the end of the vaccination period, 6 months from baseline, and then every 6 months in the following 3 years. Individuals were followed until the onset of the infection or the end of the follow-up period. While the vaccine showed evidence for the prevention of the HIV1 infection with a vaccine efficacy of 31.2\% (95\% CI: 1.1\%, 52.1\%) in the modified intention-to-treat sample, the effect of the vaccine waned over time and, therefore was not granted licensure by the FDA. 

\citet{rolland_increased_2012} conducted a sub-analysis based on the genome sequences of 110 individuals and found that, depending on the match and the mismatch of the vaccine at the amino acid positions 169 and 181 respectively, of the HIV-1 envelope variable 2, the vaccine efficacy increased to 48\% (95\% CI: 18\%, 66\%) against viruses matched to the vaccine at position 169, and to 78\% (95\% CI: 35\%, 93\%) against viruses mismatched to the vaccine at position 181. This difference compared to the overall vaccine efficacy of 31.2\% suggests the appearance of a sieve effect \citep{gilbert_statistical_1998}. However, we cannot immediately interpret these differences as being causal effects. For example, the observed differences in rates may be confounded by factors such as the number of previous infections, like-with-like mixing, or changes in public policy. Another issue is interference, which is likely to be present in a setting where the vaccine is broadly available.

\section{Data Structure and preliminary assumptions}\label{SEC: Data}
Consider data from a study where individuals $i \in \{1, \, \dots, \, n\}$ were randomly assigned to treatment or placebo, denoted by $A_i=1$ or $A_i=0$, respectively. Suppose that the individuals were drawn from a much larger super-population, such that interference between trial participants is negligible.  Thus, we assume that individuals are iid in the experiment (but not necessarily in the target population), which is also an implicit default assumption in classical vaccine trials \citep{chang_universal_1997, whitney_decline_2003, buchbinder_efficacy_2008}, see Appendix \ref{APP: interference} for more details. To avoid clutter, we hereby omit the subscript $i$ on the random variables. Moreover, let $L \in \mathcal{L}$ denote the vector of measured baseline covariates. Our effect of interest is the effect of the treatment, possibly conditional on or under interventions on exposure.  Thus, we intentionally distinguish between the terms \textit{treatment} ($A$) and \textit{exposure} to a variant $(E)$, where $E \in \{0,1,2, \mathbf{B}\}$ encodes no exposure ($E=0$), exposure to variant 1 ($E=1$), exposure to variant 2 ($E=2$), and $\mathbf{B}$ denotes exposure to both variants during the follow-up period (e.g., 42 months). The causal estimands we will consider can be identified in a setting with more than two variants, but for notational simplicity, we consider only variant 1 and variant 2. The following assumption will be used in the first parts of the manuscript:

\begin{assumption}[Unique exposure]\label{ASM: no multiple exp}
    $\mathbb{P}(E=\mathbf{B})=0.$
\end{assumption}
Assumption \ref{ASM: no multiple exp} states that exposure to both variants is a probability zero event, which implies that exposure to each variant is mutually exclusive. The plausibility of Assumption \ref{ASM: no multiple exp} depends on the context: when studying diseases with low prevalence and correspondingly low exposure rates, exposure to multiple variants in a given time interval will have a probability (close to) zero. Such an argument will justify the preliminary assumption of mutual exclusivity of exposures. 
Let $Y \in \{0,1,2\}$ denote the outcome of interest, say severe infection, where the encoding corresponds to the definition of $E$: 0 is no infection, 1 is infection by variant 1, and 2 is infection by variant 2 by the end of the follow-up period. In the case of the event $\{E=\mathbf{B}\}$, we assume that only one of the outcomes can occur. Under Assumption \ref{ASM: no multiple exp}, the mechanism of determining which of the three outcomes, $\{0,1,2\}$, are realized when $\{E=\mathbf{B}\}$ is left undefined. 

In Section \ref{SEC: TTE estimands}, we discuss the time-to-event setting when these variables are taken to be time-dependent, which also allows us to assess exposure to multiple variants over time. Consideration of the time-to-event setting requires more involved notation, which we introduce when needed.

\section{Time-fixed estimands}\label{SEC: fixed time estimands}
\begin{table}[!ht]
    \centering
    \hspace*{-1.5cm}
    {\setlength{\extrarowheight}{22pt}%
    \begin{tabular}{|c|c|c|c|c|}
        \hline
        Notation&Name&\makecell{Identification\\ Assumptions}& Estimand &\makecell{Identifying\\ formula}\\[15pt]
        \hline
        $ATE(j)$& \makecell{Average treatment effect\\ ratio} & \makecell{\ref{ASM: Standard RCT}} & $\frac{\mathbb{P}(Y^{a=1}=j)}{\mathbb{P}(Y^{a=0}=j)}$ & $\frac{\mathbb{P}(Y=j|A=1)}{\mathbb{P}(Y=j|A=0)}$ \\[15pt]
        \hline
        $CECE(j)$& \makecell{Relative causal effect\\ conditional on exposure} &\makecell{\ref{ASM: no multiple exp}, \ref{ASM: No eff exp}, \ref{ASM: Standard RCT},\\ \ref{ASM: Exp nec}, \ref{ASM: No cross}} &$\frac{\mathbb{P}(Y^{a=1}=j|E^{a=1}=j)}{\mathbb{P}(Y^{a=0}=j|E^{a=0}=j)}$ & $\frac{\mathbb{P}(Y=j|A=1)}{\mathbb{P}(Y=j|A=0)}$\\[15pt]
        \hline
        \multicolumn{5}{|c|}{\textbf{Time-fixed estimands}} \\
        \hline
        $CCS$& \makecell{Contrast conditional\\ on specific exposure} &\makecell{\ref{ASM: no multiple exp}, \ref{ASM: Standard RCT}, \ref{ASM: Exp nec}, \\ \ref{ASM: No cross}, \ref{ASM: No eff on exp ratios}} &$\frac[5pt]{\frac{\mathbb{P}(Y^{a=1}=1|E^{a=1}=1)}{\mathbb{P}(Y^{a=0}=1|E^{a=0}=1)}}{\frac{\mathbb{P}(Y^{a=1}=2|E^{a=1}=2)}{\mathbb{P}(Y^{a=0}=2|E^{a=0}=2)}}$ &$\frac[5pt]{\frac{\mathbb{P}(Y=1|A=1)}{\mathbb{P}(Y=1|A=0)}}{\frac{\mathbb{P}(Y=2|A=1)}{\mathbb{P}(Y=2|A=0)}}$ \\[15pt]
        \hline
        $CCE$& \makecell{Contrast conditional\\ on exposure}&\makecell{\ref{ASM: no multiple exp}, \ref{ASM: Standard RCT},\\ \ref{ASM: Exp nec}} & $\frac[5pt]{\frac{\mathbb{P}(Y^{a=1}=1|E^{a=1}\neq 0)}{\mathbb{P}(Y^{a=0}=1|E^{a=0}\neq0)}}{\frac{\mathbb{P}(Y^{a=1}=2|E^{a=1}\neq 0)}{\mathbb{P}(Y^{a=0}=2|E^{a=0}\neq0)}}$&$\frac[5pt]{\frac{\mathbb{P}(Y=1|A=1)}{\mathbb{P}(Y=1|A=0)}}{\frac{\mathbb{P}(Y=2|A=1)}{\mathbb{P}(Y=2|A=0)}}$ \\[15pt]
        \hline
        $EIE(l)$&\makecell{Effect with\\intervened exposure} &\makecell{\ref{ASM: no multiple exp}, \ref{ASM: Exp nec}, \ref{ASM: No cross},\\ \ref{ASM: Cond no effect on exp},  \ref{ASM: Generalised exposure RCT}} & $\frac[5pt]{\frac{\mathbb{P}(Y^{a=1, e=1}=1|L=l)}{\mathbb{P}(Y^{a=0, e=1}=1|L=l)}}{\frac{\mathbb{P}(Y^{a=1, e=2}=2|L=l)}{\mathbb{P}(Y^{a=0, e=2}=2|L=l)}}$& $\frac[5pt]{\frac{\mathbb{P}(Y=1|A=1, L=l)}{\mathbb{P}(Y=1|A=0, L=l)}}{\frac{\mathbb{P}(Y=2|A=1, L=l)}{\mathbb{P}(Y=2|A=0, L=l)}}$\\[15pt]
        \hline
        $EET(l)$& \makecell{Effect \\of exposure \\under treatment}&\makecell{\ref{ASM: no multiple exp}, \ref{ASM: Exp nec}, \ref{ASM: No cross},\\ \ref{ASM: Exposure RCT}, \ref{ASM: Equal treated exp prob}} & $\frac{\mathbb{P}(Y^{a=1, e=1}=1|L=l)}{\mathbb{P}(Y^{a=1, e=2}=2|L=l)}$& $\frac{\mathbb{P}(Y=1|A=1, L=l)}{\mathbb{P}(Y=2|A=1, L=l)}$ \\[15pt]
        \hline
        \multicolumn{5}{|c|}{\textbf{Time-to-event  estimands}} \\
        \hline
        $CCE_k$& \makecell{Time-to-event\\
         contrast \\conditional on exposure}&\makecell{\ref{ASM: no multiple exp}, \ref{ASM: TTE Standard RCT},\\ \ref{ASM: TTE original exp nec}} & $\frac[5pt]{\frac{\mathbb{P}(Y_k^{a=1}=1|E_k^{a=1}\neq0)}{\mathbb{P}(Y_k^{a=0}=1|E_k^{a=0}\neq0)}}{\frac{\mathbb{P}(Y_k^{a=1}=2|E_k^{a=1}\neq0)}{\mathbb{P}(Y_k^{a=1}=2|E_k^{a=1}\neq0)}}$
        &$\frac[5pt]{\frac{\mu_k^1(1)}{\mu_k^1(0)}}{\frac{\mu_k^2(1)}{\mu_k^2(0)}}$ \\[20pt]
        \hline
        $CSE_k$& \makecell{Challenge\\ subtype effect}
        &\makecell{\ref{ASM: TTE Standard RCT}, \ref{ASM: TTE Exp nec}, \ref{ASM: TTE No cross}, \\
        \ref{ASM: TTE exposure rat of exposed}, \ref{ASM: TTE Scaled new inf}}
        & $\frac[5pt]{\frac{\mathbb{P}(Y_k^{a=1, e_k=1, \overline{e}_{k-1}=0}=1)}{\mathbb{P}(Y_k^{a=0, e_k=1, \overline{e}_{k-1}=0}=1)}}{\frac{\mathbb{P}(Y_k^{a=1, e_k=2, \overline{e}_{k-1}=0}=2)}
        {\mathbb{P}(Y_k^{a=0, e_k=2, \overline{e}_{k-1}=0}=2)}}$
        &$\frac[5pt]{\frac{\mathbb{P}(Y_k=1|Y_{k-1}=0, A=1)}
        {\mathbb{P}(Y_k=1|Y_{k-1}=0, A=0)}}
        {\frac{\mathbb{P}(Y_k=2|Y_{k-1}=0, A=1)}
        {\mathbb{P}(Y_k=2|Y_{k-1}=0, A=0)}}$ \\[25pt]
        \hline        
    \end{tabular}}
    \caption{Estimands, sufficient identification assumptions, and corresponding identification formulas. The definition of $\mu_k^j$ is given in Appendix \ref{SUBSEC: One-shot}.}
\label{TAB: Causal estimands}
\end{table}

Our aim is to quantify the effect of treatment  $A$, say a vaccine, on the risk of developing the outcome $Y$, encoding infection by different variants. One motivation is that such estimands can inform us about the differential effects on the different variants, which, in turn, can guide future vaccine policies and development. However, there are various ways of defining such effects, and these definitions have different practical implications. To clarify these differences, we will explicitly define effects in a formal causal framework, assuming that the data were generated according to a \textit{finest fully randomized causally interpretable structured tree graph} (FFRCISTG) model \citep{robins_new_1986, richardson_single_2013}, which is a strictly weaker causal model compared to the \textit{non-parametric structural equation model with independent errors} (NPSEM-IE) \citep{pearl_causality_2009}. Let superscripts denote potential outcome variables. In particular, if the value of treatment $A$ is fixed to $a \in \{0,1\}$, then the potential outcome of the variable $Y$ is denoted by $Y^a$. Equivalently, $Y^e$ denotes the outcome of an individual, if possibly contrary to fact, they had been assigned to exposure $E=e$, for $e \in \{0,1,2\}$.

We first define a conventional estimand used in trials for assessing the effect of the vaccine on the specific outcome, that is the (relative) average treatment effect ($ATE$).
\begin{definition}[Average treatment effect]
$ATE(j)=\frac{\mathbb{P}(Y^{a=1}=j)}{\mathbb{P}(Y^{a=0}=j)} \ \forall  j \in \{1,2\}.$
\end{definition}
The relative vaccine efficacy is conventionally reported as $1-ATE(j)$ in vaccine trials. While the $ATE(j)$ is identifiable with iid assumptions from standard RCT data, the $ATE(j)$, estimated from a conventional vaccine trial, cannot be interpreted as an average effect in a target population where there is interference. This is a problem because, in many infectious disease settings, there will be interference as treatments of one individual can affect the outcomes of others, e.g., through herd immunity. This fact is rarely discussed in vaccine trials but poses a problem for the policy relevance of the parameters. 

To address the challenges in interpreting the population-level $ATE(j)$, we will define new estimands that are insensitive to interference by conditioning or intervening on exposure to variants. The rationale is that interference in vaccine settings is mediated through exposure status: when it is known that an individual is exposed to an infectious agent, their outcome is independent of the vaccine or infectious status of other individuals. Informally, we have an iid data structure conditional on, or under interventions on, exposure, even in the target population where the vaccine is rolled out. We further consider inference on these effects, even if the exposure status of an individual is unmeasured.

\subsection{Estimands conditional on exposure}
\label{subsec: conditional estimands}
Consider first the effect on the individuals who would be exposed to variant $j \in \{1,2\}$ regardless of treatment.
\begin{definition}[Principal stratum effect of the always-exposed]\label{DEF: PSE always exposed}
$$
\frac{\mathbb{P}(Y^{a=1}=j|E^{a=1}=E^{a=0}=j)}{\mathbb{P}(Y^{a=0}=j|E^{a=1}=E^{a=0}=j)} \ \forall \ j \in \{1,2\}.
$$
\end{definition}
This is a causal effect, in the sense that it contrasts average outcomes under different treatments in the same subpopulation of individuals. However, the subpopulation is defined by counterfactual exposures under two different treatment assignments. Even in a hypothetical trial, such as a challenge trial, where the investigator observes the exposure status of the participants, the conditioning set cannot be identified without relying on additional untestable assumptions. Suppose, however, that the treatment does not affect the exposure \citep{stensrud_identification_2023}, which is plausible in successfully blinded randomized experiments: 
\begin{assumption}[No effect on exposure]\label{ASM: No eff exp}

$        E^{a=1}=E^{a=0}.$
\end{assumption}
It follows from Assumption \ref{ASM: No eff exp}  that $E^{a=1}=E^{a=0}=E$. Therefore, the causal contrast in Definition \ref{DEF: PSE always exposed} can be defined on the population of individuals who were observed to be exposed, as
$\mathbb{P}(Y^{a=1}=j|E = j) \text{ vs } \mathbb{P}(Y^{a=0}=j|E = j) \ \forall  j \in \{1, 2\}$. Assumption \ref{ASM: No eff exp} is analogous to assumptions imposed by, e.g., \citet{greenwood_statistics_1915} and  \citet{halloran_design_2010}, positing that the exposure to the infectious agent should be the same for individuals, regardless of their inoculation status. Despite being of importance \citep{stensrud_distinguishing_2024, obolski_call_2024}, Assumption \ref{ASM: No eff exp} is often only implicitly supposed in practical studies \citep{walsh_safety_2024}. 

Assumption \ref{ASM: No eff exp} motivates the relative casual effect conditional on exposure (relative $CECE$), similarly to \citet{stensrud_identification_2023}, but here we generalize this contrast to the multiple variant setting.
\begin{definition}[Variant specific relative $CECE$]
    \begin{equation*}
        \begin{aligned}
            CECE(j)=\frac{\mathbb{P}(Y^{a=1}=j|E^{a=1}=j)}{\mathbb{P}(Y^{a=0}=j|E^{a=0}=j)}\\
        \end{aligned}
            \ \forall \  j \in \{1,2\}.
    \end{equation*}
\end{definition}

\begin{remark}
    Under Assumption \ref{ASM: No eff exp}, the relative $CECE(j)$ can be equivalently defined as 
    \begin{align*}
        CECE(j)&=\frac{\mathbb{P}(Y^{a=1}=j|E=j)}{\mathbb{P}(Y^{a=0}=j|E=j)} =\frac{\mathbb{P}(Y^{a=1}=j|E^{a=0}=E^{a=1}=j)}{\mathbb{P}(Y^{a=0}=j|E^{a=0}=E^{a=1}=j)},
    \end{align*}
    that is, it equals the principal stratum effect of the always-exposed (Definition \ref{DEF: PSE always exposed}).
\end{remark}

The ratio between relative $CECE(j)-s$ for $j=1,2$ quantifies heterogeneity: 
\begin{definition}[Contrast conditional on subtype-specific exposure]\label{DEF: CCS}
    \begin{equation*}
       CCS= \left.\frac{\mathbb{P}(Y^{a=1}=1|E^{a=1}=1)}{\mathbb{P}(Y^{a=0}=1|E^{a=0}=1)} \middle \slash
        \frac{\mathbb{P}(Y^{a=1}=2|E^{a=1}=2)}{\mathbb{P}(Y^{a=0}=2|E^{a=0}=2)}\right. .
    \end{equation*}
\end{definition}
The $CCS$ is a measure of the relative effect of the treatment on the outcome, among those who were exposed to the specific subtype corresponding to the outcome. The contrast compares protection against variant 1 relative to variant 2, by examining those who were exposed to variants 1 and 2, respectively. This also means that the $CCS$ compares two relative $CECE$s that are defined in distinct populations: those who were exposed to variant 1 vs those who were exposed to variant 2.
The individuals exposed to variants 1 and 2 might have different characteristics, complicating the interpretation of the $CCS$, as we discuss in more detail in Section \ref{SEC: Estimand choice}.  

Consider a different estimand, where we instead condition on $E^a\neq0$ $\forall \ a \in \{0,1\}$:
\begin{definition}[Contrast conditional on exposure]
    $$CCE= \left. \frac{\mathbb{P}(Y^{a=1}=1|E^{a=1}\neq0)}{\mathbb{P}(Y^{a=0}=1|E^{a=0}\neq0)} \middle \slash
    \frac{\mathbb{P}(Y^{a=1}=2|E^{a=1}\neq0)}{\mathbb{P}(Y^{a=0}=2|E^{a=0}\neq0)} \right. .$$
\end{definition}
Analogous to the $CCS$, the $CCE$ measures the relative effect of the treatment on the two competing variants. However, the contrast is now defined among the exposed, without specifying the exact variant. If Assumption \ref{ASM: No eff exp} holds, then the two conditioning sets are equal, allowing for a causal interpretation of the contrast. Even under a less restrictive assumption than Assumption \ref{ASM: No eff exp}, $I(E^{a=1}\neq 0)=I(E^{a=0} \neq 0)$, the $CCE$ has a causal interpretation as a contrast of outcomes in the same population of individuals.

\subsection{Estimands under interventions on exposure}
\label{subsec: intervention estimands}
Causal effects on different variants can also be defined via interventions on both the treatment and the exposure, i.e., with respect to the potential outcome $Y^{a,e}$. These effects correspond to contrasts identified in a \textit{challenge trial} \citep{lambkin-williams_human_2018}, where participants are exposed to the infectious pathogens in a controlled manner.
Because we will consider these estimands conditional on baseline covariates $l \in \mathcal{L}$, we will first include $l$ in the definitions.

\begin{definition}[Effect with intervened exposure]\label{DEF: EIE}
    \begin{equation*}
        EIE(l) =\left. \frac{\mathbb{P}(Y^{a=1, e=1}=1|L=l)}{\mathbb{P}(Y^{a=0, e=1}=1|L=l)}\middle \slash \frac{\mathbb{P}(Y^{a=1, e=2}=2|L=l)}{\mathbb{P}(Y^{a=0, e=2}=2|L=l)}\right. .
    \end{equation*}
\end{definition}

The $CCS$, $CCE$ and the $EIE(l)$ compare 
relative effect in the treated versus the untreated. We could also define a comparative estimand with respect to outcomes in the treated only:

\begin{definition}[Effect of exposure under treatment]\label{DEF: EET}
    \begin{equation*}
        EET(l)=\frac{\mathbb{P}(Y^{a=1,e=1}=1|L=l)}{\mathbb{P}(Y^{a=1,e=2}=2|L=l)}.
    \end{equation*}
\end{definition}

When we generically discuss $EIE(l)$ and $EET(l)$, we will omit the $l$ argument whenever it is implied by the context. 

The various definitions of the causal contrasts illustrate that there is no unique way of quantifying heterogeneous (sieve) effects. Instead, investigators should ask themselves, what is the relevant definition of heterogeneity in their context, and choose the effect measure accordingly, as we further discuss in Subsection \ref{SEC: Estimand choice}. Furthermore, the investigators need to evaluate the assumptions needed to identify and estimate the estimands, as we described next (see also Table \ref{TAB: Causal estimands} for an overview).

\section{Identification of the causal estimands}\label{SEC: Identification}
\subsection{Identification of the $CECE$}
We first consider sufficient assumptions for the identification of the $CECE$. We invoke conventional exchangeability, positivity, and consistency assumptions, which can be ensured by design in a classical RCT.  
\begin{assumption}\label{ASM: Standard RCT}
    \begin{enumerate}[label=\theassumption\alph*, leftmargin=*]
        \item[]
        \item \label{ASM: Exchange}  Exchangeability: $Y^a, \, E^a \independent A \ \forall \ a \in \{0,1\}.$
        \item \label{ASM: Pos}  Positivity: $\mathbb{P}(A=a)>0 \ \forall \ a \in {0,1}.$
        \item \label{ASM: Cons}  Consistency: $\text{If } A=a \text{, then } E^a=E \text{ and } Y^a=Y \ \forall \ a \in \{0,1\}.$
    \end{enumerate}
\end{assumption}
While Assumption \ref{ASM: Standard RCT} is sufficient to identify the average treatment effect, $ATE(j)$, from the observed data, the relative $CECE(j)$ of variants 1 and 2 require conditioning on exposure $E$. If the exposure status of individuals is known,  the $CECE(j)$ can be expressed as a functional of the observed variables under Assumptions  \ref{ASM: no multiple exp}, \ref{ASM: No eff exp} and \ref{ASM: Standard RCT}. However, except for challenge trials, exposure status is usually unobserved in experiments. To identify the $CECE(j)$, we will introduce an additional assumption of exposure necessity. 
\begin{assumption}[Exposure necessity]\label{ASM: Exp nec} $
        E=0 \implies Y^a=0 \ \forall \ a \in\{0,1\}.
$
\end{assumption}
Assumption \ref{ASM: Exp nec} guarantees that those who were not exposed to the infectious agent, will not develop either of the outcomes.  In the running example on HIV, this means that individuals who were not exposed to either the matched or the unmatched variant, did not develop HIV infection.
Because we consider different variants, we invoke an assumption guaranteeing that infection by either variant can only occur if individuals are exposed to that variant.
\begin{assumption}[No cross-infectivity]\label{ASM: No cross}
        $$E^a=j \implies Y^a\neq i\ \forall \ a \in \{0,1\}, \, i \neq j, \, i,j \in \{1,2\}.$$
\end{assumption}

Assumption \ref{ASM: No cross} allows the identification of the relative $CECE$: 

\begin{proposition}\label{PROP: Relative CECE estimator}
    Under Assumptions \ref{ASM: no multiple exp}-\ref{ASM: No cross} the relative $CECE$ can be expressed as
    \begin{equation*}
        CECE(j)=\frac{\mathbb{P}(Y=j|A=1)}{\mathbb{P}(Y=j|A=0)} \ \forall \ j  \in \{1,2\}
            \text{ given that } \mathbb{P}(Y=j|A=0)>0.
    \end{equation*}
\end{proposition}

\subsection{Identification of the $CCS$ and the $CCE$}
It follows from Proposition \ref{PROP: Relative CECE estimator} that, under Assumptions \ref{ASM: no multiple exp}-\ref{ASM: No cross}, the $CCS$ and the $CCE$ can be identified 
\begin{equation}\label{EQ: CCS cond on spec exposure}
    CCS=  CCE=
        \left. \frac{\mathbb{P}(Y=1|A=1)}{\mathbb{P}(Y=1|A=0)}\middle \slash
        \frac{\mathbb{P}(Y=2|A=1)}{\mathbb{P}(Y=2|A=0)}\right.
\end{equation}
    given that $\mathbb{P}(Y=1|A=0)>0$ and $ \mathbb{P}(Y=2|A=1)>0$.
    The proof of Proposition \ref{EQ: CCS cond on spec exposure} and all the following identification proofs are given in Appendices \ref{APP: Prop relative CECE estimator}-\ref{APP: Marginal EIE}.

If the relative $CECE(j)$ is identified for each $j$, then the $CCS$ and the $CCE$ are identified. However, $CCS$ and the $CCE$ can also be identified under weaker assumptions, as presented next.

\subsection{Relaxation of the identification conditions of the $CCS$ and the $CCE$}\label{SEC: relaxation of ind conds}
Assumption \ref{ASM: Standard RCT} is standard in the causal inference literature and would hold by design in a randomized experiment with iid data. Here we will consider relaxations of Assumptions \ref{ASM: No eff exp}, \ref{ASM: Exp nec}, and \ref{ASM: No cross}. These relaxations will allow identification of the $CCS$ and the $CCE$, even if the $CECE$ is unidentified.
\subsubsection{No effect on exposure}\label{subsec: no no eff on exp}
Assumption \ref{ASM: No eff exp} requires that $E^{a=1}=E^{a=0}$. Consider the following weaker assumption:
\begin{assumption}[No relative effect on exposure]\label{ASM: No eff on exp ratios}
$$
    \frac{\mathbb{P}(E^{a=1}=1)}{\mathbb{P}(E^{a=1}=2)}=
    \frac{\mathbb{P}(E^{a=0}=1)}{\mathbb{P}(E^{a=0}=2)} 
    .$$   
\end{assumption}

Assumption \ref{ASM: No eff on exp ratios} states that the relative ratio between the exposures to variant 1 and variant 2 does not change with the assigned treatment. Under Assumption \ref{ASM: Standard RCT}, Assumption \ref{ASM: No eff on exp ratios} can be equivalently formulated as 
    $$\frac{\mathbb{P}(E=1|A=1)}{\mathbb{P}(E=2|A=1)}=
    \frac{\mathbb{P}(E=1|A=0)}{\mathbb{P}(E=2|A=0)}.$$    
In particular, Assumption \ref{ASM: No eff on exp ratios} does not require successful blinding as, e.g., assumed by \citet{stensrud_identification_2023}. Successful blinding is likely to be broken, e.g., when individuals more carefully adhere to protective behaviors under no treatment compared to vaccine assignment. The relaxed assumption would allow for such behaviors so long as the relative risk of exposure is constant across the two variants. This assumption could be plausible, e.g., when unvaccinated individuals reduce their number of social contacts and are careful about social distancing. Then the assigned treatment might influence the overall exposure levels, but likely not the ratio between the two variants.

\begin{proposition}\label{PROP: No eff on exp ratios}
    Under Assumptions \ref{ASM: no multiple exp} and  \ref{ASM: Standard RCT}-\ref{ASM: No eff on exp ratios} 
    the $CCS$ is identified as 
    \begin{equation}
    \begin{aligned}\label{EQ: CCS ind exp ratio}
       CCS&=\left. \frac{\mathbb{P}(Y=1|A=1)}{\mathbb{P}(Y=1|A=0)} \middle \slash
        \frac{\mathbb{P}(Y=2|A=1)}{\mathbb{P}(Y=2|A=0)} \right. .
    \end{aligned}
    \end{equation}
\end{proposition}

\subsubsection{Exposure necessity and No cross infectivity}\label{Subsec: no nec, no no cross}
Assumptions \ref{ASM: Exp nec} and \ref{ASM: No cross} guarantee that
$$ATEr\overset{\ref{ASM: Exp nec}}{=}CCE\overset{\ref{ASM: No cross}}{=}CCS,$$
where $ATEr=ATE(1)\slash ATE(2)$. Thus, Assumption \ref{ASM: Exp nec} ensures that the identification formula for the $CCE$ defined in Table \ref{TAB: Causal estimands} not only identifies the ratio of the two $ATE$s, but it is interpretable as the ratio of the expected potential outcomes, conditional on the non-zero status of the exposure.

If Assumption \ref{ASM: No cross} further holds, then the same identification formula can be used to identify the ratio of the conditional outcomes under intervention conditional on the specific exposure that corresponds to the outcome.

\subsection{Interpretation and identification of the $EET$ and the $EIE$}\label{SEC: EET}
The $EET$ and the $EIE$ require interventions on exposure, unlike the $CCS$ and the $CCE$. Because the exposure in the $EET$ and the $EIE$ is controlled, the ratio is insensitive to exposure patterns in the observed population. Therefore the estimated quantities can directly be generalized from a trial to the target population, as opposed to the conventional vaccine effect measures, that suffer from the confounding by the different interference structures. Both the $EET$ and the $EIE$ can be identified in a challenge trial \citep{lambkin-williams_human_2018}, where the exposure of the individuals to the infectious agents is controlled by the investigators. In practice, it is difficult to conduct such trials because of ethical considerations. Thus we focus on identifying the $EET$ and the $EIE$  from standard RCT data which requires stronger assumptions. Sufficient conditions and identifying formulas are listed in Table \ref{TAB: Causal estimands}, and these assumptions are described in more detail in Appendix \ref{APP: EET and EIE}. Furthermore, we give results on $EIE$ and $EET$ marginalized over $l$ in Appendices \ref{APP: Proof EET marg} and \ref{APP: Marginal EIE}.

\section{On the choice of estimand}\label{SEC: Estimand choice}
The $ATEr$ is identifiable from standard RCT data, without further assumptions, but its relevance to the target population is often questionable due to interference. The $CCS$, $CCE$, $EIE$, and $EET$ are defined by conditioning or intervening on the exposure of individuals (See Appendix \ref{APP: interference} for details). Thus, these estimands are insensitive to interference in the target population.  However, these four estimands have different interpretations and require additional assumptions for identification. Here we suggest more details on the choice of estimand, which is context-dependent. 

\subsection{$CCS$ and $CCE$}\label{subsec: secs choice}
The $CCS$ quantifies the ratio of relative treatment effect on each of the two competing variants ($Y=1$ vs $Y=2$). In the context of the HIV1 example of Subsection \ref{subsec: running example hiv}, the $CCS$ compares the risk of infection with the matched variant ($Y=1$) and the unmatched variant ($Y=2$), among those who were exposed to the matched ($E=1$) and the unmatched ($E=2$) variants, respectively. However, in general, $CCS \neq 1$ would not necessarily mean that the vaccine has a more beneficial effect against one variant than the other because the conditioning sets differ; the individuals in the numerators and denominators might have different characteristics, which affect the risk of infection. Yet, we can conclude that the effect in the subpopulation of the exposed, $E=1$, is different from the effect among the other subpopulation, $E=2$.

In other words, the $CCS$ is not necessarily a causal effect, in the sense that it is defined by contrasting potential outcomes in different subpopulations, characterized by $I(E^{a=0}=1), I(E^{a=1}=1), I(E^{a=0}=2)$ and $ I(E^{a=1}=2)$. Assumption \ref{ASM: No eff exp} ensures that $E^{a=1}=E^{a=0}$. Assumption \ref{ASM: no multiple exp} states that being exposed to both of the variants has probability zero, hence there exists no subpopulation for which the four conditioning sets used in the $CCS$ are identical.

The $CCS$ is relevant when investigators are interested in whether the vaccine effects differ across subgroups of individuals, described by those exposed to exactly one of the two variants. In our running example on HIV, the identifying assumptions of the $CCS$ seem to be plausible. The short duration and the low prevalence of HIV justifies Assumption \ref{ASM: no multiple exp}. Assumption \ref{ASM: Standard RCT} holds by design. Assumptions \ref{ASM: Exp nec} and \ref{ASM: No cross} are expected to hold, as without being in contact with the infectious agent, HIV cannot be developed, and the variant of exposure determines the type of outcome. The plausibility of Assumption \ref{ASM: No eff on exp ratios} in RCTs even if blinding is broken is discussed in Subsection \ref{subsec: no no eff on exp}.

Unlike the $CCS$, the $CCE$ is defined conditional on any exposure $(E \neq 0)$. This ensures that the numerator and denominator have the same conditioning sets. To fix ideas about the interpretation of the $CCE$, consider the HIV1 trial, where the $CCE$ is the contrast of relative risks of infection with the two variants ($Y=1$ vs $Y=2$) among those who were exposed to either variant $(E \neq 0)$. Under Assumption \ref{ASM: No cross}, individuals who developed outcome $Y=j$ among the exposed ($E\neq 0$) were exposed to variant $E=j$. Indeed, Assumption \ref{ASM: No cross} guarantees that the identification formulas for the $CCE$ and the $CCS$ are identical. The identifying assumptions for the $CCS$ are a strict superset of the ones that are sufficient for the identification of the $CCE$.

Both the $CCS$ and the $CCE$ are estimands conditional on exposure, in the sense that they are defined in the subset of individuals who would be exposed in a given trial. This subset is context-dependent: it is possible that different people with different characteristics would be exposed at other times and locations. The context-dependence might threaten the generalizability of the $CCS$ and $CCE$ from the trial to the larger target population; the $CCS$ and $CCE$ would be equivalent in the target population if the characteristics of exposed individuals are exchangeable in the trial data and the target population.

In contrast, the $EIE$ and the $EET$ are defined via an intervention on the exposure, that does not depend on the environment in which these estimands are estimated. However, the $EIE$ and the $EET$ require stronger identification conditions compared to the $CCS$ and the $CCE$, and their interpretations are different as well, which we discuss in the next subsections.

\subsection{$EIE$}
To illustrate the relevance of the $EIE$, consider two competing HIV1 variants that are present at distinct locations at a particular point in time. Suppose also that individuals at the two locations have different distributions of baseline covariates $l$. In the future, however, it is likely that the variants will spread to different areas. A decision maker is interested in whether the vaccine effect differs across variants, potentially conditional on $l$. The $EIE$ will give a concrete answer to this question.

The HIV1 example is not contrived, as different variants of various infectious agents are often present at different locations. For example, \citet{castillo_geographical_2020} discussed the geographical distribution of the variants of the SARS-CoV-2 in Chile. They found two regions where the proportion of variants was dominated by variant S, while in the rest of the regions, the most prevalent one was variant G. In the case of two available vaccines, with the possibly qualitatively different $EIE$s with respect to these two variants, e.g., could have guided the distribution of these two hypothetical vaccines.

\subsection{$EET$}
Unlike the other estimands introduced so far, the $EET$ only quantifies outcomes in vaccinated individuals ($a=1$).   Thus, the $EET$ is particularly relevant when the vaccinated group is of special interest, like in settings where vaccination is compulsory. For example,  yellow fever vaccination is required for visa applications to various countries. 

The identification assumptions for both the $EET$ and the $EIE$ require the measurement of baseline variables to adjust for unmeasured confounding between exposure ($E$), treatment ($A$), and outcome ($Y$). However, under certain assumptions (Assumptions \ref{ASM: no multiple exp}, \ref{ASM: Exp nec}-\ref{ASM: No eff on exp ratios}, \ref{ASM: Generalised exposure RCT} and with $L=\emptyset$), the $EIE$ can also be identified by the conventionally reported vaccine estimand, as defined in Equation \eqref{EQ: CCS cond on spec exposure}. The $EET$ could be identified with the same functional under
Assumptions \ref{ASM: no multiple exp}, \ref{ASM: Exp nec}-\ref{ASM: No eff on exp ratios}, \ref{ASM: Exposure RCT}, \ref{ASM: equal inf untrt} and with $L=\emptyset$, see Appendix \ref{APP: EET and EIE}.

Our considerations illustrate that the estimands have different interpretations and require different assumptions. We have given explicit assumptions ensuring that these (relative) estimands equal a conventional relative estimand. Had we considered estimands on the additive scale, then the same equalities would not in general hold, see Appendix \ref{APP: Absolute measures}.

\section{Time-to-event estimands}\label{SEC: TTE estimands}
Many RCTs and observational studies produce longitudinal data, where events are recorded over time. Here we present time-to-event estimands and further extend the results from Section \ref{SEC: Identification}. The extensions are non-trivial, particularly because we need to consider (unmeasured) exposures and outcomes that vary over time.  

\subsection{Preliminaries}\label{SUBSEC: TTE Notation}
Let $E_k$ denote whether the individual was exposed to variant $j$ at time $k$. In contrast, let $Y_k$ indicate whether an individual has experienced the outcome $j$ by time $k$, that is $Y_k=j$ if the event $j$ has occurred at time $k$ or at any time $k'<k$, for $j \in \{1,2\}$. 
In particular, the exposure can change from time $k$ to $k+1$. For example, an individual can be exposed to variant 1 at time $k$, and then be exposed to variant 2, or be unexposed, at time $k'$, for $k'>k$.  We also consider the further (and simpler) generalization to handle censoring in Appendix \ref{APP: Censoring}. As we study the time to the first event in each individual, we arbitrarily set future exposures to zero after an event has occurred: formally $E_k\cdot Y_{k'} =0$ for all $k' \neq k$.

Consistent with the conventional causal inference literature \citep{robins_causal_1997, hernan_causal_2023}, consider discrete time intervals $k= 0, \dots, K$, and the random variables indexed with negative subscripts are considered to be equal to 0. The history of the random variables, through the current interval $k$, are denoted by an overline, for example, $\overline{Y}_k=(Y_0, \dots, Y_k)$.

\begin{figure}
        \centering
        \begin{tikzpicture}
            \tikzset{line width=1.5pt, outer sep=0pt,
            ell/.style={draw,fill=white, inner sep=2pt,
            line width=1.5pt},
            swig vsplit={gap=5pt,
            inner line width right=0.5pt},
            swig hsplit={gap=5pt}
            };
                \node[name=E1,shape=swig vsplit] at (8,-2){
                \nodepart{left}{$E_1$}
                \nodepart{right}{$e_1$}
                };
                \node[name=E2,shape=swig vsplit] at (12,-2){
                \nodepart{left}{$E_2^{e_1}$}
                \nodepart{right}{$e_2$}
                };
                \node[name=A,shape=swig vsplit] at (4,0) {
                                                  \nodepart{left}{$A$}
                                                  \nodepart{right}{$a$} };
                \node[name=Y1, ell, shape=ellipse] at (8,0){$Y_1^{a, e_1}$};
                \node[name=Y2, ell, shape=ellipse] at (12,0){$Y_2^{a, e_1, e_2}$};
                \node[name=U, ell, shape=ellipse] at (8,2){$U$};
                \node[name=Z, ell, shape=ellipse] at (12,2) {$Z$};
                \node[name=W, ell, shape=ellipse] at (10,-4){$W$};
            \begin{scope}[>={Stealth[black]},
              every node/.style={fill=white,circle},
              every edge/.style={draw=black,very thick}]
                \path[->] (U) edge (Y1);
                \path[->] (U) edge (Y2);
                \path[->] (U) edge[bend right=45] (E1);
                \path[->] (U) edge (E2);
                \path[->] (A) edge (Y1);
                \path[->] (A) edge[bend right=20] (Y2);
                \path[->] (E1.70) edge (Y1);
                \path[->] (E1) edge (E2);
                \path[->] (E2.70) edge (Y2);
                \path[->] (Z) edge (Y1);
                \path[->] (Z) edge (Y2);
                \path[->] (Y1) edge (Y2);
                \path[->] (Y1) edge (E2);
                \path[->] (W) edge (E1.220);
                \path[->] (W) edge (E2);
            \end{scope}
        \end{tikzpicture}
        \caption{SWIG describing a setting with two time points.}
        \label{FIG: Weak SWIG}
    \end{figure}
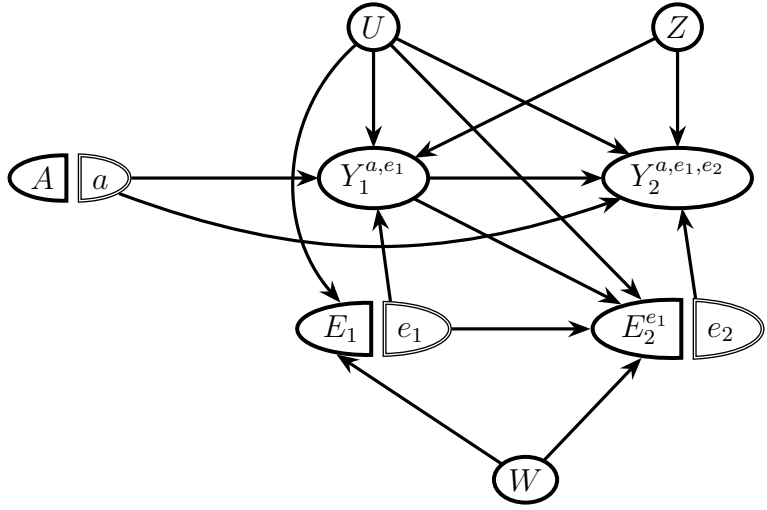

Let $U$ denote the common causes of the outcomes and the exposures, let $Z$ denote the common cause of the outcomes only, and let $W$ denote the common cause of the exposures, as illustrated by the Single World Intervention Graph (SWIG) \citep{richardson_single_2013} in Figure \ref{FIG: Weak SWIG}. Here, $U$, $Z$ and $W$ might be unmeasured.

To motivate the extension to multiple exposures over time, consider the HIV1 example and suppose that at time $k$ sufficient contact for infection was made between a susceptible ($Y_k=0$) individual and someone infected by the matched variant ($Y_k=1$), e.g. needle sharing during injection drug use \citep{patel_estimating_2014}. It is plausible that a second sufficient contact follows at time $k'>k$, with someone who is infected by the unmatched strain ($Y_{k'}=2$).

\subsection{Challenge subtype effect}\label{SUBSEC: CSE}
 The challenge subtype effect quantifies the differential effect of the vaccine on the two competing variants in a setting where the exposure is administered at a particular time:
\begin{definition}[Challenge subtype effect]\label{DEF: CSE}
\begin{equation*}
    CSE_k=\left. \frac{\bbP(Y_k^{a=1, e_k=1, \overline{e}_{k-1}=0}=1)}{\bbP(Y_k^{a=0, e_k=1, \overline{e}_{k-1}=0}=1)}
    \middle \slash \frac{\bbP(Y_k^{a=1, e_k=2, \overline{e}_{k-1}=0}=2)}{\bbP(Y_k^{a=0, e_k=2, \overline{e}_{k-1}=0}=2)} \right. 
\end{equation*}
for all $ \ k \in \{1, \dots, K\}.$
\end{definition}
The $CSE_k$ is insensitive to interference by, e.g., mixing patterns, as it is defined under (time-varying) interventions on exposure, $E_k$. We also consider an additional time-to-event estimand in Appendix \ref{SUBSEC: One-shot}, which can be conceptualized as a straight-forward extension of the time-fixed estimand $CCE$ to the time-to-event data.

\section{Time-to-event identification}\label{SEC: TTE ident}
Consider the following assumptions, which generalize Assumption \ref{ASM: Standard RCT} and hold by design in a properly executed randomized trial.

\begin{assumption}\label{ASM: TTE Standard RCT}
    \begin{enumerate}[label=\theassumption\alph*, leftmargin=*]
        \item[]
        \item \label{ASM: TTE EXCH} TTE Exchangeability: $\overline{E}_K^{a} \independent A|U, Z; \; \overline{Y}_K^{a, \overline{e}_K} \independent A| U, Z; \newline \text{ and }
        \overline{Y}_K^{a, \overline{e}_K} \independent \overline{E}_K^{a}|Y_{K-1}^{a, \overline{e}_{K-1}}, U, Z, A.$
        \item \label{ASM: TTE Pos} TTE Positivity: $\bbP(A=a, E_k=j| U=u, Z=z)>0 \\ \forall \ a \in \{0,1\}, j \in \{0,1,2\}, u \in \mathcal{U}, z \in \mathcal{Z}.$
        \item \label{ASM: TTE Cons} TTE Consistency: $\text{if } A=a \text{ then } \overline{E}_k^{a}=\overline{E}_k; \;
        \text{if } A=a \text{ and } \overline{E}_k=\overline{e}_k \text{ then } \overline{Y}_k^{a, \overline{e}_k}=\overline{Y}_k.$
    \end{enumerate}
\end{assumption}

We also generalize Assumptions \ref{ASM: Exp nec} and \ref{ASM: No cross} to the time-to-event setting.

\begin{assumption}[TTE exposure necessity]\label{ASM: TTE Exp nec}
    \begin{enumerate}[label=\theassumption\alph*, leftmargin=*]
        \item[]
        \item \label{ASM: TTE original exp nec} $Y_k^{a, e_k=0}=0 \ \forall \ k, \ a \in \{0,1\}.$
        \item \label{ASM: TTE ignorable non-exposure} $I(\overline{E}_{k-1}=0) \independent Y_k|Y_{k-1}=0, A, E_k, U, Z \, .$
    \end{enumerate}
\end{assumption}
 
\begin{remark}
    If Assumption \ref{ASM: TTE Cons} and Assumption \ref{ASM: TTE original exp nec} hold and
    $
    E^a_k=0
    $
    then\\
    $0=Y^{a, e_k=0}_k=Y^a_k$
    for all $a \in \{0,1\}$.
\end{remark}

Under Assumption \ref{ASM: TTE ignorable non-exposure}, the history of non-exposure until time $k-1$ is independent of the outcome at time $k$, given the exposure at time $k$, the history of no outcome until time $k-1$ and the value of the possibly unmeasured variables $U$ and $Z$.

\begin{assumption}[TTE no cross-infectivity]\label{ASM: TTE No cross}

    \begin{equation*}
        Y_k^{a, e_k=j}\neq i \  \forall \ k, \ a \in \{0,1\}, \, i \neq j, \, i,j \in \{1,2\}.
    \end{equation*}
\end{assumption}

\begin{remark}
     If Assumption \ref{ASM: TTE Cons} and Assumption \ref{ASM: TTE No cross} hold and
    $
    E^a_k=j,
    $
    then\\ 
    $i \neq Y^{a, e_k=j}_k=Y^a_k$,
    that is,
    $
    E_k^a=j \implies Y_k^{a} \neq i
    $
    for all $a \in \{0,1\}\, i \neq j, \, i,j \in \{1,2\}$.
\end{remark}

Furthermore, consider the following assumption that describes how exposures to different variants are related. 

\begin{assumption}[Exposure ratio of the exposed]\label{ASM: TTE exposure rat of exposed}
    \begin{equation*}
        \bbP(E_k=1|E_k\neq 0, U=u, Z=z)= \alpha_k \bbP(E_k=2|E_k\neq0, U=u, Z=z)
    \end{equation*}
    for all $k$, $u \in \mathcal{U}$ and $z \in \mathcal{Z}$, with $\alpha_k$ being constant across the values $u$ and $z$.
\end{assumption}
Assumption \ref{ASM: TTE exposure rat of exposed} guarantees that, among those exposed to any of the variants, the ratio of probabilities of being exposed to variant 1 and variant 2 is constant in $u$ and $z$. For example, let the variables $U$ and $Z$ represent age and urban vs non-urban residency respectively, both of which can be associated with the social activity of an individual. Suppose that socially active people have a higher chance of exposure compared to those who limit social interactions. Then we assume that both the socially active and inactive have the same \textit{relative} rate of being exposed to variant 1 and variant 2. However, we do allow that the relative prevalence of the two variants might change over time and, thus, we also allow that the relative exposure ratios change with $k$, see the Single World Intervention Graph (SWIG) in Figure \ref{FIG: Weak SWIG} as an example, consistent with Assumptions \ref{ASM: TTE Standard RCT}-\ref{ASM: TTE exposure rat of exposed}.  Assumption \ref{ASM: TTE exposure rat of exposed} can fail if either $U$ and $Z$ are associated with geographical regions where variants appear with different prevalence \citep{castillo_geographical_2020}.

\begin{proposition}[Identifiability of $CSE_k$ conditional on $U$ and $Z$]\label{PROP: CSE ident given U}
    Under Assumptions \ref{ASM: TTE Standard RCT}-\ref{ASM: TTE exposure rat of exposed},
    \begin{align*}
    &CSE_k(u,z) \\ 
    & = \left. \frac{\bbP(Y_k^{a=1, e_k=1, \overline{e}_{k-1}=0}=1|U=u, Z=z)}{\bbP(Y_k^{a=0, e_k=1, \overline{e}_{k-1}=0}=1|U=u, Z=z)}
     \middle \slash \frac{\bbP(Y_k^{a=1, e_k=2, \overline{e}_{k-1}=0}=2|U=u, Z=z)}{\bbP(Y_k^{a=0, e_k=2, \overline{e}_{k-1}=0}=2|U=u, Z=z)} \right.\\
    &=\left. \frac{\bbP(Y_k=1|Y_{k-1}=0, U=u, Z=z, A=1)}{\bbP(Y_k=2|Y_{k-1}=0, U=u, Z=z, A=1)} \middle \slash \frac{\bbP(Y_k=1|Y_{k-1}=0, U=u, Z=z, A=0)}{\bbP(Y_k=2|Y_{k-1}=0, U=u, Z=z, A=0)} \right.
    \end{align*}
    for all $k$, $u \in \mathcal{U}$ and $z \in \mathcal{Z}$.
\end{proposition}

Proposition \ref{PROP: CSE ident given U} gives an identification formula of the $CSE_k(u,z)$ based on factual variables. However, $U$ and $Z$ capture all common causes of the time-variant outcomes and exposures or outcomes only, that may, e.g., include social activity, age, sex, general health status, socio-economic status, and genetic factors. All of these factors are usually impossible to measure in an RCT. To identify the population $CSE_k$, without measuring $U$ and $Z$, consider the following assumption.
\begin{assumption}[Scaled new infection]\label{ASM: TTE Scaled new inf}
    \begin{equation*}
        \bbP(Y^{a, e_k=1}_k=1|Y^{a, e_k=1}_{k-1}=0, U=u, Z=z)= \gamma_{a,k} \bbP(Y^{a, e_k=2}_k=2|Y^{a, e_k=2}_{k-1}=0, U=u, Z=z)
    \end{equation*}
    for all $k$, $u \in \mathcal{U}$ and $z \in \mathcal{Z}$, where we define $\gamma_{a, k} = 0$ whenever\\
 $\bbP(Y^{a, e_k=2}_k=2|Y^{a, e_k=2}_{k-1}=0, U=u, Z=z)=0$.
\end{assumption}

Assumption \ref{ASM: TTE Scaled new inf} is encoding a particular heterogeneity in infectivity on the ratio scale. This assumption will be plausible if the infectious pathways are similar for the two variants across all the subpopulations defined by the different values of $U$ and $Z$. A falsification test of Assumption \ref{ASM: TTE Scaled new inf}, based on the observed subset of $U$ and $Z$, is presented in Appendix \ref{APP sub: comment on testing scaled inf}.
\begin{remark}
Assumption \ref{ASM: TTE Scaled new inf} can equivalently be formulated such that the conditioning set includes the history of the outcome variable under interventions $E_k=j$ and $A=a$, until time $k-1$,
    $$
        \bbP(Y^{a, e_k=1}_k=1|\overline{Y}^{a, e_k=1}_{k-1}=0, U=u, Z=z)= \gamma_{a,k} \bbP(Y^{a, e_k=2}_k=2|\overline{Y}^{a, e_k=2}_{k-1}=0, U=u, Z=z)
    $$
    for all $k$, $u \in \mathcal{U}$ and $z \in \mathcal{Z}$. Under the FFRCISTG model, factorizing according to the SWIG in Figure \ref{FIG: Weak SWIG}, Assumption \ref{ASM: TTE Scaled new inf} can be rewritten using minimal labeling as 
    $$
        \bbP(Y^{a, e_k=1}_k=1|Y^{a}_{k-1}=0, U=u, Z=z)= \gamma_{a,k} \bbP(Y^{a, e_k=2}_k=2|Y^{a}_{k-1}=0, U=u, Z=z).
    $$
\end{remark}

The next corollary and proposition give convenient identification results for $CSE_k$. 

\begin{corollary}[General $CSE_k$]\label{PROP: all across CSE}
    Under Assumptions \ref{ASM: TTE Standard RCT}-\ref{ASM: TTE Scaled new inf}, the $CSE_k(u,z)= \gamma_{1,k}/\gamma_{0,k}$, for all $u \in \mathcal{U}, z \in \mathcal{Z}$ and $k \in \{1, \dots, K\}$, and the $CSE_k=\gamma_{1,k}/\gamma_{0,k}$ for each $k \in \{1, \dots, K\}$.
\end{corollary}

Using the equality between the marginal and the conditional $CSE_k$s the marginal $CSE_k$ can be identified as a function of observed variables.

\begin{proposition}\label{PROP: marginal CSE ident}
    Under Assumptions \ref{ASM: TTE Standard RCT}-\ref{ASM: TTE Scaled new inf} the marginal $CSE_k$ can be identified as 
    \begin{equation*}
        CSE_k=\left. \frac{\mathbb{P}(Y_k=1|Y_{k-1}=0, A=1)}{\mathbb{P}(Y_k=1|Y_{k-1}=0, A=0)}  \middle \slash \frac{\mathbb{P}(Y_k=2|Y_{k-1}=0, A=1)}{\mathbb{P}(Y_k=2|Y_{k-1}=0, A=0)}\right .
    \end{equation*}
    $\forall \ k \in \{1, \dots, K\}.$ 
\end{proposition}

\subsection{On testing of a null effect}
Consider the sharp null hypothesis that the vaccine has the same effect on developing the two types of outcomes at time $k$, under controlled exposure conditions:
\begin{equation}\label{EQ: sharp CSE null}
    \begin{aligned}
    &H_0^k:
            &&\quad I(Y_{k, i}^{a=1, e_k=1, \overline{e}_{k-1}=0}=1)-I(Y_{k, i}^{a=0, e_k=1, \overline{e}_{k-1}=0}=1)\\
            &
            &&=I(Y_{k, i}^{a=1, e_k=2, \overline{e}_{k-1}=0}=2)-I(Y_{k, i}^{a=0, e_k=2, \overline{e}_{k-1}=0}=2) \ \forall \ i.
    \end{aligned}
\end{equation}

Alternatively, consider the stronger sharp null hypothesis

\begin{equation}\label{EQ: Strong sharp null CSE}
    \begin{aligned}
    &H_0^{k, strong}:
            &&\quad I(Y_{k, i}^{a, e_k=1, \overline{e}_{k-1}=0}=1)=I(Y_{k, i}^{a, e_k=2, \overline{e}_{k-1}=0}=2) \ \forall \ a \in \{0,1\}, \ \forall \ i.
    \end{aligned}
\end{equation}

Suppose that an individual had their first exposure $j$ at time $k$. Suppose that this exposure led to infection with variant $j$. Then, $H_0^{k, strong}$ implies that, exposure to variant $j' \neq j$ would lead to infection with variant $j'$.  It is easy to show that
$H_0^{k, strong}$ implies $H_0^k$. The distinction between  $H_0^{k, strong}$ and $H_0^k$ can be illustrated by an individual with characteristics: \\
\begin{tabular}{|c|c|c|c|}
\hline
$Y_{k,i}^{a=1, e_k=1, \overline{e}_{k-1}=0}=1$ & 
$Y_{k,i}^{a=0, e_k=1, \overline{e}_{k-1}=0}=1$ & 
$Y_{k,i}^{a=1, e_k=2, \overline{e}_{k-1}=0}=0$ &
$Y_{k,i}^{a=0, e_k=2, \overline{e}_{k-1}=0}=0$ \\
\hline
\end{tabular}
\\
This individual is doomed if exposed to variant 1, but immune to variant 2, consistent with $H_0^k$ but not $H_0^{k, strong}$. Under $H_0^k$ Assumption \ref{ASM: TTE Scaled new inf}, the potential outcome probabilities under different treatment interventions are related:  

\begin{proposition}[Proportionality under the null]\label{PROP: Prop hazards under null}
    When  $H_0^k$ and Assumption \ref{ASM: TTE Scaled new inf} hold for $a=0$ for all $k$, then Assumption \ref{ASM: TTE Scaled new inf} holds for $a=1$. Analogously under $H_0^k$ and Assumption \ref{ASM: TTE Scaled new inf} for $a=1$, Assumption \ref{ASM: TTE Scaled new inf} is implied for $a=0$.
\end{proposition}

A practical implication of Proposition \ref{PROP: Prop hazards under null} is that by assuming proportional potential outcomes amongst the untreated, and Assumptions \ref{ASM: TTE Standard RCT}-\ref{ASM: TTE exposure rat of exposed}, under the null hypothesis $H_0^k$, $\gamma_{1,k} /\gamma_{0,k}$ is identified as
\begin{equation}\label{EQ: hazards as gammas}
1=    \frac{\gamma_{1,k}}{\gamma_{0,k}}= \left. \frac{\bbP(Y_k=1|Y_{k-1}=0, A=1)}{\bbP(Y_k=2|Y_{k-1}=0, A=1)} \middle \slash \frac{\bbP(Y_k=1|Y_{k-1}=0, A=0)}{\bbP(Y_k=2|Y_{k-1}=0, A=0)} \right. .
\end{equation}

Under the strong null hypothesis $H_0^{k, strong}$, we have that Assumption \ref{ASM: TTE Scaled new inf} holds, with the particular value of $\gamma_{a,k}=1$.

\begin{lemma}\label{LEM: H0 strong means scaled hazard}
Under the strong sharp null hypothesis \eqref{EQ: Strong sharp null CSE} and Assumption \ref{ASM: TTE Exp nec}, Assumption \ref{ASM: TTE Scaled new inf} is implied with the particular value of $\gamma_{a,k}=1$ for all $k$ and $a \in \{0,1\}$.
\end{lemma}

By Lemma \ref{LEM: H0 strong means scaled hazard} and Proposition \ref{PROP: Prop hazards under null} the fraction of the probabilities of the factual variables in \eqref{EQ: hazards as gammas} can be used as a valid test of the strong null hypothesis $H_0^{k, strong}$. Under Assumption \ref{ASM: TTE Scaled new inf}, then the marginal $CSE_k$ can also be interpreted as the estimated challenge subtype effect, even under the alternative hypothesis.

\section{Testing and estimation}\label{SEC: Estimation}
\subsection{Time-fixed estimation}
The identifying formulas in Table \ref{TAB: Causal estimands} are all ratios of estimable conditional probabilities. Denote the  estimators for conditional probabilities $\mathbb{P}(Y=j|A=a)$ and $\mathbb{P}(Y=j|A=a, L=l)$ by $\hat{\tau}_j(a)$ and $\hat{\tau}_j(a, l)$, respectively. Furthermore, let $\hat{\tau}_j^*(a)$ and $\hat{\tau}_j^*(a,l)$ be the estimators of $\mathbb{P}(Y=j|A=a, E=j)$ and $\mathbb{P}(Y=j|A=a, E=j, L=l)$ for all $j \in \{1,2\}$ and $a \in \{0,1\}$, respectively. Suppose first that we have unbiased estimators of these quantities, for example, empirical means of indicator functions. Then, under Assumptions \ref{ASM: Standard RCT} the $CCS$ can be estimated by 
$$
\mathbf{\hat{T}^*}=\left. \frac{\hat{\tau}_1^*(1)}{\hat{\tau}_1^*(0)} \middle \slash \frac{\hat{\tau}^*_2(1)}{\hat{\tau}_2^*(0)} \right. .
$$
However, the $\hat{\tau}^*$-s cannot be calculated in most practical settings because $E$ is unobserved.
If we further impose Assumptions \ref{ASM: Exp nec}-\ref{ASM: No eff on exp ratios}, then 
$$
\mathbf{\hat{T}}=\left. \frac{\hat{\tau}_1(1)}{\hat{\tau}_1(0)} \middle \slash \frac{\hat{\tau}_2(1)}{\hat{\tau}_2(0)} \right.
$$
is an asymptotically unbiased estimator of the $CCS$. Estimators for the causal estimands in Table \ref{TAB: Causal estimands} follow analogously, and the corresponding estimators of the $EET$ are denoted as $\hat{T}_{EET}$ and $\hat{T}_{EET}^*$ respectively. Estimated confidence intervals for the $CCS$, the $CCE$, and the $EIE$ are can, e.g., be derived using the log-normal approximation discussed by \citet{katz_obtaining_1978}, and for the $EET$ the confidence intervals are calculated based on the work of \citet{nelson_statistical_1972}.

We provide the details of these derivations in Appendix \ref{APP: Var calc}, and the properties of the estimators are illustrated in simulation studies in Appendix \ref{APP: Simulation}.

\subsection{Time-to-event estimation}
The identification formula of the marginal $CSE_k$ in Proposition \ref{PROP: marginal CSE ident} is a functional of two time-varying hazard ratios, one for each variant. These hazard ratios can, e.g., be estimated semi-parametrically using Cox regression \citep{cox_regression_1972}, or non-parametrically by sample means of the event rates among the \textit{at risk} population, see Appendix \ref{APP: Estimation} for details. 

Assumption \ref{ASM: TTE Scaled new inf} imposes strong proportionality conditions, conditional on the unobserved variables $U$ and $Z$. Let us consider a subset of these two variables denoted as $L \in \{U, Z\}$, that is observed by the investigators. Then under Assumptions \ref{ASM: TTE Standard RCT}-\ref{ASM: TTE Scaled new inf},
$$
\frac{\bbP(Y_k=1|Y_{k-1}=0, A=a, L=l_1)}{\bbP(Y_k=1|Y_{k-1}=0, A=a, L=l_2)}=
\frac{\bbP(Y_k=2|Y_{k-1}=0, A=a, L=l_1)}{\bbP(Y_k=2|Y_{k-1}=0, A=a, L=l_2)}
$$
for all $l_1, l_2 \in \mathcal{L}$. This equality can, e.g., be tested by inverting confidence intervals of the corresponding coefficients in two Cox proportional hazard models fitted to estimate the cause-specific hazards of outcomes with variant 1 and 2, respectively.   Details are provided in Appendix \ref{APP sub: comment on testing scaled inf}. The use of different estimators of the $CSE_k$ is further illustrated in Appendix \ref{APP sub: CSE comment}.

\section{Effects of the ALVAC/AIDSVAX vaccine on HIV1 variants}\label{SEC: real data}
We use publicly available data from \citet{benkeser_benkesersurvtmle_2017} based on the RV144 trial that studied the effect of the ALVAC/AIDSVAX vaccine on the risk of HIV1 infection \citep{rerks-ngarm_vaccination_2009}, as described in Subsection \ref{subsec: running example hiv}. For data privacy reasons, our individual-level dataset is synthetically constructed to mimic the outcomes in the original trial \citep{benkeser_benkesersurvtmle_2017}. Participants received placebo ($A=0$) or active vaccine ($A=1$). The outcome of interest is the presence of HIV infection matched ($Y=1$) or mismatched ($Y=2$) to the vaccine at the amino acid position 169.  We also had access to categorical baseline covariates $L$ encoding \textit{year of the enrollment}, \textit{sex}, \textit{age}, \textit{risk profile}. In our analysis, we assumed uninformative censoring (Assumption \ref{ASM: Indep censoring}, see Appendix \ref{APP: Censoring}).

\subsection{Time-fixed estimates}
We estimated the $CCS$ to be $0.423$ ($95\% $ CI: $[0.169,1.061]$)  based on Method C described by \citet{katz_obtaining_1978}.
Under the identification assumptions of the $EIE$ (Assumptions \ref{ASM: no multiple exp}, \ref{ASM: Exp nec}, \ref{ASM: No cross},\ref{ASM: Cond no effect on exp}, and \ref{ASM: Generalised exposure RCT}), which we suppose to hold conditional on the baseline covariate \textit{risk profile}, the $EIE$ is identified by the same functional as the $CCS$ in the \textit{High risk} and \textit{Non-high risk} populations. The corresponding estimates are $0.208$ ($95\% $ CI: $[0.045,0.956]$) and $0.811$ ($95\% $ CI: $[0.230, 2.860]$), respectively. We estimated effects under alternative assumptions in Appendix \ref{APP: data EIE vs EET}.

\subsection{Time-to-event estimates}
Based on the developments in Section \ref{SEC: Estimation}, we estimated the marginal $CSE_k$ using two marginal Cox models, for the two subtypes respectively, where individuals were censored when they either experienced infection with the competing subtype or early drop-out \citep{young_causal_2020}. That is,

\begin{equation*}
        \mathbb{P}(Y_k^{\overline{c}_k=0}=j|Y^{\overline{c}_{k-1}}_{k-1}=0, A=a)=\lambda_0(k)\cdot exp(A \cdot \beta^j_a) \ \forall \ k,  \ j \in \{1,2\},
\end{equation*}
where $Y^{c=0}$ denotes the potential outcomes under no censoring, see for more details Appendix \ref{APP: Censoring}.
Under this model, the identification formula of the $CSE_k$ simplifies to $exp(\beta_a^1)/exp(\beta_a^2)$, which was estimated to be $\widehat{CSE_k}=0.423 \,(95\% CI: [0.000, 0.696])$, where the 95\% confidence interval was estimated by non-parametric Bootstrap in 10,000 samples. Homogeneity in effects across strains corresponds to $CSE_k=1$. Thus, we conclude that there is heterogeneity, i.e., variant-specific effects.

We also used the conventional Nelson-Aalen estimator \citep{aalen_nonparametric_1978} to non-parametrically estimate the cumulative hazards of being infected with each variant. We can use these estimates to test a modified null hypothesis, as formally defined in Appendix \ref{APP sub: comment on testing scaled inf}.
The ratio of the cumulative hazards is estimated to be 0.230 (95\% CI: [0, 0.613]), and 0.427 (95\% CI: [0.000, 0.876]) in the first and the second half of the trial, respectively. 

\subsection{Remark on rare events}

The estimated $CCS$ in a time-fixed setting is $0.423$, and the estimated $CSE_k$ via Cox regression is $0.423$, for all $k \in \{1, \dots, K\}$, while the cumulative hazard using the non-parametric Nelson-Aalen estimator at the end of the follow-up period is estimated to be $0.427$. The similarity between the estimates is expected because of the small number of events; when events are rare, it is well-known that the hazard ratio approximates the risk ratio \citep{symons_hazard_2002}. For the details of alternative estimators of the sieve effect when events are rare, see Appendix \ref{APP sub: rare CI}.

\section{Discussion}\label{SEC: discussion}
We have formally defined various causal estimands that are relevant in studies of subtype-specific outcomes. Under explicit assumptions, we showed that the estimands can be identified by simple functionals in conventional RCTs. These formalizations clarify the interpretation of commonly estimated vaccine parameters used in large-scale randomized trials. It is practically important that our results can justify the use of conventional relative vaccine estimands, calculated under iid assumptions from RCT data, even if there is interference in the target populations. While we have given sufficient conditions for the identification of various estimands, alternative identification strategies might also be useful and plausible in certain settings. In future research, we will particularly study (sharp) bounds, i.e., partial identification under weaker identifiability assumptions.

\bibliographystyle{biometrika_mod.bst}
\bibliography{references}

\begin{thebibliography}{58}
\expandafter\ifx\csname natexlab\endcsname\relax\def\natexlab#1{#1}\fi

\bibitem[{noa(2023)}]{noauthor_volume_2023}
 (2023).
\newblock Volume 28, {Number} 3 {\textbar} {HIV} {Surveillance} {\textbar} {Reports} {\textbar} {Resource} {Library} {\textbar} {HIV}/{AIDS} {\textbar} {CDC}.

\bibitem[{Aalen(1978)}]{aalen_nonparametric_1978}
\textsc{Aalen, O.} (1978).
\newblock Nonparametric {Inference} for a {Family} of {Counting} {Processes}.
\newblock \textit{The Annals of Statistics} \textbf{6}, 701--726.

\bibitem[{Aronow(2012)}]{aronow_general_2012}
\textsc{Aronow, P.~M.} (2012).
\newblock A {General} {Method} for {Detecting} {Interference} {Between} {Units} in {Randomized} {Experiments}.
\newblock \textit{Sociological Methods \& Research} \textbf{41}, 3--16.

\bibitem[{Barouch(2008)}]{barouch_challenges_2008}
\textsc{Barouch, D.~H.} (2008).
\newblock Challenges in the development of an {HIV}-1 vaccine.
\newblock \textit{Nature} \textbf{455}, 613--619.

\bibitem[{Benkeser et~al.(2019)Benkeser, Gilbert \& Carone}]{benkeser_estimating_2019}
\textsc{Benkeser, D.}, \textsc{Gilbert, P.~B.}, \textsc{Carone, M.} (2019).
\newblock Estimating and {Testing} {Vaccine} {Sieve} {Effects} {Using} {Machine} {Learning}.
\newblock \textit{Journal of the American Statistical Association} \textbf{114}, 1038--1049.

\bibitem[{Benkeser \& Hejazi(2017)}]{benkeser_benkesersurvtmle_2017}
\textsc{Benkeser, D.}, \textsc{Hejazi, N.} (2017).
\newblock benkeser/survtmle: survtmle -- first {CRAN} release.

\bibitem[{Benkeser et~al.(2020)Benkeser, Juraska \& Gilbert}]{benkeser_assessing_2020}
\textsc{Benkeser, D.}, \textsc{Juraska, M.}, \textsc{Gilbert, P.~B.} (2020).
\newblock Assessing trends in vaccine efficacy by pathogen genetic distance.
\newblock \textit{Journal De La Societe Francaise De Statistique (2009)} \textbf{161}, 164--175.

\bibitem[{Buchbinder et~al.(2008)Buchbinder, Mehrotra, Duerr, Fitzgerald, Mogg, Li, Gilbert, Lama, Marmor, Rio, McElrath, Casimiro, Gottesdiener, Chodakewitz, Corey \& Robertson}]{buchbinder_efficacy_2008}
\textsc{Buchbinder, S.~P.}, \textsc{Mehrotra, D.~V.}, \textsc{Duerr, A.} et~al. (2008).
\newblock Efficacy assessment of a cell-mediated immunity {HIV}-1 vaccine (the {Step} {Study}): a double-blind, randomised, placebo-controlled, test-of-concept trial.
\newblock \textit{Lancet} \textbf{372}, 1881--1893.

\bibitem[{Castillo et~al.(2020)Castillo, Parra, Tapia, Lagos, Arata, Acevedo, Andrade, Leal, Tambley, Bustos, Fasce \& Fernández}]{castillo_geographical_2020}
\textsc{Castillo, A.~E.}, \textsc{Parra, B.}, \textsc{Tapia, P.} et~al. (2020).
\newblock Geographical {Distribution} of {Genetic} {Variants} and {Lineages} of {SARS}-{CoV}-2 in {Chile}.
\newblock \textit{Frontiers in Public Health} \textbf{8}, 562615.

\bibitem[{Chang et~al.(1997)Chang, Chen, Lai, Hsu, Wu, Kong, Liang, Shau \& Chen}]{chang_universal_1997}
\textsc{Chang, M.-H.}, \textsc{Chen, C.-J.}, \textsc{Lai, M.-S.} et~al. (1997).
\newblock Universal {Hepatitis} {B} {Vaccination} in {Taiwan} and the {Incidence} of {Hepatocellular} {Carcinoma} in {Children}.
\newblock \textit{New England Journal of Medicine} \textbf{336}, 1855--1859.

\bibitem[{Cox(1972)}]{cox_regression_1972}
\textsc{Cox, D.~R.} (1972).
\newblock Regression {Models} and {Life}-{Tables}.
\newblock \textit{Journal of the Royal Statistical Society. Series B (Methodological)} \textbf{34}, 187--220.

\bibitem[{Garnett et~al.(1996)Garnett, Hughes, Anderson, Stoner, Aral, Whittington, Handsfield \& Holmes}]{garnett_sexual_1996}
\textsc{Garnett, G.~P.}, \textsc{Hughes, J.~P.}, \textsc{Anderson, R.~M.} et~al. (1996).
\newblock Sexual {Mixing} {Patterns} of {Patients} {Attending} {Sexually} {Transmitted} {Diseases} {Clinics}.
\newblock \textit{Sexually Transmitted Diseases} \textbf{23}, 248--257.

\bibitem[{Gaschen et~al.(2002)Gaschen, Taylor, Yusim, Foley, Gao, Lang, Novitsky, Haynes, Hahn, Bhattacharya \& Korber}]{gaschen_diversity_2002}
\textsc{Gaschen, B.}, \textsc{Taylor, J.}, \textsc{Yusim, K.} et~al. (2002).
\newblock Diversity {Considerations} in {HIV}-1 {Vaccine} {Selection}.
\newblock \textit{Science} \textbf{296}, 2354--2360.

\bibitem[{Gilbert et~al.(2001)Gilbert, Self, Rao, Naficy \& Clemens}]{gilbert_sieve_2001}
\textsc{Gilbert, P.}, \textsc{Self, S.}, \textsc{Rao, M.} et~al. (2001).
\newblock Sieve analysis: methods for assessing from vaccine trial data how vaccine efficacy varies with genotypic and phenotypic pathogen variation.
\newblock \textit{Journal of Clinical Epidemiology} \textbf{54}, 68--85.

\bibitem[{Gilbert(2000)}]{gilbert_comparison_2000}
\textsc{Gilbert, P.~B.} (2000).
\newblock Comparison of competing risks failure time methods and time-independent methods for assessing strain variations in vaccine protection.
\newblock \textit{Statistics in Medicine} \textbf{19}, 3065--3086.

\bibitem[{Gilbert(2001)}]{gilbert_interpretability_2001}
\textsc{Gilbert, P.~B.} (2001).
\newblock Interpretability and robustness of sieve analysis models for assessing {HIV} strain variations in vaccine efficacy.
\newblock \textit{Statistics in Medicine} \textbf{20}, 263--279.

\bibitem[{Gilbert et~al.(2008)Gilbert, McKeague \& Sun}]{gilbert_2-sample_2008}
\textsc{Gilbert, P.~B.}, \textsc{McKeague, I.~W.}, \textsc{Sun, Y.} (2008).
\newblock The 2-sample problem for failure rates depending on a continuous mark: an application to vaccine efficacy.
\newblock \textit{Biostatistics (Oxford, England)} \textbf{9}, 263--276.

\bibitem[{Gilbert et~al.(1998)Gilbert, Self \& Ashby}]{gilbert_statistical_1998}
\textsc{Gilbert, P.~B.}, \textsc{Self, S.~G.}, \textsc{Ashby, M.~A.} (1998).
\newblock Statistical {Methods} for {Assessing} {Differential} {Vaccine} {Protection} {Against} {Human} {Immunodeficiency} {Virus} {Types}.
\newblock \textit{Biometrics} \textbf{54}, 799--814.

\bibitem[{Greenwood \& Yule(1915)}]{greenwood_statistics_1915}
\textsc{Greenwood}, \textsc{Yule, G.~U.} (1915).
\newblock The {Statistics} of {Anti}-{Typhoid} and {Anti}-{Cholera} {Inoculations}, and the {Interpretation} of {Such} {Statistics} in {General}.
\newblock \textit{Proceedings of the Royal Society of Medicine} \textbf{8}, 113--194.

\bibitem[{Halloran et~al.(2010)Halloran, Longini \& Struchiner}]{halloran_design_2010}
\textsc{Halloran, M.~E.}, \textsc{Longini, I.~M.}, \textsc{Struchiner, C.~J.} (2010).
\newblock \textit{Design and {Analysis} of {Vaccine} {Studies}}.
\newblock Statistics for {Biology} and {Health}. New York, NY: Springer New York.

\bibitem[{Hayes et~al.(2000)Hayes, Alexander, Bennett \& Cousens}]{hayes_design_2000}
\textsc{Hayes, R.~J.}, \textsc{Alexander, N.~D.}, \textsc{Bennett, S.} et~al. (2000).
\newblock Design and analysis issues in cluster-randomized trials of interventions against infectious diseases.
\newblock \textit{Statistical Methods in Medical Research} \textbf{9}, 95--116.

\bibitem[{Hernan \& Robins(2023)}]{hernan_causal_2023}
\textsc{Hernan, M.~A.}, \textsc{Robins, J.~M.} (2023).
\newblock \textit{Causal {Inference}: {What} {If}}.
\newblock CRC Press.

\bibitem[{Hertz et~al.(2016)Hertz, Logan, Rolland, Magaret, Rademeyer, Fiore-Gartland, Edlefsen, DeCamp, Ahmed, Ngandu, Larsen, Frahm, Marais, Thebus, Geraghty, Hural, Corey, Kublin, Gray, McElrath, Mullins, Gilbert \& Williamson}]{hertz_study_2016}
\textsc{Hertz, T.}, \textsc{Logan, M.~G.}, \textsc{Rolland, M.} et~al. (2016).
\newblock A study of vaccine-induced immune pressure on breakthrough infections in the {Phambili} phase 2b {HIV}-1 vaccine efficacy trial.
\newblock \textit{Vaccine} \textbf{34}, 5792--5801.

\bibitem[{Hu et~al.(2022)Hu, Li \& Wager}]{hu_average_2022}
\textsc{Hu, Y.}, \textsc{Li, S.}, \textsc{Wager, S.} (2022).
\newblock Average direct and indirect causal effects under interference.
\newblock \textit{Biometrika} \textbf{109}, 1165--1172.

\bibitem[{Huang et~al.(2023)Huang, Luedtke \& {THE AMP INVESTIGATOR GROUP}}]{huang_improved_2023}
\textsc{Huang, T.-J.}, \textsc{Luedtke, A.}, \textsc{{THE AMP INVESTIGATOR GROUP}} (2023).
\newblock Improved efficiency for cross-arm comparisons via platform designs.
\newblock \textit{Biostatistics} \textbf{24}, 1106--1124.

\bibitem[{Johnston \& Fauci(2008)}]{johnston_hiv_2008}
\textsc{Johnston, M.~I.}, \textsc{Fauci, A.~S.} (2008).
\newblock An {HIV} {Vaccine} — {Challenges} and {Prospects}.
\newblock \textit{New England Journal of Medicine} \textbf{359}, 888--890.

\bibitem[{Juraska \& Gilbert(2013)}]{juraska_mark-specific_2013}
\textsc{Juraska, M.}, \textsc{Gilbert, P.~B.} (2013).
\newblock Mark-specific hazard ratio model with multivariate continuous marks: an application to vaccine efficacy.
\newblock \textit{Biometrics} \textbf{69}, 328--337.

\bibitem[{Katz et~al.(1978)Katz, Baptista, Azen \& Pike}]{katz_obtaining_1978}
\textsc{Katz, D.}, \textsc{Baptista, J.}, \textsc{Azen, S.~P.} et~al. (1978).
\newblock Obtaining {Confidence} {Intervals} for the {Risk} {Ratio} in {Cohort} {Studies}.
\newblock \textit{Biometrics} \textbf{34}, 469--474.

\bibitem[{Lambkin-Williams et~al.(2018)Lambkin-Williams, Noulin, Mann, Catchpole \& Gilbert}]{lambkin-williams_human_2018}
\textsc{Lambkin-Williams, R.}, \textsc{Noulin, N.}, \textsc{Mann, A.} et~al. (2018).
\newblock The human viral challenge model: accelerating the evaluation of respiratory antivirals, vaccines and novel diagnostics.
\newblock \textit{Respiratory Research} \textbf{19}, 123.

\bibitem[{Longini~Jr. et~al.(2002)Longini~Jr., Halloran \& Nizam}]{longini_jr_model-based_2002}
\textsc{Longini~Jr., I.~M.}, \textsc{Halloran, M.~E.}, \textsc{Nizam, A.} (2002).
\newblock Model-based estimation of vaccine effects from community vaccine trials.
\newblock \textit{Statistics in Medicine} \textbf{21}, 481--495.

\bibitem[{Neafsey et~al.(2015)Neafsey, Juraska, Bedford, Benkeser, Valim, Griggs, Lievens, Abdulla, Adjei, Agbenyega, Agnandji, Aide, Anderson, Ansong, Aponte, Asante, Bejon, Birkett, Bruls, Connolly, D’Alessandro, Dobaño, Gesase, Greenwood, Grimsby, Tinto, Hamel, Hoffman, Kamthunzi, Kariuki, Kremsner, Leach, Lell, Lennon, Lusingu, Marsh, Martinson, Molel, Moss, Njuguna, Ockenhouse, Ogutu, Otieno, Otieno, Otieno, Owusu-Agyei, Park, Pellé, Robbins, Russ, Ryan, Sacarlal, Sogoloff, Sorgho, Tanner, Theander, Valea, Volkman, Yu, Lapierre, Birren, Gilbert \& Wirth}]{neafsey_genetic_2015}
\textsc{Neafsey, D.~E.}, \textsc{Juraska, M.}, \textsc{Bedford, T.} et~al. (2015).
\newblock Genetic {Diversity} and {Protective} {Efficacy} of the {RTS},{S}/{AS01} {Malaria} {Vaccine}.
\newblock \textit{New England Journal of Medicine} \textbf{373}, 2025--2037.

\bibitem[{Nelson(1972)}]{nelson_statistical_1972}
\textsc{Nelson, W.} (1972).
\newblock Statistical {Methods} for the {Ratio} of {Two} {Multinomial} {Proportions}.
\newblock \textit{The American Statistician} \textbf{26}, 22--27.

\bibitem[{Obolski et~al.(2024)Obolski, Stensrud \& Nevo}]{obolski_call_2024}
\textsc{Obolski, U.}, \textsc{Stensrud, M.~J.}, \textsc{Nevo, D.} (2024).
\newblock A call for blinding assessments in dengue vaccine trials.
\newblock \textit{The Lancet Infectious Diseases} \textbf{24}, e10.

\bibitem[{Ouattara et~al.(2020)Ouattara, Niangaly, Adams, Coulibaly, Kone, Traore, Laurens, Tolo, Kouriba, Diallo, Doumbo, Plowe, Djimdé, Thera, Laufer, Takala-Harrison \& Silva}]{ouattara_epitope-based_2020}
\textsc{Ouattara, A.}, \textsc{Niangaly, A.}, \textsc{Adams, M.} et~al. (2020).
\newblock Epitope-based sieve analysis of \textit{plasmodium falciparum} sequences from a fmp2.1/as02a vaccine trial is consistent with differential vaccine efficacy against immunologically relevant ama1 variants.
\newblock \textit{Vaccine} \textbf{38}, 5700--5706.

\bibitem[{Patel et~al.(2014)Patel, Borkowf, Brooks, Lasry, Lansky \& Mermin}]{patel_estimating_2014}
\textsc{Patel, P.}, \textsc{Borkowf, C.~B.}, \textsc{Brooks, J.~T.} et~al. (2014).
\newblock Estimating per-act {HIV} transmission risk: a systematic review.
\newblock \textit{AIDS} \textbf{28}, 1509--1519.

\bibitem[{Pearl(2009)}]{pearl_causality_2009}
\textsc{Pearl, J.} (2009).
\newblock \textit{Causality}.
\newblock Cambridge University Press.

\bibitem[{Poole et~al.(2015)Poole, Shrier \& VanderWeele}]{poole_is_2015}
\textsc{Poole, C.}, \textsc{Shrier, I.}, \textsc{VanderWeele, T.~J.} (2015).
\newblock Is the {Risk} {Difference} {Really} a {More} {Heterogeneous} {Measure}?
\newblock \textit{Epidemiology} \textbf{26}, 714.

\bibitem[{Rerks-Ngarm et~al.(2009)Rerks-Ngarm, Pitisuttithum, Nitayaphan, Kaewkungwal, Chiu, Paris, Premsri, Namwat, de~Souza, Adams, Benenson, Gurunathan, Tartaglia, McNeil, Francis, Stablein, Birx, Chunsuttiwat, Khamboonruang, Thongcharoen, Robb, Michael, Kunasol \& Kim}]{rerks-ngarm_vaccination_2009}
\textsc{Rerks-Ngarm, S.}, \textsc{Pitisuttithum, P.}, \textsc{Nitayaphan, S.} et~al. (2009).
\newblock Vaccination with {ALVAC} and {AIDSVAX} to {Prevent} {HIV}-1 {Infection} in {Thailand}.
\newblock \textit{New England Journal of Medicine} \textbf{361}, 2209--2220.

\bibitem[{Richardson \& Robins(2013)}]{richardson_single_2013}
\textsc{Richardson, T.~S.}, \textsc{Robins, J.~M.} (2013).
\newblock Single world intervention graphs ({SWIGs}): {A} unification of the counterfactual and graphical approaches to causality.
\newblock \textit{Center for the Statistics and the Social Sciences, University of Washington Series. Working Paper} \textbf{128}, 2013.

\bibitem[{Robins(1986)}]{robins_new_1986}
\textsc{Robins, J.} (1986).
\newblock A new approach to causal inference in mortality studies with a sustained exposure period—application to control of the healthy worker survivor effect.
\newblock \textit{Mathematical modelling} \textbf{7}, 1393--1512.

\bibitem[{Robins(1997)}]{robins_causal_1997}
\textsc{Robins, J.~M.} (1997).
\newblock Causal {Inference} from {Complex} {Longitudinal} {Data}.
\newblock In \textit{Latent {Variable} {Modeling} and {Applications} to {Causality}}, M.~Berkane, ed. New York, NY: Springer.

\bibitem[{Robins \& Greenland(1992)}]{robins_identifiability_1992}
\textsc{Robins, J.~M.}, \textsc{Greenland, S.} (1992).
\newblock Identifiability and {Exchangeability} for {Direct} and {Indirect} {Effects}.
\newblock \textit{Epidemiology} \textbf{3}, 143--155.

\bibitem[{Rolland et~al.(2012)Rolland, Edlefsen, Larsen, Tovanabutra, Sanders-Buell, Hertz, deCamp, Carrico, Menis, Magaret, Ahmed, Juraska, Chen, Konopa, Nariya, Stoddard, Wong, Zhao, Deng, Maust, Bose, Howell, Bates, Lazzaro, O’Sullivan, Lei, Bradfield, Ibitamuno, Assawadarachai, O’Connell, deSouza, Nitayaphan, Rerks-Ngarm, Robb, McLellan, Georgiev, Kwong, Carlson, Michael, Schief, Gilbert, Mullins \& Kim}]{rolland_increased_2012}
\textsc{Rolland, M.}, \textsc{Edlefsen, P.~T.}, \textsc{Larsen, B.~B.} et~al. (2012).
\newblock Increased {HIV}-1 vaccine efficacy against viruses with genetic signatures in {Env} {V2}.
\newblock \textit{Nature} \textbf{490}, 417--420.

\bibitem[{Rolland \& Gilbert(2021)}]{rolland_sieve_2021}
\textsc{Rolland, M.}, \textsc{Gilbert, P.~B.} (2021).
\newblock Sieve analysis to understand how {SARS}-{CoV}-2 diversity can impact vaccine protection.
\newblock \textit{PLOS Pathogens} \textbf{17}, e1009406.

\bibitem[{Rolland et~al.(2011)Rolland, Tovanabutra, deCamp, Frahm, Gilbert, Sanders-Buell, Heath, Magaret, Bose, Bradfield, O'Sullivan, Crossler, Jones, Nau, Wong, Zhao, Raugi, Sorensen, Stoddard, Maust, Deng, Hural, Dubey, Michael, Shiver, Corey, Li, Self, Kim, Buchbinder, Casimiro, Robertson, Duerr, McElrath, McCutchan \& Mullins}]{rolland_genetic_2011}
\textsc{Rolland, M.}, \textsc{Tovanabutra, S.}, \textsc{deCamp, A.~C.} et~al. (2011).
\newblock Genetic impact of vaccination on breakthrough {HIV}-1 sequences from the {STEP} trial.
\newblock \textit{Nature Medicine} \textbf{17}, 366--371.

\bibitem[{Rosenbaum(2007)}]{rosenbaum_interference_2007}
\textsc{Rosenbaum, P.~R.} (2007).
\newblock Interference between {Units} in {Randomized} {Experiments}.
\newblock \textit{Journal of the American Statistical Association} \textbf{102}, 191--200.

\bibitem[{Stensrud et~al.(2024)Stensrud, Nevo \& Obolski}]{stensrud_distinguishing_2024}
\textsc{Stensrud, M.~J.}, \textsc{Nevo, D.}, \textsc{Obolski, U.} (2024).
\newblock Distinguishing {Immunologic} and {Behavioral} {Effects} of {Vaccination}.
\newblock \textit{Epidemiology} \textbf{35}, 154.

\bibitem[{Stensrud \& Smith(2023)}]{stensrud_identification_2023}
\textsc{Stensrud, M.~J.}, \textsc{Smith, L.} (2023).
\newblock Identification of {Vaccine} {Effects} {When} {Exposure} {Status} {Is} {Unknown}.
\newblock \textit{Epidemiology} \textbf{34}, 216--224.

\bibitem[{Sun et~al.(2009)Sun, Gilbert \& McKeague}]{sun_proportional_2009}
\textsc{Sun, Y.}, \textsc{Gilbert, P.~B.}, \textsc{McKeague, I.~W.} (2009).
\newblock Proportional {Hazards} {Models} with {Continuous} {Marks}.
\newblock \textit{The Annals of Statistics} \textbf{37}, 394--426.

\bibitem[{Symons \& Moore(2002)}]{symons_hazard_2002}
\textsc{Symons, M.~J.}, \textsc{Moore, D.~T.} (2002).
\newblock Hazard rate ratio and prospective epidemiological studies.
\newblock \textit{Journal of Clinical Epidemiology} \textbf{55}, 893--899.

\bibitem[{Sävje et~al.(2021)Sävje, Aronow \& Hudgens}]{savje_average_2021}
\textsc{Sävje, F.}, \textsc{Aronow, P.~M.}, \textsc{Hudgens, M.~G.} (2021).
\newblock Average treatment effects in the presence of unknown interference.
\newblock \textit{The Annals of Statistics} \textbf{49}, 673--701.

\bibitem[{Tsiatis \& Davidian(2022)}]{tsiatis_estimating_2022}
\textsc{Tsiatis, A.~A.}, \textsc{Davidian, M.} (2022).
\newblock Estimating vaccine efficacy over time after a randomized study is unblinded.
\newblock \textit{Biometrics} \textbf{78}, 825--838.

\bibitem[{VanderWeele \& Knol(2014)}]{vanderweele_tutorial_2014}
\textsc{VanderWeele, T.~J.}, \textsc{Knol, M.~J.} (2014).
\newblock A {Tutorial} on {Interaction}.
\newblock \textit{Epidemiologic Methods} \textbf{3}.

\bibitem[{Walsh et~al.(2024)Walsh, Alam, Pierce, Carmolli, Alam, Dickson, Bak, Afreen, Nazib, Golam, Qadri, Diehl, Durbin, Whitehead, Haque \& Kirkpatrick}]{walsh_safety_2024}
\textsc{Walsh, M.-C.~R.}, \textsc{Alam, M.~S.}, \textsc{Pierce, K.~K.} et~al. (2024).
\newblock Safety and durable immunogenicity of the {TV005} tetravalent dengue vaccine, across serotypes and age groups, in dengue-endemic {Bangladesh}: a randomised, controlled trial.
\newblock \textit{The Lancet Infectious Diseases} \textbf{24}, 150--160.

\bibitem[{Whitney et~al.(2003)Whitney, Farley, Hadler, Harrison, Bennett, Lynfield, Reingold, Cieslak, Pilishvili, Jackson, Facklam, Jorgensen \& Schuchat}]{whitney_decline_2003}
\textsc{Whitney, C.~G.}, \textsc{Farley, M.~M.}, \textsc{Hadler, J.} et~al. (2003).
\newblock Decline in {Invasive} {Pneumococcal} {Disease} after the {Introduction} of {Protein}–{Polysaccharide} {Conjugate} {Vaccine}.
\newblock \textit{New England Journal of Medicine} \textbf{348}, 1737--1746.

\bibitem[{Yang et~al.(2022)Yang, Balzer \& Benkeser}]{yang_causal_2022}
\textsc{Yang, G.}, \textsc{Balzer, L.~B.}, \textsc{Benkeser, D.} (2022).
\newblock Causal inference methods for vaccine sieve analysis with effect modification.
\newblock \textit{Statistics in Medicine} \textbf{41}, 1513--1524.

\bibitem[{Young et~al.(2020)Young, Stensrud, Tchetgen~Tchetgen \& Hernán}]{young_causal_2020}
\textsc{Young, J.~G.}, \textsc{Stensrud, M.~J.}, \textsc{Tchetgen~Tchetgen, E.~J.} et~al. (2020).
\newblock A causal framework for classical statistical estimands in failure‐time settings with competing events.
\newblock \textit{Statistics in Medicine} \textbf{39}, 1199--1236.

\bibitem[{Zolla-Pazner et~al.(2014)Zolla-Pazner, Edlefsen, Rolland, Kong, deCamp, Gottardo, Williams, Tovanabutra, Sharpe-Cohen, Mullins, deSouza, Karasavvas, Nitayaphan, Rerks-Ngarm, Pitisuttihum, Kaewkungwal, O'Connell, Robb, Michael, Kim \& Gilbert}]{zolla-pazner_vaccine-induced_2014}
\textsc{Zolla-Pazner, S.}, \textsc{Edlefsen, P.~T.}, \textsc{Rolland, M.} et~al. (2014).
\newblock Vaccine-induced {Human} {Antibodies} {Specific} for the {Third} {Variable} {Region} of {HIV}-1 gp120 {Impose} {Immune} {Pressure} on {Infecting} {Viruses}.
\newblock \textit{eBioMedicine} \textbf{1}, 37--45.

\end{thebibliography}
\appendix

\appendix

\addcontentsline{toc}{section}{Online Appendix}
\section*{Online Appendix}
\addcontentsline{toc}{section}{Appendices}
\renewcommand{\thesubsection}{\Alph{subsection}}

\subsection{Concordance between notions of sieve effects}\label{APP: connect two sieves}
Assume that data were collected from a randomized clinical trial. Specifically, if an individual was affected by any outcome, $Y\neq0$, then the genetic distance from the infectious variant to the vaccine insert $D \in \mathbb{R}^+$, was recorded.

Suppose that the outcome was classified as matched, $Y=1$, if the distance from the vaccine insert was less than a threshold value $t$. Based on this threshold value, the distance variable $D$ can be dichotomized to $D_d$, and the one-to-one relationship between the outcome and the discrete distance can be established as $Y=D_d$.

Using a statistical sieve effect approach \citep{gilbert_statistical_1998,gilbert_interpretability_2001, gilbert_2-sample_2008, sun_proportional_2009, juraska_mark-specific_2013, benkeser_estimating_2019, benkeser_assessing_2020, yang_causal_2022}, see for example Table \ref{TAB: Causal estimands}, the effect of the vaccine on the two variants would be contrasted, for instance by calculating the respective relative risks from a trial where $A$ is randomly assigned. That is, 
$$
\frac{\mathbb{P}(Y=1|A=1)}{\mathbb{P}(Y=1|A=0)} \text{ vs } \frac{\mathbb{P}(Y=2|A=1)}{\mathbb{P}(Y=2|A=0)}.
$$
The ratios of the dichotomized distances are contrasted between the treatment and the control arms, that is
$$
\frac{\mathbb{P}(D_d=1|A=1)}{\mathbb{P}(D_d=2|A=1)} \text{ vs } \frac{\mathbb{P}(D_d=1|A=0)}{\mathbb{P}(D_d=2|A=0)}.
$$
In our setting, $Y=D_d$ by definition, and there is a bijective map between these two contrasts.

\subsection{Interference and vaccine trials}\label{APP: interference}

The assumption of no interference is often invoked in vaccine trials.

No interference implies that the potential outcomes of individuals are independent of the treatment received by other trial participants. More explicitly, $Y_i^{a_i, \mathbf{a}_{-i}}=Y_i^{a_i, \mathbf{a^*}_{-i}}$ for all $\mathbf{a}_{-i}$ and $\mathbf{a^*}_{-i}$, where
$\mathbf{a}_{-i}=(a_1, \dots, a_{i-1}, a_{i+i}, \dots, a_n)$ are the treatment assignments of all but the $i$-th individual. The recurring reasoning in vaccine trials is that the infectious contacts made between the trial participants are negligible. Therefore the treatment $A_j$ of individual $j$  does not affect the outcome $Y_i^\mathbf{a}$ for any $i \neq j$. This allows us to write $Y_i^\mathbf{a}=Y_i^{a_i}$, where $\mathbf{a}=\{a_1, \dots, a_n\}$. Following this argument, suppose that a vaccine trial is analyzed with conventional frequentist (superpopulation) methods under iid assumptions, in particular, no interference. Then we can interpret the results as valid for a hypothetical population where the no interference assumption holds, for example, a superpopulation of individuals that, say, are sparsely embedded in a larger population so that they do not interact. 
 
 Yet, decision-makers will frequently use the trial results to inform policies in populations where there is interference.  In realistic target populations, the treatment assignments of one unit can affect outcomes of other units, often referred to as a \textit{spillover effect}, or in the setting of vaccines against infectious diseases, as, e.g., \textit{herd immunity}. It follows that, in this target population, the potential outcome $Y^a$ is no longer well-defined because of interference; the assumption
$Y_i^{a_i, \mathbf{a}_{-i}}=Y_i^{a_i, \mathbf{a^*}_{-i}}$
does not hold for every $\mathbf{a}_{-i}$ and $\mathbf{a^*}_{-i}$, thus the potential outcomes cannot be defined at individual level interventions only. Thus, even if individuals are iid in an experiment, the use of conventional iid methods will not necessarily target estimands that are of broader policy interest. 

A potential solution could be defining the causal effects in terms of interventions on the whole study population, for example as
$\mathbb{E}[Y^\mathbf{a}_1]-\mathbb{E}[Y^\mathbf{a}_0]$. However, identification and estimation of this quantity would not be possible from the conventional trial data, without imposing strong assumptions about the interference structure.

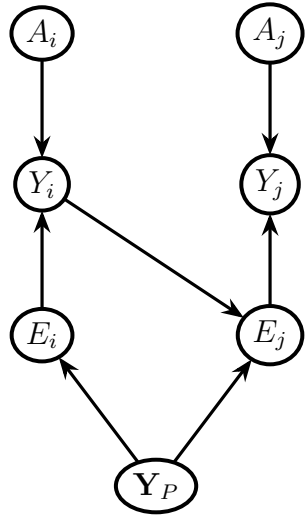
\begin{figure}
        \centering
        \begin{tikzpicture}
            \tikzset{line width=1.5pt, outer sep=0pt,
            ell/.style={draw,fill=white, inner sep=2pt,
            line width=1.5pt},
            swig vsplit={gap=5pt,
            inner line width right=0.5pt},
            swig hsplit={gap=5pt}
            };
                \node[name=Ai, ell, shape=ellipse] at (3,6){$A_i$};
                \node[name=Aj, ell, shape=ellipse] at (6,6){$A_j$};

                \node[name=Yi, ell, shape=ellipse] at (3,4){$Y_i$};
                \node[name=Yj, ell, shape=ellipse] at (6,4){$Y_j$};
                \node[name=Ei, ell, shape=ellipse] at (3,2){$E_i$};
                \node[name=Ej, ell, shape=ellipse] at (6,2) {$E_j$};
                \node[name=YP, ell, shape=ellipse] at (4.5,0) {$\mathbf{Y}_P$};
            \begin{scope}[>={Stealth[black]},
              every node/.style={fill=white,circle},
              every edge/.style={draw=black,very thick}]
                \path[->] (Ai) edge (Yi);
                \path[->] (Aj) edge (Yj);
                \path[->] (Yi) edge (Ej);
                
                \path[->] (Ei) edge (Yi);
                \path[->] (Ej) edge (Yj);
                \path[->] (YP) edge (Ei);
                \path[->] (YP) edge (Ej);
            \end{scope}
        \end{tikzpicture}
        \caption{Illustrative graph of outcome interference between individuals, with the topological and temporal ordering $\{i,j\}$ between the individuals; we consider the exposure and outcome of individual $i$ to happen just before the exposure of any other individual $j$. For simplicity, $i$ can be considered as \textit{patient 0}. $\mathbf{Y}_P$ denotes the outcomes of all individuals outside of the trial, in the population where the trial is conducted. In the case of targeting the whole population, $\mathbf{Y}_P$ can be removed from the graph.}
        \label{FIG: Interference SWIG}
    \end{figure}

To consider another solution, suppose that the treatment assignment of units affects other units' outcomes mediated through exposure. In specific, the single directed causal path between

$A_i$ and $Y_j$ is $A_i \rightarrow Y_i \rightarrow E_{j} \rightarrow Y_j$.
This structure is plausible when considering communicable infectious diseases, which precisely can spread from infectious to susceptible subjects. This is, e.g., reflected in conventional infectious disease models, although we emphasize that our causal structure is less restrictive than, say, SIR or SIER models.  

By conditioning or intervening on the value of exposure to the infectious agent, there will be no dependence between the treatment of individual $i$ and outcome of individual $j$. That is,
$Y_j^{a_j, \mathbf{a}_{-j}}=Y_j^{a_j, \mathbf{a^*}_{-j}}$
for all $\mathbf{a}_{-j}$ and $\mathbf{a^*}_{-j}$, such that for all $j$, $Y_j^{a_j, \mathbf{a}_{-j}}=Y_j^{a_j}$.
Thus, when we condition or intervene on exposure, we can assume no interference between the potential outcomes of individuals in the realistic target population. This is one rationale for considering our target estimands that are explicitly defined with respect to exposure status.


Our work contributes to the existing rich literature on the statistical issue of interference \citep{rosenbaum_interference_2007, aronow_general_2012, savje_average_2021, hu_average_2022}. While some of these works study specifically vaccine effects as well \citep{hayes_design_2000, longini_jr_model-based_2002}, we do not impose constraints on the interference structure, or consider special types of randomized designs. Our results are valid under the usual frequentist superpopulation framework, without considering design based inference, where the estimands are explicit functions of randomization probabilities.

\subsection{Proofs}\label{APP: Section of proofs}
\setcounter{assumption}{0}
    \renewcommand{\theassumption}{S\arabic{assumption}}
\subsubsection{Proof of Proposition \ref{PROP: Relative CECE estimator}}\label{APP: Prop relative CECE estimator}
\begin{proof}
By the law of total probability, and by Assumptions \ref{ASM: no multiple exp}, \ref{ASM: Exp nec}, and \ref{ASM: No cross} $\forall \ a \in \{0,1\}, \, j \in \{1,2\}$
\begin{align}\label{EQ: always helper}
    \mathbb{P}(Y=j|A=a)&=\mathbb{P}(Y=j|A=a, E=j) \cdot \mathbb{P}(E=j|A=a)\\ \nonumber
    &+ \mathbb{P}(Y=j|A=a, E\neq j) \cdot \mathbb{P}(E\neq j|A=a)\\ 
    &=\mathbb{P}(Y=j|A=a, E=j) \cdot \mathbb{P}(E=j|A=a). \nonumber
\end{align}
Furthermore,
    \begin{align*}
        \mathbb{P}(Y^{a}=j|E^{a}=j)&=\mathbb{P}(Y^{a}=j|E^{a}=j, A=a)\\
        &=\mathbb{P}(Y=j|E=j, A=a)\\
        &=\frac{\mathbb{P}(Y=j|A=a)}{\mathbb{P}(E=j|A=a)},
    \end{align*}
    where the first equality follows from assumption, \ref{ASM: Exchange}, the second from Assumptions \ref{ASM: Pos}- \ref{ASM: Cons} and the last from Equation \eqref{EQ: always helper}.
    Finally, by Assumption \ref{ASM: No eff exp}
    \begin{align*}
        \frac{\mathbb{P}(Y^{a=1}=j|E^{a=1}=j)}{\mathbb{P}(Y^{a=0}=j|E^{a=0}=j)}&=
        \left. \frac{\mathbb{P}(Y=j|A=1)}{\mathbb{P}(E=j|A=1)}\middle \slash
        \frac{\mathbb{P}(Y=j|A=0)}{\mathbb{P}(E=j|A=0)} \right. \\
        &=\frac{\mathbb{P}(Y=j|A=1)}{\mathbb{P}(Y=j|A=0)}
    \end{align*}
    $\forall \ j \in \{1,2\}$.
\end{proof}

\subsubsection{Proof of Proposition \ref{PROP: No eff on exp ratios}}\label{APP: Prop no eff on exp ratios proof}
\begin{proof}
    \begin{align*}
        & \left.\frac{\mathbb{P}(Y^{a=1}=1|E^{a=1}=1)}{\mathbb{P}(Y^{a=0}=1|E^{a=0}=1)}\middle \slash
        \frac{\mathbb{P}(Y^{a=1}=2|E^{a=1}=2)}{\mathbb{P}(Y^{a=0}=2|E^{a=0}=2)}\right.\\
        &=\left. \frac{\mathbb{P}(Y^{a=1}=1|E^{a=1}=1, A=1)}{\mathbb{P}(Y^{a=0}=1|E^{a=0}=1, A=0)}\middle \slash
        \frac{\mathbb{P}(Y^{a=1}=2|E^{a=1}=2, A=1)}{\mathbb{P}(Y^{a=0}=2|E^{a=0}=2, A=0)}\right.\\
        &\cdot \left. \frac{\mathbb{P}(E^{a=1}=1|A=1)}{\mathbb{P}(E^{a=0}=1|A=0)}\middle \slash
        \frac{\mathbb{P}(E^{a=1}=2|A=1)}{\mathbb{P}(E^{a=0}=2|A=0)} \right. \\
        &= \left. \frac{\mathbb{P}(Y=1|E=1, A=1)}{\mathbb{P}(Y=1|E=1, A=0)}\middle \slash 
        \frac{\mathbb{P}(Y=2|E=2, A=1)}{\mathbb{P}(Y=2|E=2, A=0)} \right. \cdot 
        \left. \frac{\mathbb{P}(E=1|A=1)}{\mathbb{P}(E=1|A=0)}\middle \slash
        \frac{\mathbb{P}(E=2|A=1)}{\mathbb{P}(E=2|A=0)} \right. \\
        &=\left. \frac{\mathbb{P}(Y=1|A=1)}{\mathbb{P}(Y=1|A=0)}\middle \slash
        \frac{\mathbb{P}(Y=2|A=1)}{\mathbb{P}(Y=2|A=0)} ,\right. 
    \end{align*}
    where the first line follows from Assumption \ref{ASM: Exchange} and \ref{ASM: No eff on exp ratios}. The second line follows from Assumption \ref{ASM: Cons} and \ref{ASM: Pos}. For the last line by the law of total probability
    \begin{align*}
        \mathbb{P}(Y=j|A=a)&=\mathbb{P}(Y=j|A=a, E=j)\cdot \mathbb{P}(E=j|A=a)\\
        &+\mathbb{P}(Y=j|A=a, E\neq j)\cdot \mathbb{P}(E\neq j|A=a),
    \end{align*}
    and under Assumption \ref{ASM: Exp nec} and \ref{ASM: No cross}, $\mathbb{P}(Y=j|A=a, E\neq j)=0$, hence the last equality follows.
\end{proof}

\subsubsection{Proof of $CCE$ identification}\label{APP: CCS star ident}
Similarly to the proof of Proposition \ref{PROP: No eff on exp ratios}, we can show that the $CCE$ can be identified even when Assumption \ref{ASM: No cross} does not hold.
\begin{proof}
    \begin{align*}
        \frac{\frac{\mathbb{P}(Y^{a=1}=1|E^{a=1}\neq 0)}{\mathbb{P}(Y^{a=0}=1|E^{a=0}\neq 0)}}
        {\frac{\mathbb{P}(Y^{a=1}=2|E^{a=1}\neq 0)}{\mathbb{P}(Y^{a=0}=2|E^{a=0}\neq 0)}}&=
        \frac{\frac{\mathbb{P}(Y^{a=1}=1|E^{a=1}\neq 0, A=1)}{\mathbb{P}(Y^{a=0}=1|E^{a=0}\neq 0, A=0)}}
        {\frac{\mathbb{P}(Y^{a=1}=2|E^{a=1}\neq 0, A=1)}{\mathbb{P}(Y^{a=0}=2|E^{a=0}\neq 0, A=0)}} \cdot 
        \frac{\frac{\mathbb{P}(E^{a=1}\neq 0|A=1)}{\mathbb{P}(E^{a=0}\neq 0|A=0)}}{\frac{\mathbb{P}(E^{a=1}\neq 0|A=1)}{\mathbb{P}(E^{a=0}\neq 0|A=0)}}\\
        &= \frac{\frac{\mathbb{P}(Y=1|E\neq 0, A=1)}{\mathbb{P}(Y=1|E\neq 0, A=0)}}
        {\frac{\mathbb{P}(Y=2|E\neq 0, A=1)}{\mathbb{P}(Y=2|E\neq 0, A=0)}}\cdot 
        \frac{\frac{\mathbb{P}(E\neq 0|A=1)}{\mathbb{P}(E\neq 0|A=0)}}{\frac{\mathbb{P}(E\neq 0|A=1)}{\mathbb{P}(E\neq 0|A=0)}}\\
        &=\frac{\frac{\mathbb{P}(Y=1|A=1)}{\mathbb{P}(Y=1|A=0)}}
        {\frac{\mathbb{P}(Y=2|A=1)}{\mathbb{P}(Y=2|A=0)}},
    \end{align*}    
    where the first line follows from Assumption \ref{ASM: Exchange} and by multiplying by an expression equal to 1. The second line follows from Assumption \ref{ASM: Cons} and \ref{ASM: Pos}. For the last line by the law of total probability
    \begin{align*}
        \mathbb{P}(Y=j|A=a)&=\mathbb{P}(Y=j|A=a, E \neq 0)\cdot \mathbb{P}(E \neq 0|A=a)\\
        &+\mathbb{P}(Y=j|A=a, E=0)\cdot \mathbb{P}(E=0|A=a),
    \end{align*}

    and under Assumption \ref{ASM: Exp nec}, $\mathbb{P}(Y=j|A=a, E =0 )=0$, hence the last equality follows.
\end{proof}

\subsubsection{Proof of Proposition \ref{PROP: CSE ident given U}}\label{APP: CSE ident}
\begin{proof}
    First note that for all $a \in \{0,1\}$ and $j \in \{1,2\}$

        \begin{align*}
        &\bbP(Y_k^{a, e_k=j, \overline{e}_{k-1}=0}=j|U=u, Z=z)\\
        &=\bbP(Y_k^{a, e_k=j, \overline{e}_{k-1}=0}=j|Y_{k-1}^{a, e_k=j, \overline{e}_{k-1}=0}=0, U=u, Z=z)\\
        &\cdot \bbP(Y_{k-1}^{a, e_k=j, \overline{e}_{k-1}=0}=0|U=u, Z=z)\\
        &=\bbP(Y_k^{a, e_k=j, \overline{e}_{k-1}=0}=j|Y_{k-1}^{a, e_k=j, \overline{e}_{k-1}=0}=0, U=u, Z=z)\\
        &=\bbP(Y_k^{a, e_k=j, \overline{e}_{k-1}=0}=j|Y_{k-1}^{a, e_k=j, \overline{e}_{k-1}=0}=0, A=a, E_k^a=j, \overline{E}^a_{k-1}=0, U=u, Z=z)\\
        &=\bbP(Y_k=j|Y_{k-1}=0, A=a, E_k=j, \overline{E}_{k-1}=0, U=u, Z=z)\\
        &=\bbP(Y_k=j|Y_{k-1}=0, A=a, E_k=j, U=u, Z=z)\\
        &=\bbP(Y^{a, e_k=j}_k=j|Y^{a, e_k=j}_{k-1}=0, U=u, Z=z),
    \end{align*}
where for the second line we used laws of probability and Assumption \ref{ASM: TTE original exp nec} in both lines two and three. In the fourth line, we used 
  Assumption \ref{ASM: TTE EXCH},
  while in the fifth line, we used Assumptions \ref{ASM: TTE Pos} and \ref{ASM: TTE Cons}. The sixth line follows from Assumptions \ref{ASM: TTE ignorable non-exposure}, and the last line follows from Assumption \ref{ASM: TTE Standard RCT}.

    Then for a given level of treatment $a$
        \begin{align*}
        &\quad \frac{\bbP(Y_k^{e_k=1, a}=1|Y_{k-1}^{e_k=1, a}=0, U=u, Z=z)}{\bbP(Y_k^{e_k=2, a}=2|Y_{k-1}^{e_k=2, a}=0, U=u, Z=z)}\\
        &=\frac{\bbP(Y_k^{e_k=1, a}=1|Y_{k-1}^{e_k=1, a}=0, U=u, Z=z, E_k^a=1, A=a)}{\bbP(Y_k^{e_k=2, a}=2|Y_{k-1}^{e_k=2, a}=0, U=u, Z=z, E_k^a=2, A=a)}\\
        &=\frac{\bbP(Y_k=1|Y_{k-1}=0, U=u, Z=z, E_k=1, A=a)}{\bbP(Y_k=2|Y_{k-1}=0, U=u, Z=z, E_k=2, A=a)}\\
        &=\frac{\bbP(Y_k=1|Y_{k-1}=0, U=u, Z=z, A=a)}{\bbP(E_k=1|Y_{k-1}=0, U=u, Z=z, A=a)}\\
        &\cdot \left(\frac{\bbP(Y_k=2|Y_{k-1}=0, U=u, Z=z, A=a)}{\bbP(E_k=2|Y_{k-1}=0, U=u, Z=z, A=a)} \right)^{-1}\\
        &=\frac[5pt]{\bbP(Y_k=1|Y_{k-1}=0, U=u, Z=z, A=a)}{\frac{\bbP(E_k=1|E_k \neq 0, Y_{k-1}=0, U=u, Z=z, A=a)}{\bbP(E_k \neq 0|Y_{k-1}=0, U=u, Z=z, A=a)}}\\
        &\cdot \left(\frac[5pt]{\bbP(Y_k=2|Y_{k-1}=0, U=u, Z=z, A=a)}{\frac{\bbP(E_k=2|E_k \neq 0, Y_{k-1}=0, U=u, Z=z, A=a)}{\bbP(E_k \neq 0|Y_{k-1}=0, U=u, Z=z, A=a}} \right)^{-1}\\
        &=\frac{\bbP(Y_k=1|Y_{k-1}=0, U=u, Z=z, A=a)}{\bbP(E_k=1|E_k \neq 0, Y_{k-1}=0, U=u, Z=z)}\\
        &\cdot \left(\frac{\bbP(Y_k=2|Y_{k-1}=0, U=u, Z=z, A=a)}{\bbP(E_k=2|E_k \neq 0, Y_{k-1}=0, U=u, Z=z)} \right)^{-1}\\
        &=\frac{\bbP(Y_k=1|Y_{k-1}=0, U=u, Z=z, A=a)}{\bbP(E_k=1|E_k \neq 0, U=u, Z=z)\cdot (\bbP(Y_{k-1}=0|E_k \neq 0, U=u, Z=z))^{-1}}\\
        &\cdot \left(\frac{\bbP(Y_k=2|Y_{k-1}=0, U=u, Z=z, A=a)}{\bbP(E_k=2|E_k \neq 0, U=u, Z=z) \cdot (\bbP(Y_{k-1}=0|E_k \neq 0, U=u, Z=z))^{-1}} \right)^{-1}\\
        &=\frac{\bbP(Y_k=1|Y_{k-1}=0, U=u, Z=z, A=a)}{\alpha_k \bbP(E_k=2|E_k \neq 0, U=u, Z=z)}\\
        &\cdot \left(\frac{\bbP(Y_k=2|Y_{k-1}=0, U=u, Z=z, A=a)}{\bbP(E_k=2|E_k \neq 0, U=u, Z=z)} \right)^{-1}\\
        &=\frac{1}{\alpha_k} \cdot \frac{\bbP(Y_k=1|Y_{k-1}=0, U=u, Z=z, A=a)}{\bbP(Y_k=2|Y_{k-1}=0, U=u, Z=z, A=a)},
    \end{align*}
    where in the second line we used Assumption \ref{ASM: TTE EXCH}, and in the third line Assumptions \ref{ASM: TTE Pos} and \ref{ASM: TTE Cons}. The fourth line follows from the laws of probability and Assumption \ref{ASM: TTE No cross}. The fifth line is by the laws of probability and that $\bbP(E_k=j|E_k=0)=0 \ \forall \ j \in \{1,2\}$. The sixth line follows using cancellation and from $A\independent E_k|\overline{Y}_{k-1}, U=u, Z=z$ as well as that $Y_k=0 \Leftrightarrow \overline{Y}_k=0$ by the irreversibility of the outcome. The seventh line follows from the laws of probability and $E_k \cdot Y_j=0 \, , k\neq j$. The last two lines follow from Assumption \ref{ASM: TTE exposure rat of exposed} and cancellation of identical terms.

    Finally, by dividing the expressions corresponding to $a=1$ with the one corresponding to $a=0$ we have the stated result, as $1/\alpha_k$ cancels out.
\end{proof}

\subsubsection{Proof of Corollary \ref{PROP: all across CSE}}\label{APP: all across CSE}
\begin{proof}
By the first part of the proof in \ref{APP: CSE ident} for any $u$ and $z$
    \begin{align*}
    \bbP(Y_k^{a, e_k=j, \overline{e}_{k-1}=0}=j|U=u, Z=z)=\bbP(Y^{a, e_k=j}_k=j|Y^{a, e_k=j}_{k-1}=0, U=u, Z=z)
    \end{align*}
Then the $CSE_k(u,z)$ in the sub-population defined by $U$ and $Z$ is
    Then 
    \begin{align*}
    & \frac{\bbP(Y_k^{a=1, e_k=1, \overline{e}_{k-1}=0}=1|U=u, Z=z)}{\bbP(Y_k^{a=0, e_k=1, \overline{e}_{k-1}=0}=1|U=u, Z=z)}
    \cdot \left(\frac{\bbP(Y_k^{a=1, e_k=2, \overline{e}_{k-1}=0}=2|U=u, Z=z)}{\bbP(Y_k^{a=0, e_k=2, \overline{e}_{k-1}=0}=2|U=u, Z=z)} \right)^{-1}\\
    &=\frac{\bbP(Y_k^{a=1, e_k=1}=1|Y_{k-1}^{a=1 e_k=1}=0, U=u, Z=z)}{\bbP(Y_k^{a=0, e_k=1}=1|Y_{k-1}^{a=0, e_k=1}=0, U=u, Z=z)}\\
    &\cdot \left(\frac{\bbP(Y_k^{a=1, e_k=2}=2|Y_{k-1}^{a=1, e_k=2}=0, U=u, Z=z)}{\bbP(Y_k^{a=0, e_k=2}=2|Y_{k-1}^{a=0, e_k=2}=0, U=u, Z=z)} \right)^{-1}\\
    &=\frac{\gamma_{1,k} \bbP(Y_k^{a=1, e_k=2}=2|Y_{k-1}^{a=1 e_k=2}=0, U=u, Z=z)}{\gamma_{0,k} \bbP(Y_k^{a=0, e_k=2}=2|Y_{k-1}^{a=0, e_k=2}=0, U=u, Z=z)}\\
    &\cdot \left(\frac{\bbP(Y_k^{a=1, e_k=2}=2|Y_{k-1}^{a=1, e_k=2}=0, U=u, Z=z)}{\bbP(Y_k^{a=0, e_k=2}=2|Y_{k-1}^{a=0, e_k=2}=0, U=u, Z=z)} \right)^{-1}\\
    &=\frac{\gamma_{1,k}}{\gamma_{0,k}},
    \end{align*}
where the second equality follows from Assumption \ref{ASM: TTE Scaled new inf}. Then the claim follows from cancellations of equal terms.  

For marginal $CSE_k$ we can proceed analogously.
    \begin{align*}
        CSE_k&=\frac{\bbP(Y_k^{a=1, e_k=1, \overline{e}_{k-1}=0}=1)}{\bbP(Y_k^{a=0, e_k=1, \overline{e}_{k-1}=0}=1)} \left(\frac{\bbP(Y_k^{a=1, e_k=2, \overline{e}_{k-1}=0}=2)}{\bbP(Y_k^{a=0, e_k=2, \overline{e}_{k-1}=0}=2)}\right)^{-1}\\
        &=\frac{\sum_{u,z} \bbP(Y_k^{a=1, e_k=1, \overline{e}_{k-1}=0}=1|U=u, Z=z) \cdot \bbP(U=u, Z=z)}
        {\sum_{u,z} \bbP(Y_k^{a=0, e_k=1, \overline{e}_{k-1}=0}=1|U=u, Z=z) \cdot \bbP(U=u, Z=z)}\\
    &\cdot \left(\frac{\sum_{u,z} \bbP(Y_k^{a=1, e_k=2, \overline{e}_{k-1}=0}=2|U=u, Z=z) \cdot \bbP(U=u, Z=z)}
    {\sum_{u,z} \bbP(Y_k^{a=0, e_k=2, \overline{e}_{k-1}=0}=2|U=u, Z=z) \cdot \bbP(U=u, Z=z)} \right)^{-1}\\
    &=\frac{\sum_{u,z} \gamma_{1,k} \bbP(Y_k^{a=1, e_k=2}=2|Y_{k-1}^{a=1 e_k=2}=0, U=u, Z=z) \cdot \bbP(U=u, Z=z)}{\sum_{u,z} \gamma_{0,k} \bbP(Y_k^{a=0, e_k=2}=2|Y_{k-1}^{a=0, e_k=2}=0, U=u, Z=z) \cdot \bbP(U=u, Z=z)}\\
    &\cdot \left(\frac{ \sum_{u,z} \bbP(Y_k^{a=1, e_k=2}=2|Y_{k-1}^{a=1, e_k=2}=0, U=u, Z=z) \cdot \bbP(U=u, Z=z)}{\sum_{u,z} \bbP(Y_k^{a=0, e_k=2}=2|Y_{k-1}^{a=0, e_k=2}=0, U=u, Z=z) \cdot \bbP(U=u, Z=z)} \right)^{-1}\\
    &=\frac{\gamma_{1,k}}{\gamma_{0,k}}
    \end{align*}
    where the first equality is from the laws of probability and the second follows by the first half of Appendix \ref{APP: CSE ident}. The result follows trivially by cancellations of equal terms.
\end{proof}

\subsubsection{Proof of Proposition \ref{PROP: marginal CSE ident}}\label{APP: marginal CSE ident}
\begin{proof}
For all $a \in \{0,1\}$
    \begin{align*}
        &\frac{\mathbb{P}(Y_k=1|Y_{k-1}=0, A=a)}{\mathbb{P}(Y_k=2|Y_{k-1}=0, A=a)}\\
        &\quad =\frac{\sum_{u,z} \mathbb{P}(Y_k=1|Y_{k-1}=0, A=a, U=u, Z=z) \cdot \mathbb{P}(U=u, Z=z|Y_{k-1}=0, A=a)}{\sum_{u,z} \mathbb{P}(Y_k=2|Y_{k-1}=0, A=a, U=u, Z=z) \cdot \mathbb{P}(U=u, Z=z|Y_{k-1}=0, A=a)}\\
        &\quad  =\frac{\sum_{u,z} \frac{\gamma_{a,k}}{\alpha_k} \mathbb{P}(Y_k=2|Y_{k-1}=0, A=a, U=u, Z=z) \cdot \mathbb{P}(U=u, Z=z|Y_{k-1}=0, A=a)}{\sum_{u,z} \mathbb{P}(Y_k=2|Y_{k-1}=0, A=a, U=u, Z=z) \cdot \mathbb{P}(U=u, Z=z|Y_{k-1}=0, A=a)}\\
        &\quad = \frac{\gamma_{a,k}}{\alpha_k}
    \end{align*}
    where the first line is by the laws of probability and the second line follows from the proof Proposition \ref{PROP: CSE ident given U} and Corollary \ref{PROP: all across CSE}.\\
    The claim follows from dividing $\frac{\gamma_{1,k}}{\alpha_k}$ by $\frac{\gamma_{0,k}}{\alpha_k}$
\end{proof}

\subsubsection{Proof of Proposition \ref{PROP: Prop hazards under null}}\label{APP: null scaling}
\begin{proof}
    As it is a sharp null hypothesis, equality among the probability of the intervened outcomes holds in the subpopulations characterized by the patient characteristics $U$ and $Z$. That is 
    $$
    \frac{\bbP(Y_{k}^{a=1, e_k=1, \overline{e}_{k-1}=0}=1|U=u, Z=z)}{\bbP(Y_{k}^{a=0, e_k=1, \overline{e}_{k-1}=0}=1|U=u, Z=z)} =\frac{\bbP(Y_{k}^{a=1, e_k=2, \overline{e}_{k-1}=0}=2|U=u, Z=z)}{\bbP(Y_{k}^{a=0, e_k=2, \overline{e}_{k-1}=0}=2|U=u, Z=z)}
    $$
    implying 
    \begin{align*}
    \bbP(Y_{k}^{a=1, e_k=1, \overline{e}_{k-1}=0}=1|U=u, Z=z)
    &=\frac{\bbP(Y_{k}^{a=0, e_k=1, \overline{e}_{k-1}=0}=1|U=u, Z=z)}{\bbP(Y_{k}^{a=0, e_k=2, \overline{e}_{k-1}=0}=2|U=u, Z=z)}\\
    &\cdot\bbP(Y_{k}^{a=1, e_k=2, \overline{e}_{k-1}=0}=2|U=u, Z=z)
    \end{align*}
    By identical derivation as in the first half of the proof of Proposition \ref{PROP: CSE ident given U} it follows that
    \begin{align*}
        &\bbP(Y_k^{a=1, e_k=1}=1|Y_{k-1}^{a=1, e_k=1}=0, U=u, Z=z)\\
        &= \frac{\bbP(Y_k^{a=0, e_k=1}=1|Y_{k-1}^{a=0, e_k=1}=0, U=u, Z=z)}{\bbP(Y_k^{a=0, e_k=2}=2|Y_{k-1}^{a=0, e_k=2}=0, U=u, Z=z)}\\
        &\cdot \bbP(Y_k^{a=1, e_k=2}=2|Y_{k-1}^{a=1, e_k=2}=0, U=u, Z=z)
    \end{align*}
    Assuming that the proportionality holds independent of $U$ and $Z$ for $a=0$, it is implied that the ratio for the treated is also independent of $U$ and $Z$. It further means that this ratio for the two hazards for $a=1$ is equal to $\gamma_{0,k}$, that is $\gamma_{0,k}=\gamma_{1,k}$.

    To prove that Assumption \ref{ASM: TTE Scaled new inf} holds for $a=0$, under $H_0^k$ and Assumption \ref{ASM: TTE Scaled new inf} for $a=1$, analogous steps should be followed that is omitted.
\end{proof}

\subsubsection{Proof of Lemma \ref{LEM: H0 strong means scaled hazard}}\label{APP: strong null means scaled}
\begin{proof}
    Under the strong null hypothesis, $H_0^{k, strong}$ the two probabilities of the outcome with respect to the two levels of exposures $e_k=1$ and $e_k=2$ in the subpopulations characterized by $U$ and $Z$ are equal.
    \begin{align*}
        \mathbb{P}(Y_{k}^{a, e_k=1, \overline{e}_{k-1}=0}=1|U=u, Z=z)=\mathbb{P}(Y_{k}^{a, e_k=2, \overline{e}_{k-1}=0}=2|U=u, Z=z)\\
    \end{align*}
    for $a \in \{0,1\}$
    Then, 
    \begin{align*}
        &\mathbb{P}(Y_{k}^{a, e_k=1, \overline{e}_{k-1}=0}=1|U=u, Z=z)\\
        &=\mathbb{P}(Y_{k}^{a, e_k=1, \overline{e}_{k-1}=0}=1|Y_{k-1}^{a, e_k=1, \overline{e}_{k-1}=0}=0, U=u, Z=z)\\
        &\cdot \mathbb{P}(Y_{k-1}^{a, e_k=1, \overline{e}_{k-1}=0}=0| U=u, Z=z)\\
        &=\mathbb{P}(Y_{k}^{a, e_k=1, \overline{e}_{k-1}=0}=1|Y_{k-1}^{a, e_k=1, \overline{e}_{k-1}=0}=0, U=u, Z=z),
    \end{align*}
    where both lines follow from exposure necessity.
    Therefore by analogous rewriting
    \begin{align*}
        &\mathbb{P}(Y_{k}^{a, e_k=1, \overline{e}_{k-1}=0}=1|Y_{k-1}^{a, e_k=1, \overline{e}_{k-1}=0}=0, U=u, Z=z)\\
        &=\mathbb{P}(Y_{k}^{a, e_k=1, \overline{e}_{k-1}=0}=1|U=u, Z=z)=\mathbb{P}(Y_{k}^{a, e_k=2, \overline{e}_{k-1}=0}=2|U=u, Z=z)\\
        &=1 \cdot \mathbb{P}(Y_{k}^{a, e_k=2, \overline{e}_{k-1}=0}=2|Y_{k-1}^{a, e_k=2, \overline{e}_{k-1}=0}=0, U=u, Z=z)
        .
    \end{align*}
    for all $a \in \{0,1\}$ and $k \in \{1, \dots , K\}$.
\end{proof}

\subsubsection{Proof of Proposition \ref{PROP: EET identifiability}}\label{APP: PROP EET ident proof}
\begin{proof}
    \begin{align*}
        \frac{\mathbb{P}(Y^{a=1, e=1}=1|L=l)}{\mathbb{P}(Y^{a=1, e=2}=2|L=l)}&=
        \frac{\mathbb{P}(Y^{a=1, e=1}=1|L=l, A=1, E^{a=1}=1)}{\mathbb{P}(Y^{a=1, e=2}=2|L=l, A=1, E^{a=1}=2)}\\
        &=\frac{\mathbb{P}(Y=1|L=l, A=1, E=1)}{\mathbb{P}(Y=2|L=l, A=1, E=2)}\\
        &=\frac{\mathbb{P}(Y=1|A=1, L=l)}{\mathbb{P}(Y=2|A=1, L=l)} \cdot \left(\frac{\mathbb{P}(E=1|A=1, L=l)}{\mathbb{P}(E=2|A=1, L=l)}\right)^{-1}
    \end{align*}
    where the first line follows from Assumption \ref{ASM: Exp exch}, the second line from Assumptions \ref{ASM: Exp pos} and \ref{ASM: Exp cons}, and the last line from using Assumptions \ref{ASM: Exp nec} and \ref{ASM: No cross} and the law of total probability.
\end{proof}

\subsubsection{Proof of Proposition \ref{PROP: Marginal EET}}\label{APP: Proof EET marg}
\begin{proof}
    \begin{align*}
        \frac{\mathbb{P}(Y^{a=1, e=1}=1)}{\mathbb{P}(Y^{a=1, e=2}=2)}&=\frac{\sum_l \mathbb{P}(Y^{a=1, e=1}=1|L=l)\cdot \mathbb{P}(L=l)}{\sum_l \mathbb{P}(Y^{a=1, e=2}=2|L=l)\cdot \mathbb{P}(L=l)}\\
        &=\frac{\beta_1 \sum_l \mathbb{P}(Y^{a=1, e=2}=2|L=l)\cdot \mathbb{P}(L=l)}{\sum_l \mathbb{P}(Y^{a=1, e=2}=2|L=l)\cdot \mathbb{P}(L=l)}\\
        &=\frac{\mathbb{P}(Y^{a=1, e=1}=1|L=l)}{\mathbb{P}(Y^{a=1, e=2}=2|L=l)}\\
        &=\frac{\mathbb{P}(Y=1|A=1, L=l)}{\mathbb{P}(Y=2|A=1, L=l)}
    \end{align*}
    for all $l \in \mathcal{L}$. the first line follows from the laws of probability, the second and the third from Assumption \ref{ASM: Prop potential outcome}, while the last line from Proposition \ref{PROP: EET identifiability} and Assumption \ref{ASM: Equal treated exp prob}.
    Then by the laws of probability and the findings above
    \begin{align*}
        \frac{\mathbb{P}(Y=1|A=1)}{\mathbb{P}(Y=2|A=1)}&=\frac{\sum_l \mathbb{P}(Y=1|A=1, L=l)\cdot \mathbb{P}(L=l)}{\sum_l \mathbb{P}(Y=2|A=1, L=l)\cdot \mathbb{P}(L=l)}\\
        &=\frac{\beta_1 \sum_l \mathbb{P}(Y=2|A=1, L=l)\cdot \mathbb{P}(L=l)}{\sum_l \mathbb{P}(Y=2|A=1, L=l)\cdot \mathbb{P}(L=l)}\\
        &=\frac{\mathbb{P}(Y^{a=1, e=1}=1)}{\mathbb{P}(Y^{a=1, e=2}=2)}
    \end{align*}
    
\end{proof}

\subsubsection{Proof of Proposition \ref{PROP: EIE ident}}\label{APP: EIE ident}
\begin{proof}
    \begin{align*}
        \frac{\frac{\mathbb{P}(Y^{a=1, e=1}=1|L=l)}{\mathbb{P}(Y^{a=0, e=1}=1|L=l)}}
        {\frac{\mathbb{P}(Y^{a=1, e=2}=2|L=l)}{\mathbb{P}(Y^{a=0, e=2}=2|L=l)}}&=
        \frac{\frac{\mathbb{P}(Y^{a=1, e=1}=1|L=l, A=1, E^{a=1}=1)}{\mathbb{P}(Y^{a=0, e=1}=1|L=l, A=0, E^{a=0}=1)}}
        {\frac{\mathbb{P}(Y^{a=1, e=2}=2|L=l, A=1, E^{a=1}=2)}{\mathbb{P}(Y^{a=0, e=2}=2|L=l, A=0, E^{a=0}=2)}}\\
        &=\frac{\frac{\mathbb{P}(Y=1|L=l, A=1, E=1)}{\mathbb{P}(Y=1|L=l, A=0, E=1)}}{\frac{\mathbb{P}(Y=2|L=l, A=1, E=2)}{\mathbb{P}(Y=2|L=l, A=0, E=2)}}\\
        &=\frac{\frac{\mathbb{P}(Y=1|L=l, A=1)}{\mathbb{P}(Y=1|L=l, A=0)}}{\frac{\mathbb{P}(Y=2|L=l, A=1)}{\mathbb{P}(Y=2|L=l, A=0)}} \cdot 
        \left(\frac{\frac{\mathbb{P}(E=1|L=l, A=1)}{\mathbb{P}(E=1|L=l, A=0)}}{\frac{\mathbb{P}(E=2|L=l, A=1)}{\mathbb{P}(E=2|L=l, A=0)}}\right)^{-1}\\
        &=\frac{\frac{\mathbb{P}(Y=1|L=l, A=1)}{\mathbb{P}(Y=1|L=l, A=0)}}{\frac{\mathbb{P}(Y=2|L=l, A=1)}{\mathbb{P}(Y=2|L=l, A=0)}}
    \end{align*}
\end{proof}
where the first line follows by Assumption \ref{ASM: Gen exp exch}, the second line from Assumptions \ref{ASM: Gen exp pos} and \ref{ASM: Gen exp cons}.
For the third line the law of total probability and Assumptions \ref{ASM: Exp nec} and \ref{ASM: No cross} were used to show
$$
\mathbb{P}(Y=j|L=l, A=a)=\mathbb{P}(Y=j|L=l, A=a, E=j) \cdot \mathbb{P}(E=j|L=l, A=a).
$$
Finally, by Assumptions \ref{ASM: Cond no effect on exp} the last line follows.

\subsubsection{Proof of Proposition \ref{PROP: Marginal EIE}}\label{APP: Marginal EIE}

\begin{proof}
    \begin{align*}
            \frac{\frac{\mathbb{P}(Y^{a=1,e=1}=1)}{\mathbb{P}(Y^{a=0,e=1}=1)}}{\frac{\mathbb{P}(Y^{a=1,e=2}=2)}{\mathbb{P}(Y^{a=0,e=2}=2)}}&=
            \frac{\frac{\sum_l \mathbb{P}(Y^{a=1,e=1}=1|L=l)\cdot \mathbb{P}(L=l)}{\sum_l\mathbb{P}(Y^{a=0,e=1}=1|L=l)\cdot \mathbb{P}(L=l)}}{\frac{\sum_l\mathbb{P}(Y^{a=1,e=2}=2|L=l)\cdot \mathbb{P}(L=l)}{\sum_l \mathbb{P}(Y^{a=0,e=2}=2|L=l)\cdot \mathbb{P}(L=l)}}\\
            &=\frac{\frac{\beta_1 \sum_l \mathbb{P}(Y^{a=1,e=2}=2|L=l)\cdot \mathbb{P}(L=l)}{\beta_0 \sum_l\mathbb{P}(Y^{a=0,e=2}=2|L=l)\cdot \mathbb{P}(L=l)}}{\frac{\sum_l\mathbb{P}(Y^{a=1,e=2}=2|L=l)\cdot \mathbb{P}(L=l)}{\sum_l \mathbb{P}(Y^{a=0,e=2}=2|L=l)\cdot \mathbb{P}(L=l)}}\\
            &=\frac{\frac{\mathbb{P}(Y^{a=1,e=1}=1|L=l)}{\mathbb{P}(Y^{a=0,e=1}=1|L=l)}}{\frac{\mathbb{P}(Y^{a=1,e=2}=2|L=l)}{\mathbb{P}(Y^{a=0,e=2}=2|L=l)}}\\
            &=\frac{\frac{\mathbb{P}(Y=1|A=1, L=l)}{\mathbb{P}(Y=1|A=0, L=l)}}{\frac{\mathbb{P}(Y=2|A=1, L=l)}{\mathbb{P}(Y=2|A=0, L=l)}}
    \end{align*}
    for all $l \in \mathcal{L}$. The first line follows from the laws of probability, the second and the third lines from Assumption \ref{ASM: General proportional potential outcomes}, and the last line from Proposition \ref{PROP: EIE ident}.
    Then,
    \begin{align*}
        \frac{\frac{\mathbb{P}(Y=1|A=1)}{\mathbb{P}(Y=1|A=0)}}{\frac{\mathbb{P}(Y=2|A=1)}{\mathbb{P}(Y=2|A=0)}}&=
        \frac{\frac{\sum_l \mathbb{P}(Y=1|A=1,L=l)\cdot \mathbb{P}(L=l)}{\sum_l\mathbb{P}(Y=1|A=0,L=l)\cdot \mathbb{P}(L=l)}}{\frac{\sum_l\mathbb{P}(Y=2|A=1,L=l)\cdot \mathbb{P}(L=l)}{\sum_l\mathbb{P}(Y=2|A=0,L=l)\cdot \mathbb{P}(L=l)}}\\
        &=\frac{\frac{\beta_1 \sum_l \mathbb{P}(Y=2|A=1,L=l)\cdot \mathbb{P}(L=l)}{\beta_0 \sum_l\mathbb{P}(Y=2|A=0,L=l)\cdot \mathbb{P}(L=l)}}{\frac{\sum_l\mathbb{P}(Y=2|A=1,L=l)\cdot \mathbb{P}(L=l)}{\sum_l\mathbb{P}(Y=2|A=0,L=l)\cdot \mathbb{P}(L=l)}}\\
        &=\frac{\frac{\mathbb{P}(Y^{a=1,e=1}=1)}{\mathbb{P}(Y^{a=0,e=1}=1)}}{\frac{\mathbb{P}(Y^{a=1,e=2}=2)}{\mathbb{P}(Y^{a=0,e=2}=2)}}
    \end{align*}
    by the laws of probability and the findings above.
\end{proof}

\subsubsection{Proof of Proposition \ref{PROP: Ident when cens}}\label{APP: Prop ident when cens proof}
\begin{proof}
    \begin{align*}
        \frac{\frac{\mathbb{P}(Y^{a=1, c=0}=1|E^{a=1, c=0}=1)}{\mathbb{P}(Y^{a=0, c=0}=1|E^{a=0, c=0}=1)}}
        {\frac{\mathbb{P}(Y^{a=1, c=0}=2|E^{a=1, c=0}=2)}{\mathbb{P}(Y^{a=0, c=0}=2|E^{a=0, c=0}=2)}}&=
        \frac{\frac{\mathbb{P}(Y^{a=1, c=0}=1|E^{a=1, c=0}=1, A=1)}{\mathbb{P}(Y^{a=0, c=0}=1|E^{a=0, c=0}=1, A=0)}}
        {\frac{\mathbb{P}(Y^{a=1, c=0}=2|E^{a=1, c=0}=2, A=1)}{\mathbb{P}(Y^{a=0, c=0}=2|E^{a=0, c=0}=2, A=0)}} \cdot 
        \frac{\frac{\mathbb{P}(E^{a=1, c=0}=1|A=1)}{\mathbb{P}(E^{a=0, c=0}=1|A=0)}}{\frac{\mathbb{P}(E^{a=1, c=0}=2|A=1)}{\mathbb{P}(E^{a=0, c=0}=2|A=0)}}\\
        &= \frac{\frac{\mathbb{P}(Y=1|E=1, A=1)}{\mathbb{P}(Y=1|E=1, A=0)}}
        {\frac{\mathbb{P}(Y=2|E=2, A=1)}{\mathbb{P}(Y=2|E=2, A=0)}}\cdot 
        \frac{\frac{\mathbb{P}(E=1|A=1)}{\mathbb{P}(E=1|A=0)}}{\frac{\mathbb{P}(E=2|A=1)}{\mathbb{P}(E=2|A=0)}}\\
        &=\frac{\frac{\mathbb{P}(Y=1|A=1)}{\mathbb{P}(Y=1|A=0)}}
        {\frac{\mathbb{P}(Y=2|A=1)}{\mathbb{P}(Y=2|A=0)}}
    \end{align*}
    where the first line follows from Assumption \ref{ASM: Cens exch} and \ref{ASM: cens no eff exp}. The second line follows from Assumptions \ref{ASM: Cens pos} and \ref{ASM: Cens cons}, while the last line follows as in the proof \ref{APP: Prop no eff on exp ratios proof}, using Assumptions \ref{ASM: cens exp nec} and \ref{ASM: cens no cross}.
\end{proof}

\subsubsection{Proof of Proposition \ref{PROP: CCE TTE ident}}\label{APP: Proof CCE TTE}

\begin{proof}
    \begin{align*}
        \frac{\bbP(Y_k^{a=1}=j)}{\bbP(Y_k^{a=0}=j)}&=
        \frac{\bbP(Y_k^{a=1}=j|E_k^{a=1}\neq0)\bbP(E_k^{a=1}\neq0)}{\bbP(Y_k^{a=0}=j|E_k^{a=0}\neq0)\bbP(E_k^{a=0}\neq0)}
    \end{align*}
    for all $j \in \{1,2\}$,
    where the equality follows from the laws of probability and Assumption \ref{ASM: TTE original exp nec}.
    Then
    \begin{align*}
        &\frac{\bbP(Y_k^{a=1}=1)}{\bbP(Y_k^{a=0}=1)} \cdot \left(\frac{\bbP(Y_k^{a=1}=2)}{\bbP(Y_k^{a=0}=2)} \right)^{-1}\\
        &=\frac{\bbP(Y_k^{a=1}=1|E_k^{a=1}\neq0)\bbP(E_k^{a=1}\neq0)}{\bbP(Y_k^{a=0}=1|E_k^{a=0}\neq0)\bbP(E_k^{a=0}\neq0)} 
        \cdot
        \left(\frac{\bbP(Y_k^{a=1}=2|E_k^{a=1}\neq0)\bbP(E_k^{a=1}\neq0)}{\bbP(Y_k^{a=0}=2|E_k^{a=0}\neq0)\bbP(E_k^{a=0}\neq0)}\right)^{-1}\\
        &=\frac{\bbP(Y_k^{a=1}=1|E_k^{a=1}\neq0)}{\bbP(Y_k^{a=0}=1|E_k^{a=0}\neq0)}
        \cdot
        \left(\frac{\bbP(Y_k^{a=1}=2|E_k^{a=1}\neq0)}{\bbP(Y_k^{a=0}=2|E_k^{a=0}\neq0)}\right)^{-1}
    \end{align*}
    Lastly, under Assumptions \ref{ASM: TTE EXCH}-\ref{ASM: TTE Cons}, $\bbP(Y_k^{a}=j)$ can be expressed in terms of the observed cumulative incidence.
\end{proof}

\subsection{Simulation studies}\label{APP: Simulation}
This section will illustrate the performance of the proposed estimator under different sample sizes in simulated examples. The simulations also illustrate the behaviour of the estimator in certain special cases where identifying assumptions are violated.

\subsubsection{Data-generation}
In the first scenario, the data is generated from a setting where Assumptions \ref{ASM: no multiple exp}, \ref{ASM: No eff exp}, \ref{ASM: Exp nec}, and \ref{ASM: No cross} hold. Let us consider individual-level data generating as follows
\begin{equation}\label{EQ: data gen 1}
    \begin{aligned}
        A \sim Ber\left(\frac{1}{2}\right)\\
        E \sim Unif\{0,1,2\}.
    \end{aligned}
\end{equation}
Let the probability of the outcome be determined by 
\begin{equation}\label{EQ: data gen 2}
    \begin{aligned}
        \mathbb{P}(Y=j|E=e, A=a) &=
        \begin{cases}
        \text{expit}(\beta_0+\beta_e e)\cdot \text{expit}(\beta_{e, a} a) & \text{if } j=e, \; e \neq 0 ,\\
        1-\text{expit}(\beta_0+\beta_e e)\cdot \text{expit}(\beta_{e, a} a) & \text{if } j=0, \, e\neq 0,\\
        1 & \text{if } j=e=0,\\
        0 & \text{otherwise,}
        \end{cases}
    \end{aligned}
\end{equation}
where the treatment term is included in the model for the probability of the outcome in a separate logistic sigmoid function, to guarantee that the treatment has a multiplicative effect on the probability of developing the outcome.
If $\beta_{e,a}=\beta_a$ $\forall \ e \in {1,2}$, then the vaccine effect is identical for both variants, hence we expect the estimate of the $CCS$ to be 1. In the following assume that\\ $\beta_0=-2, \, \beta_{e=0}=-1, \, \beta_{e=1}=2, \, \beta_{e=2}=1$ and $\beta_{1,a}=\beta_{2,a}=\beta_a=-3$.

\subsubsection{Estiamtion of the $CCS$ under Assumptions \ref{ASM: No eff exp}-\ref{ASM: No cross}} 
\begin{figure}[!h]
    \centering
    \includegraphics[width=14cm]{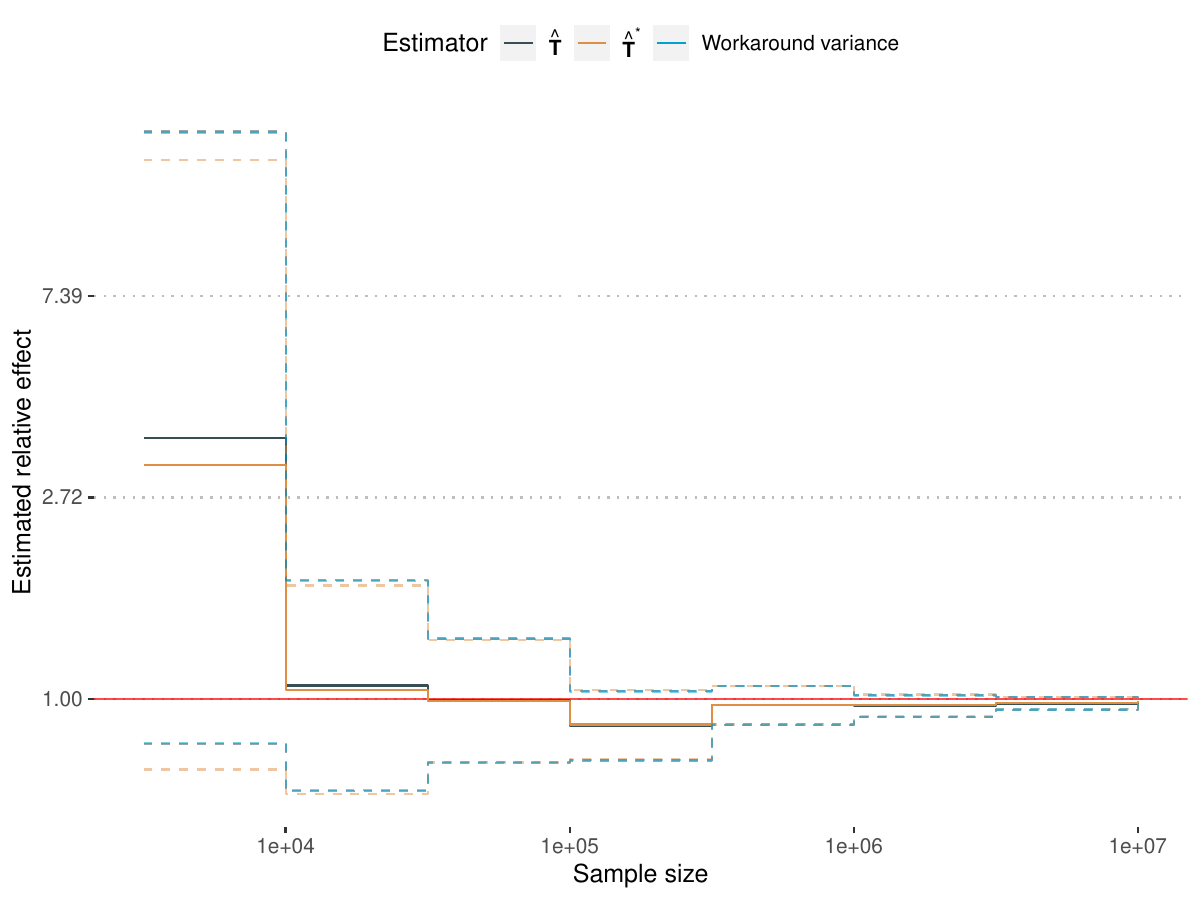}
    \caption{If Assumptions \ref{ASM: No eff exp}-\ref{ASM: No cross} hold, then the estimator of the $CCS$ converge to the truth, regardless of whether the conditioning set includes $E$ or not with approximately equal variance.}
    \label{FIG: Ideal estimator}
\end{figure}

See Figure \ref{FIG: Ideal estimator}

\subsubsection{Estimation of $CCS$ under relaxed Assumption \ref{ASM: No eff exp}}
While the data-generating process (see Equations \ref{EQ: data gen 1}) was designed to adhere to the assumption of no effect on exposure, as it was shown in Section \ref{SEC: Identification}, it is not a necessary condition for the identification of the $CCS$, it can be relaxed to Assumption \ref{ASM: No eff on exp ratios} instead.
Thus, let us assume that
\begin{equation}\label{EQ: Exp CCS ratio}
    \begin{aligned}
        \mathbb{P}(E=e|A=a)&=
        \begin{cases}
        \text{if } a=1
            \begin{cases}
                \frac{10}{20}  &\text{if } e = 0 \\
                \frac{5}{20}  &\text{if } e = 1 \\ 
                \frac{5}{20}  &\text{if } e = 2 \\
            \end{cases}\\
        \text{if } a=0
                \begin{cases}
                \frac{4}{20}  &\text{ if } e = 0 \\
                \frac{8}{20}  &\text{ if } e = 1 \\ 
                \frac{8}{20}  &\text{ if } e = 2 \\
                \end{cases}
        \end{cases}
    \end{aligned}
\end{equation}
and leave the other elements of the data-generating mechanism unchanged.
\begin{figure}
    \centering
    \includegraphics[width=14cm]{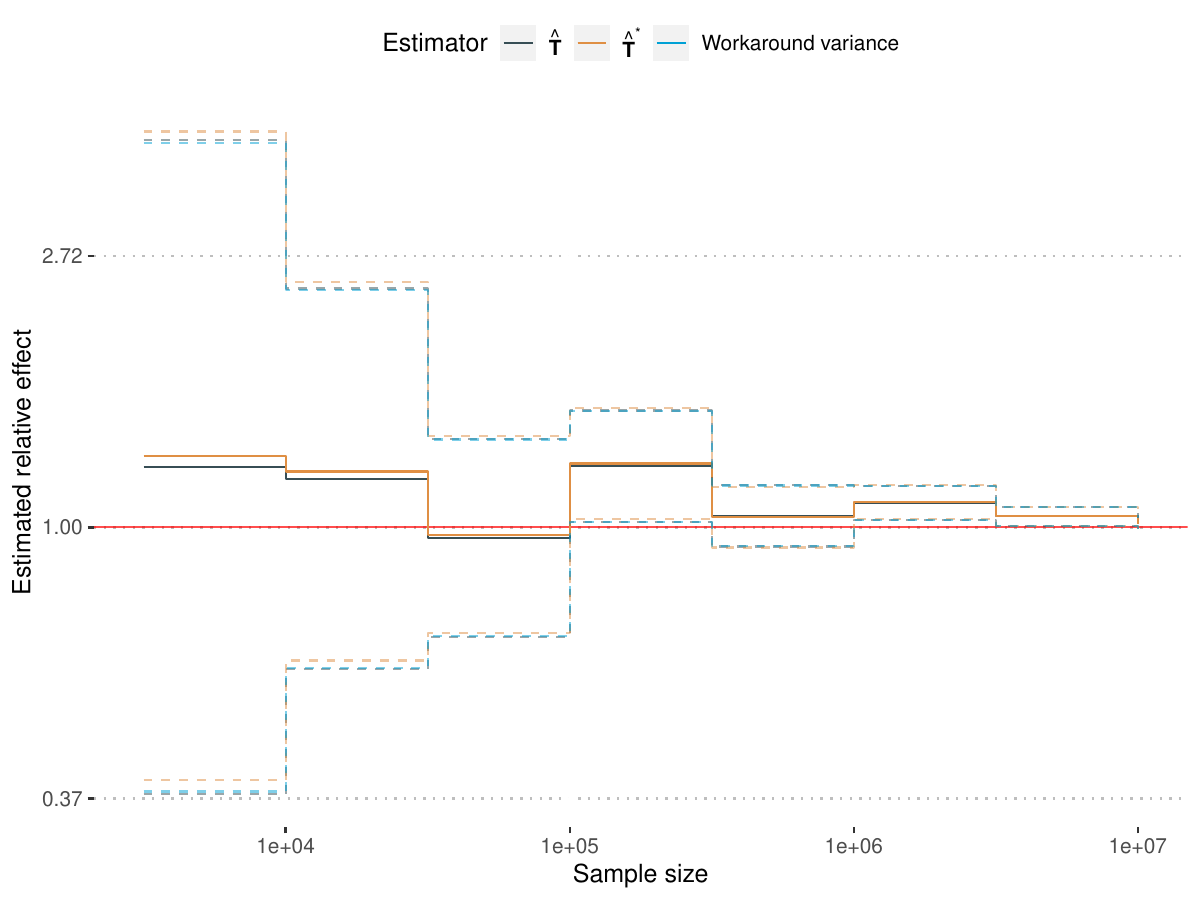}
    \caption{Estimator of the $CCS$, when Assumption \ref{ASM: No eff exp}, but \ref{ASM: No eff on exp ratios} holds. While the estimand is still identifiable, the change in the proportion of the unexposed between the two levels of the treatment can increase the variance of the estimator.}
    \label{FIG: Changing non-exp prob}
\end{figure}

In Figure \ref{FIG: Changing non-exp prob}, we can see that even though the estimator converges to the true contrast conditional on specific exposure, the variance increases, regardless of whether the exposure status information was used or not.

\subsubsection{Estimation of $CCS$ when Assumption \ref{ASM: No eff on exp ratios} fails}
Consider the setting where neither Assumption \ref{ASM: No eff exp} nor \ref{ASM: No eff on exp ratios} holds. Then the formula used for the estimation of the $CCS$ can no longer guarantee identification. However, under the standard identifiability Assumptions \ref{ASM: Standard RCT}, the estimator $\hat{T}^*$ still identifies the ratio between the two vaccine effect estimands.
Let us assume that
\begin{equation}\label{EQ: Exp CCS no ratio}
    \begin{aligned}
        \mathbb{P}(E=e|A=a)&=
        \begin{cases}
        \text{if } a=1
            \begin{cases}
                \frac{3}{6}  &\text{if } e = 0 \\
                \frac{1}{6}  &\text{if } e = 1 \\ 
                \frac{2}{6}  &\text{if } e = 2 \\
            \end{cases}\\
        \text{if } a=0
                \begin{cases}
                \frac{3}{6}  &\text{ if } e = 0 \\
                \frac{2}{6}  &\text{ if } e = 1 \\ 
                \frac{1}{6}  &\text{ if } e = 2 \\
                \end{cases}
        \end{cases}
    \end{aligned}
\end{equation}
thus, the the relative exposure probabilities between the two variants, change with the treatment assignment. 
\begin{figure}
    \centering
    \includegraphics[width=14 cm]{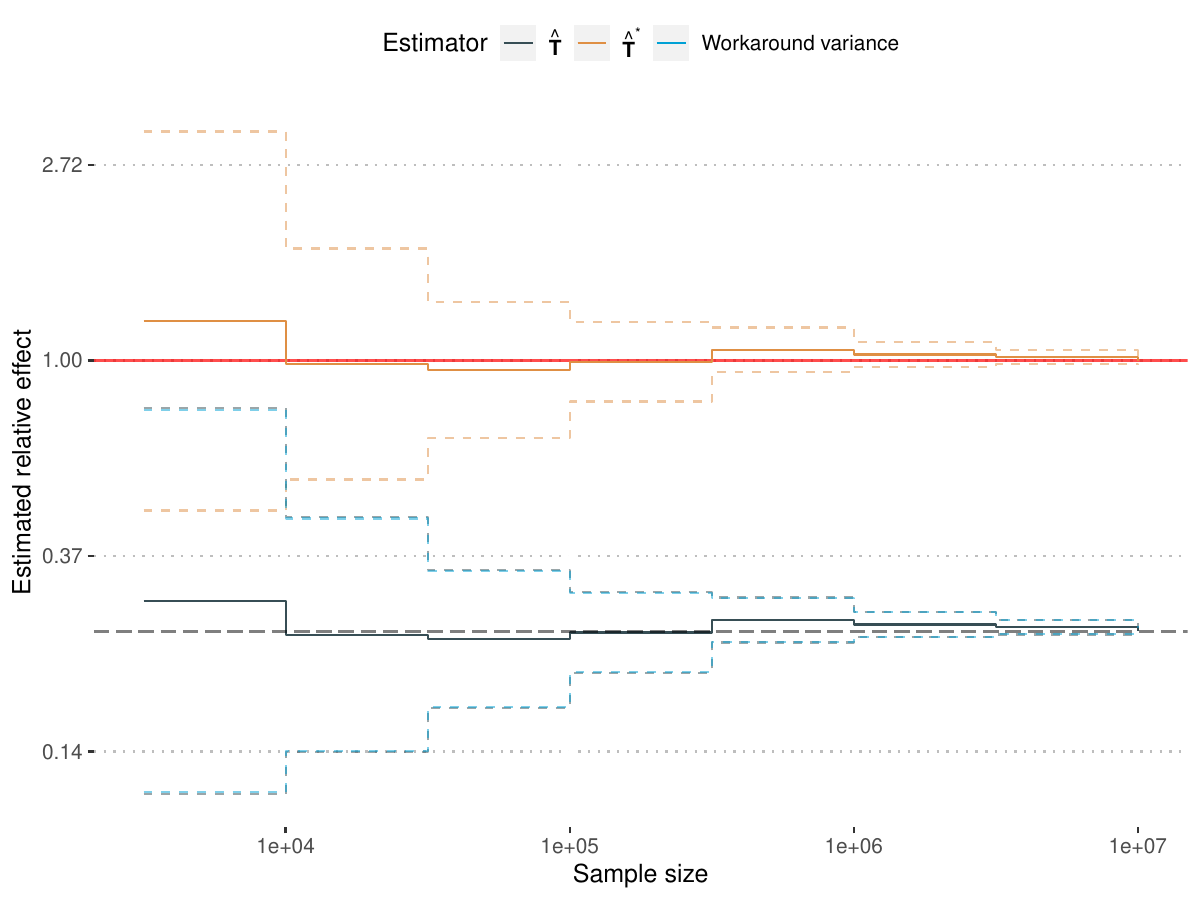}
    \caption{$CCS$ estimator with and without conditioning on the exposure status when \ref{ASM: No eff on exp ratios} fails. If the conditioning set includes $E$, then the estimator converge to the true value of the $CCS$, otherwise, the observable estimator is scaled by the relative change of the exposure ratios between the two levels of the treatment.}
    \label{FIG: Changing exp ratio}
\end{figure}

In Figure \ref{FIG: Changing exp ratio}, the estimator conditional on the exposure status converges to the true ratio between the vaccine effects for the two variants. In contrast, using the estimator based on the observed data, it converges to the ratio of the relative exposure that depends on the treatment assignments ($(\frac{1}{6}/\frac{2}{6})/(\frac{2}{6}/\frac{1}{6})=\frac{1}{4}$). Thus if we were to make an assumption about this ratio, such as that it is constrained between given bounds, we could use our estimator, without conditioning on the exposure, to derive statements; such as hypothesis tests about the $CCS$.

\subsubsection{Estimation of the $EET$ under Assumption \ref{ASM: Equal treated exp prob}}
Consider the same data generation Equations \eqref{EQ: data gen 1} and \eqref{EQ: data gen 2}. Modify the distribution of $\mathbb{P}(E|A)$ such that it satisfies Assumption \ref{ASM: Equal treated exp prob}, namely 
\begin{equation}\label{EQ: Exp EET equal}    
    \begin{aligned}
        \mathbb{P}(E=e|A=a)&=
        \begin{cases}
        \text{if } a=1
            \begin{cases}
                \frac{1}{5}  &\text{if } e = 0 \\
                \frac{2}{5}  &\text{if } e = 1 \\ 
                \frac{2}{5}  &\text{if } e = 2 \\
            \end{cases}\\
        \text{if } a=0
                \begin{cases}
                \frac{1}{7}  &\text{ if } e = 0 \\
                \frac{4}{7}  &\text{ if } e = 1 \\ 
                \frac{2}{7}  &\text{ if } e = 2 \\
                \end{cases}
        \end{cases}
    \end{aligned}
\end{equation}
\begin{figure}
    \centering
    \includegraphics[width=14 cm]{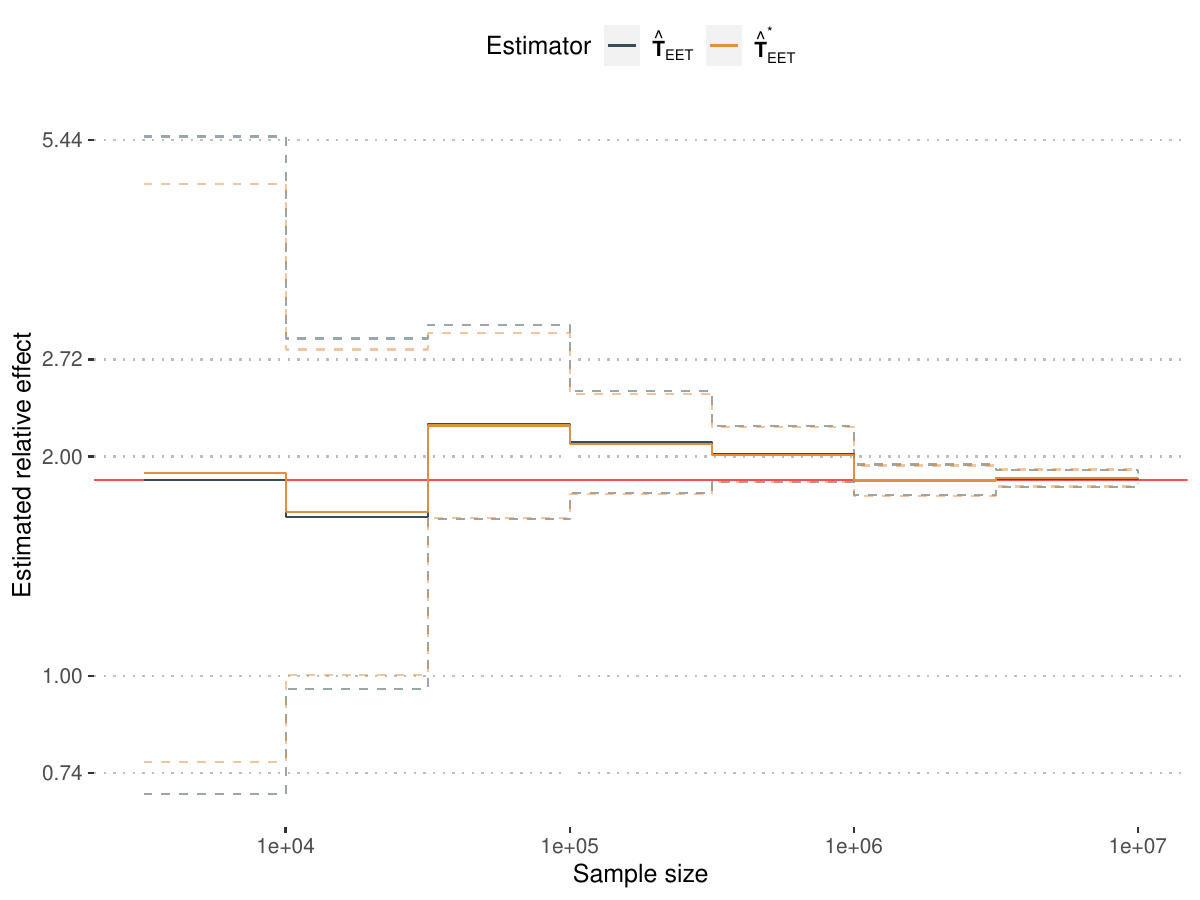}
    \caption{Unconditional $EET$, when data was generated under Assumption \ref{ASM: Equal treated exp prob}. Regardless of whether the estimator was conditioned on the exposure status or not, we observed that the  estimators converge to the true value, $\sigma(0)/\sigma(-1)\approx 1.86$, with increasing $n$.}
    \label{FIG: EET ideal}
\end{figure}
Regardless of whether we condition on the exposure status, both estimators converge to the true $EET$, with the increasing sample size, even though there are no constraints on the exposure status of the untreated, see Figure \ref{FIG: EET ideal}. Informally, since the estimators are based on the data obtained from the treated individuals, knowledge of exposures of the untreated should not change our estimates in the treated.
The true $EET$ is calculated based on the function for $\mathbb{P}(Y=j|E=e, A=a)$ defined in Equation \eqref{EQ: data gen 2}. Then the individual level $EET$ is $\sigma(0)/\sigma(-1)\approx 1.86$

\subsubsection{Estimation of the $EET$ when Assumption \ref{ASM: Equal treated exp prob} fails}
Assume that \ref{ASM: Equal treated exp prob} fails, and the two variants are present with different incidence rates among the population.
\begin{equation}\label{EQ: Exp EET not equal}
    \begin{aligned}
        \mathbb{P}(E=e|A=a)&=
        \begin{cases}
        \text{if } a=1
            \begin{cases}
                \frac{1}{7}  &\text{if } e = 0 \\
                \frac{4}{7}  &\text{if } e = 1 \\ 
                \frac{2}{7}  &\text{if } e = 2 \\
            \end{cases}\\
        \text{if } a=0
                \begin{cases}
                \frac{1}{7}  &\text{ if } e = 0 \\
                \frac{4}{7}  &\text{ if } e = 1 \\ 
                \frac{2}{7}  &\text{ if } e = 2 \\
                \end{cases}
        \end{cases}
    \end{aligned}
\end{equation}
\begin{figure}
    \centering
    \includegraphics[width=14 cm]{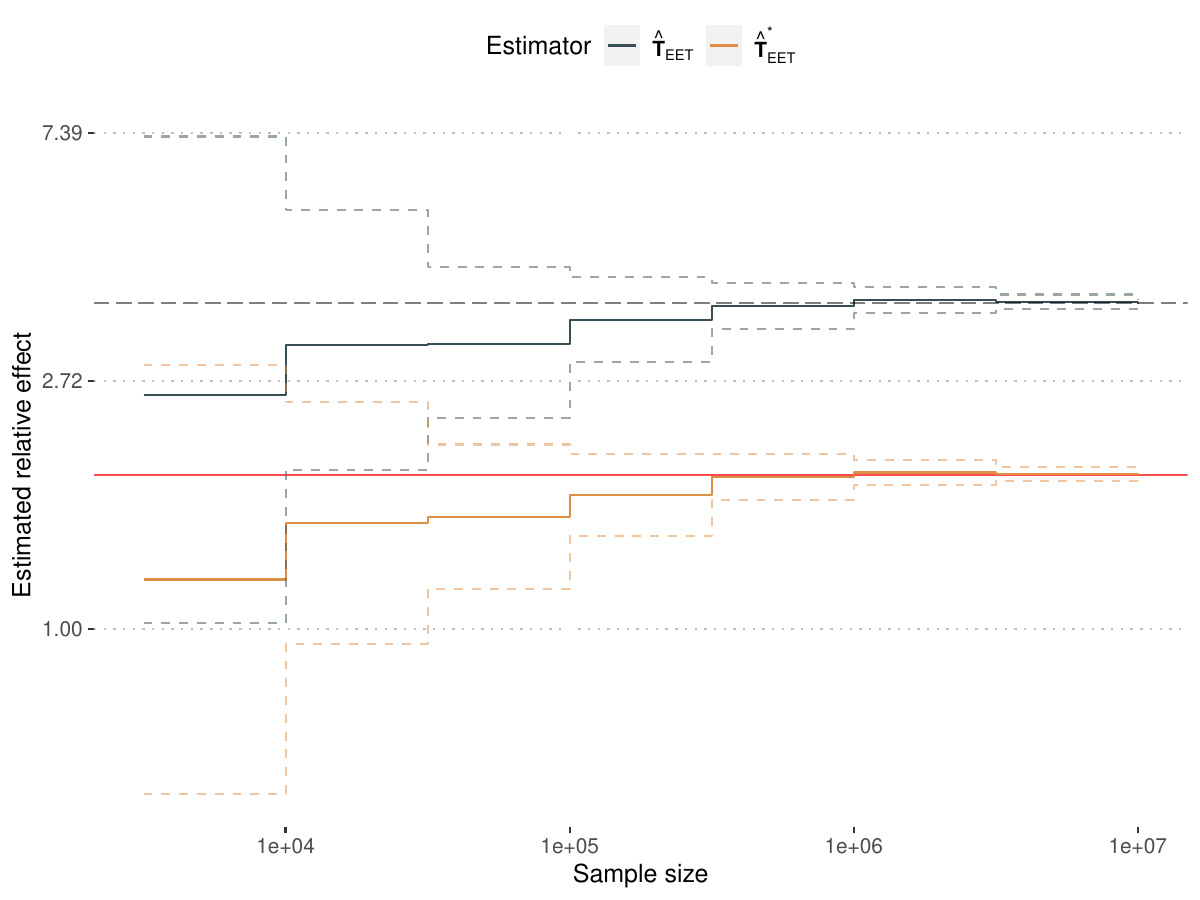}
    \caption{Unconditional $EET$ when Assumption \ref{ASM: Equal treated exp prob} fails, for increasing sample size $n$. The $EET$ estimator conditional on the exposure $E$ converges to the true value of the $EET$ ($\sigma(0)/\sigma(-1)\approx 1.86$). The $EET$ estimator based on the observed data converges to the scaled version of the true $EET$, which is multiplied by the factor of the ratio of exposures ($\approx 1.86 \cdot \frac{2}{1}=3.72$)}
    \label{FIG: EET not equal treated exposure}
\end{figure}
The estimator conditional on exposure status converges to the true $EET$, while it seems that the estimators based on the observed data are the systematically scaled version of that, see Figure \ref{FIG: EET not equal treated exposure}. This scaling factor exactly corresponds to the unobserved ratio of exposure among the treated $\mathbb{P}(E=1|A=1)/\mathbb{P}(E=2|A=1)$, that is now based on Equation \ref{EQ: Exp EET not equal} $\frac{4}{7}\slash \frac{2}{7}=2$. If conditional exposure probabilities are unknown, then $\mathbb{P}(E=1|A=1)/\mathbb{P}(E=2|A=1)$ can be obtained through the assumptions discussed in Appendix \ref{APP: EET and EIE}. Then after the successful estimation of this ratio, we can scale our estimator, to derive the desired ratio for the $EET$.

\subsection{Identification assumptions and identifying functions for the $EET$ and the $EIE$}\label{APP: EET and EIE}
\subsubsection{Effect of exposure under treatment}\label{APP: EET def}
\begin{assumption}\label{ASM: Exposure RCT}
    \begin{enumerate}[label=\theassumption\alph*, leftmargin=*]
        \item[]
        \item \label{ASM: Exp exch}  Exposure exchangeability:\\
        $E^{a=1} \independent A|L, \; Y^{a=1, e} \independent A| L, \text{ and } Y^{a=1, e} \independent E^{a=1}|L, A=1.$
        \item \label{ASM: Exp pos}  Exposure positivity:\\
        $\mathbb{P}(A=1, E=e|L)>0, \ \forall \ e \in \{1,2\}.$
        \item \label{ASM: Exp cons}  Exposure consistency:\\
        If $A=1$ then $E=E^{a=1}$ and if $A=1$ and $E=e$ then $Y=Y^{a=1, e}, \ \forall \ e \in \{1,2\}$. 
    \end{enumerate}
\end{assumption}
Assumption \ref{ASM: Exp exch} requires the measurement of common causes of the outcome and the exposure, hence imposing stronger restrictions on the data-generating mechanism, than the usual exchangeability, that is Assumption \ref{ASM: Exchange}. Similarly, the positivity and consistency are adjusted for the measurement of the baseline variables.

These identification assumptions are similar to the ones considered in the Appendix of \citet{stensrud_identification_2023} for the identification of the controlled direct effect (CDE). However, since the $EET$ is the contrast of two potential outcomes in which the treatment was intervened as $a=1$, Assumption \ref{ASM: Exposure RCT} is defined for those variables. 

\begin{proposition}\label{PROP: EET identifiability}
    Under Assumptions \ref{ASM: no multiple exp}, \ref{ASM: Exp nec}, \ref{ASM: No cross}, and \ref{ASM: Exposure RCT} the ratio between the probabilities of the counterfactual outcomes, conditional on baseline covariates $L$ from Definition \ref{DEF: EET} can be identified.
    $$EET(l)=\frac{\mathbb{P}(Y^{a=1, e=1}=1|L=l)}{\mathbb{P}(Y^{a=1, e=2}=2|L=l)}=
    \frac{\mathbb{P}(Y=1|A=1, L=l)}{\mathbb{P}(Y=2|A=1, L=l)} \cdot \left(\frac{\mathbb{P}(E=1|A=1, L=l)}{\mathbb{P}(E=2|A=1, L=l)}\right)^{-1}.$$
\end{proposition} 
See Appendix \ref{APP: PROP EET ident proof} for a proof.

If the baseline covariates confounding the outcome $Y$ and the $E$ are measured, then the first term in the identification formula can be straightforwardly approximated. However, estimating the second term requires the measurement of exposure, which is often not the case. Therefore to be able to identify the $EET$ conditional on $L$, if exposures remain unmeasured, further assumptions are required.
\begin{assumption}[Equal exposure of the treated]\label{ASM: Equal treated exp prob}
    $$\mathbb{P}(E=1|A=1, L=l)=\mathbb{P}(E=2|A=1, L=l)$$
\end{assumption}
\begin{remark}
    Under Assumptions \ref{ASM: Exposure RCT}, Assumption \ref{ASM: Equal treated exp prob} is identical to 
    $\mathbb{P}(E^{a=1}=1|L=l)= \mathbb{P}(E^{a=1}=2|L=l).$
\end{remark}
Assumption \ref{ASM: Equal treated exp prob} is significantly stronger than Assumptions \ref{ASM: No eff exp} and \ref{ASM: No eff on exp ratios}. While previously we only required some level of correspondence between the probabilities under the different levels of treatment assignments, Assumption \ref{ASM: Equal treated exp prob} requires that treated individuals are exposed to both of the competing variants at the same rate. 
Alternatively, Assumption \ref{ASM: Equal treated exp prob} can be experimentally assessed, by measuring the prevalence of the two competing variants, and assuming homogeneous mixing, investigators could potentially derive that the exposure probability of the treated is equal for the two variants. However, to deduce that this holds for the treated from separately collected observed data, we would need to modify our previous Assumption \ref{ASM: No eff on exp ratios}.
\begin{assumption}[A conditional version of no effect on exposure ratios]\label{ASM: Cond no effect on exp}
    $$\frac{\mathbb{P}(E^{a=1}=1|L=l)}{\mathbb{P}(E^{a=1}=2|L=l)}= \frac{\mathbb{P}(E^{a=0}=1|L=l)}{\mathbb{P}(E^{a=0}=2|L=l)}.$$
\end{assumption}
Under Assumption \ref{ASM: Cond no effect on exp}, if the observed data satisfies $\mathbb{P}(E=1)=\mathbb{P}(E=2)$, then it can be concluded that Assumption \ref{ASM: Equal treated exp prob} holds, and hence the conditional $EET$ can be identified.

Lastly, data extracted from the untreated population could be used to estimate the ratio between exposure probabilities. Under the Assumptions \ref{ASM: Exp nec} and \ref{ASM: No cross}, the equality holds:
\begin{equation*}\label{EQ: Untreated exposure prob}
    \frac{\mathbb{P}(Y=1|A=0, L=l)}{\mathbb{P}(Y=2|A=0, L=l)} \cdot \left(\frac{\mathbb{P}(Y=1|E=1, A=0, L=l)}{\mathbb{P}(Y=2|E=2, A=0, L=l)} \right)^{-1}= \frac{\mathbb{P}(E=1|A=0, L=l)}{\mathbb{P}(E=2|A=0, L=l)}.
\end{equation*}

Then, invoking Assumptions \ref{ASM: Exposure RCT} and \ref{ASM: Cond no effect on exp}, the right-hand side is equal to 
$$
\frac{\mathbb{P}(E=1|A=1, L=l)}{\mathbb{P}(E=2|A=1, L=l)}.
$$
However, the ratio between $\mathbb{P}(Y=1|E=1, A=0, L=l)$  and\\
$\mathbb{P}(Y=2|E=2, A=0, L=l)$ is unknown. Thus we have to make an assumption about its value (potentially being equal to 1), to be able to estimate the conditional $EET$ from the observed data.

\begin{remark}\label{REM: collapse EET}
    For the stronger identification assumptions to hold, Assumptions \ref{ASM: Exposure RCT}, values of the baseline covariates $L$ must be conditioned on. Consequently, the $EET$ is expressed conditional on these baseline covariates. However, if we further assume proportionality between the potential outcomes that hold constant across the strata of $L$, the marginal $EET$ can be expressed in a simple form. We will state the assumption formally, and then provide some explanation and the identification result for the marginal $EET$.
    \begin{assumption}\label{ASM: Prop potential outcome}
        $\mathbb{P}(Y^{a=1, e=1}=1|L=l)=\beta_1 \mathbb{P}(Y^{a=1, e=2}=2|L=l)$
        for all $l \in \mathcal{L}$.
    \end{assumption}
Assumption \ref{ASM: Prop potential outcome} requires that under a given level of treatment and intervention on exposure, the probability of developing the outcome from variant 1, is proportional to the probability of developing the outcome from variant 2, such that the ratio between the two does not depend on the baseline covariates $L$. Assumption \ref{ASM: Prop potential outcome} does not mean sieve effect or the absence of it, as it imposes no constraints on the value of $\beta_1$. However, note that if $\beta_1=1$, then the $EET=1$ for all $l \in \mathcal{L}$.

\begin{proposition}[Identification of the marginal $EET$]\label{PROP: Marginal EET}
    Under Assumptions \ref{ASM: Exp nec}, \ref{ASM: No cross}, \ref{ASM: Exposure RCT},\ref{ASM: Equal treated exp prob} and Assumption \ref{ASM: Prop potential outcome} the marginal $EET$ is identified as 
    $$
    EET=\frac{\mathbb{P}(Y=1|A=1)}{\mathbb{P}(Y=2|A=1)}
    $$
\end{proposition}
The proof is provided in Appendix \ref{APP: Proof EET marg}.
\end{remark}

\subsubsection{Effect with intervened exposure}
First, the modified identification assumptions are presented, and then the identification formula is stated.

\begin{assumption}\label{ASM: Generalised exposure RCT}
    \begin{enumerate}[label=\theassumption\alph*, leftmargin=*]
        \item[]
        \item \label{ASM: Gen exp exch}  Generalised exposure exchangeability:\\
        $E^{a} \independent A|L, \; Y^{a, e} \independent A| L, \text{ and } Y^{a, e} \independent E^{a}|L, A.$
        \item \label{ASM: Gen exp pos}  Generalised exposure positivity:\\
        $\mathbb{P}(A=a, E=e|L=l)>0, \ \forall \ e \in \{1,2\}, \ a  \in \{0,1\}, \ l \in \mathcal{L} .$
        \item \label{ASM: Gen exp cons}  Generalised exposure consistency:\\
        If $A=a$ then $E=E^{a}$ and if $A=a$ and $E=e$ then $Y=Y^{a=a, e}, \ \forall \ e \in \{1,2\}, \ \forall \ a  \in \{0,1\}$. 
    \end{enumerate}
\end{assumption}
Assumption \ref{ASM: Gen exp exch} is an extension of Assumption \ref{ASM: Exp exch}, as it requires conditional exchangeability for the untreated as well.\\
Similarly, the following two assumptions extend \ref{ASM: Exp pos} and \ref{ASM: Exp cons} for $A=0$ too.

Positing the stronger Assumption \ref{ASM: Generalised exposure RCT}, instead of \ref{ASM: Standard RCT}, and if
Assumptions \ref{ASM: Cond no effect on exp}, \ref{ASM: Exp nec}, and \ref{ASM: No cross}
holds, we can identify the conditional effect with intervened exposure.

\begin{proposition}\label{PROP: EIE ident}
    Under Assumptions \ref{ASM: no multiple exp}, \ref{ASM: Exp nec}, \ref{ASM: No cross}, \ref{ASM: Cond no effect on exp} and \ref{ASM: Generalised exposure RCT}, the effect with intervened exposure can be identified.
    \begin{align*}
        EIE(l)
        &=\left. \frac{\mathbb{P}(Y=1|A=1, L=l)}{\mathbb{P}(Y=1|A=0, L=l)} \middle \slash
    \frac{\mathbb{P}(Y=2|A=1, L=l)}{\mathbb{P}(Y=2|A=0, L=l)} \right.
    \end{align*}  
\end{proposition}
    
\begin{remark}
    Assumption \ref{ASM: Prop potential outcome} can be extended to hold for $a=0$ as well. Under this extension, the marginal $EIE$ can be identified as the fraction of conditional probabilities of the observed variables.
    \begin{assumption}\label{ASM: General proportional potential outcomes}
        $\mathbb{P}(Y^{a, e=1}=1|L=l)=\beta_a \mathbb{P}(Y^{a, e=2}=2|L=l)$
        for all $a \in \{0,1\}$ and $l \in \mathcal{L}$.
    \end{assumption}
Assumption \ref{ASM: General proportional potential outcomes} in itself does not imply a sieve effect either, as no constraints are applied to the relationship between $\beta_1$ and $\beta_0$. If it is further set that $\beta_1=\beta_0$, then $EIE=1$, hence there is no (sub)population level sieve effect.

\begin{proposition}\label{PROP: Marginal EIE}
    Under Assumptions \ref{ASM: No eff exp}, \ref{ASM: Exp nec}, \ref{ASM: No cross}, \ref{ASM: Generalised exposure RCT} and \ref{ASM: General proportional potential outcomes} the marginal $EIE$ can be identified
    \begin{align*}
        EIE
        &=\left. \frac{\mathbb{P}(Y=1|A=1)}{\mathbb{P}(Y=1|A=0)} \middle \slash
    \frac{\mathbb{P}(Y=2|A=1)}{\mathbb{P}(Y=2|A=0)} \right.
    \end{align*}  
\end{proposition}
Proof is provided in Appendix \ref{APP: Marginal EIE}
\end{remark}

\subsubsection{Connection between $EIE$ and $EET$}
As presented in the previous subsections, the exposure ratio of the untreated can be identified from the observed data, as long as the infectivity rate is known. Then under Assumption \ref{ASM: Cond no effect on exp}, and using Proposition \ref{PROP: EET identifiability}, the conditional $EET$ can be identified as
$$EET(l)=\left. \frac{\mathbb{P}(Y=1|A=1, L=l)}{\mathbb{P}(Y=2|A=1, L=l)} \middle \slash \frac{\mathbb{P}(Y=1|A=0, L=l)}{\mathbb{P}(Y=2|A=0, L=l)} \right. \cdot \frac{\mathbb{P}(Y=1|E=1, A=0, L=l)}{\mathbb{P}(Y=2|E=2, A=0, L=l)}$$
That is equal to the identification formula for the conditional $EIE$ under Assumptions \ref{ASM: Exp nec}, \ref{ASM: No cross}, \ref{ASM: Cond no effect on exp} and \ref{ASM: Generalised exposure RCT}, re-scaled by the infectivity rates of the untreated.
Hence by using the notation $$IR_0=\frac{\mathbb{P}(Y=1|E=1, A=0, L=l)}{\mathbb{P}(Y=2|E=2, A=0, L=l)}$$ for the infectivity rate of the untreated, the connection between the conditional $EET$ and the conditional $EIE$ can be summarised as 
$$
EET(l)=EIE(l)\cdot IR_0.
$$

The $EET(l)$ and the $EIE(l)$ can be equated under the following assumption:
\begin{assumption}[Equal infectivity of the untreated]\label{ASM: equal inf untrt}
    $$IR_0=\frac{\mathbb{P}(Y=1|E=1, A=0, L=l)}{\mathbb{P}(Y=2|E=2, A=0, L=l)}=1$$
\end{assumption}

\subsubsection{Data example of the $EET$ and the $EIE$}\label{APP: data EIE vs EET}
Suppose that the identification conditions in Assumption \ref{ASM: Exposure RCT} hold, conditional on the baseline variable \textit{risk profile}. Thus we partition the population into \textit{High risk} and \textit{Non-high risk} individuals. The estimated $EET$s were $1.11$ ($95\% $ CI: $[0.405, 3.090]$) and $4.4$ ($95\% $ CI: $[1.626,14.873]$) respectively.

However, as shown in Appendix \ref{APP: EET def}, we can replace Assumption \ref{ASM: Equal treated exp prob}, which in this setting is implausible, with Assumption \ref{ASM: Cond no effect on exp}. Then the $EET$ is identified by the identification functional of the $EIE$. Thus, the estimated $EIE$s, and correspondingly the estimated $EET$s,  for the \textit{High risk} and \textit{Non-high risk} are $0.208$ ($95\% $ CI: $[0.045,0.956]$) and $0.811$ ($95\% $ CI: $[0.23, 2.86]$), respectively.

There is a qualitative difference between the two sets of estimates, in the sense that the first estimates of the $EET$ show a stronger protective effect against the unmatched strain (variant 2), while the second estimates show a stronger effect against the matched strain (variant 1). Based on knowledge of vaccine mechanisms the latter is more reasonable.    

Nevertheless, both estimators are designed to estimate the heterogeneity of the vaccine effect among the treated.

Here we articulate an assumption for the value of the untreated infectivity rate, to connect the $EIE$ and the $EET$.
\begin{assumption}[Equal infectivity of the untreated]\label{ASM: equal infectivity untreated}
    $$IR_0=\frac{\mathbb{P}(Y=1|E=1, A=0, L=l)}{\mathbb{P}(Y=2|E=2, A=0, L=l)} \ \forall \ l \in \mathcal{L}.$$
\end{assumption}

\subsection{Identification when individuals are lost to follow-up}\label{APP: Censoring}
Our results can be extended to a setting where individuals are possibly censored due to loss of follow-up.
To ensure the identifiability of the $CCS$, it must be assumed that individuals are censored independent of their counterfactual outcomes. Formally, in addition to the censoring adjusted versions of Assumptions \ref{ASM: no multiple exp}, \ref{ASM: No eff exp}, \ref{ASM: Exp nec}, and \ref{ASM: No cross}, the following must hold.

\begin{assumption}[Independent censoring]\label{ASM: Indep censoring}
        \begin{enumerate}[label=\theassumption\alph*, leftmargin=*]
        \item[]
        \item \label{ASM: Cens exch}  $Y^{a, c=0} \independent C | A \ \forall \ a \in \{0,1\}.$
        \item \label{ASM: Cens pos}  $\mathbb{P}(A=a, Y=j, C=0) \ \forall \ a \in \{0,1\} \text{ and } j \in \{0,1,2\}.$
        \item \label{ASM: Cens cons}  $\text{If } A=a \text{ and } C=0\text{, then} Y^{a,c=0}=Y, E^{a, c=0}.$
    \end{enumerate}
\end{assumption}

Assumptions \ref{ASM: no multiple exp}, \ref{ASM: No eff exp}, \ref{ASM: Exp nec}, and \ref{ASM: No cross} are extended as follows:

\begin{assumption}[Negligible multiple exposure under censoring]\label{ASM: cens no multiple exp}
    $\mathbb{P}(E^{c=0}=\mathbf{B})=0.$
\end{assumption}

\begin{assumption}[No effect on exposure under censoring]\label{ASM: cens no eff exp}
    $E^{a=1, c=0}=E^{a=0, c=0}.$
\end{assumption}

\begin{assumption}[Exposure necessity under censoring]\label{ASM: cens exp nec}
    $E^{c=0}=0 \implies Y^{a, c=0}=0 \ \\\forall \ a \in\{0,1\}.$
\end{assumption}

\begin{assumption}[No cross-infectivity under censoring]\label{ASM: cens no cross}
        $E^{a, c=0}=j \implies Y^{a, c=0}\neq i\ \\ \forall \ a \in \{0,1\}, \, i \neq j, \, i,j \in \{1,2\}.$
\end{assumption}

Assumption \ref{ASM: Indep censoring} guarantees that the censoring event and the outcome are independent, that there is uncensored data from all levels of the treatment and the outcome, and that the observed data can be interpreted as the counterfactual outcome. Under these assumptions, the $CCS$ for the uncensored is identified.
\begin{proposition}[Contrast conditional on specific exposure under censoring]\label{PROP: Ident when cens}
    Under Assumptions \ref{ASM: Indep censoring}, \ref{ASM: cens no multiple exp}, \ref{ASM: cens no eff exp}, \ref{ASM: cens exp nec}, and \ref{ASM: cens no cross}, 
    \begin{align*}
        CCS^{c=0}&=\frac{\frac{\mathbb{P}(Y^{a=1, c=0}=1|E^{a=1, c=0} =1)}{\mathbb{P}(Y^{a=0, c=0}=1|E^{a=0, c=0}=1)}}
        {\frac{\mathbb{P}(Y^{a=1, c=0}=2|E^{a=1, c=0}=2)}{\mathbb{P}(Y^{a=0, c=0}=2|E^{a=0, c=0}=2)}}\\
        &=\frac{\frac{\mathbb{P}(Y=1|A=1)}{\mathbb{P}(Y=1|A=0)}}
        {\frac{\mathbb{P}(Y=2|A=1)}{\mathbb{P}(Y=2|A=0)}}.
    \end{align*}
\end{proposition}
See Appendix \ref{APP: Prop ident when cens proof} for a proof.

Extension to to time-to-event settings follows analogously, see for example the appendix of \citet{stensrud_identification_2023}.

\subsection{Time-to-event $CCE$}\label{SUBSEC: One-shot}
Following the structure introduced in \citet{stensrud_identification_2023}, denote with $Y_k$ and $E_k$ whether an individual has experienced the event $j$ by time $k$, and being exposed to variant $j$ first, by time $k$ respectively. Hence once someone is exposed to variant $j$, we will assume that the exposure to this given variant persists, while exposure to the competing variant is prohibited in the future. In this setting, we will define the time-to-event contrast conditional on specific exposure. 
\begin{definition}[Time-to-event contrast conditional on exposure]
    $$
    CCE_k=\frac{\bbP(Y_k^{a=1}=1|E_k^{a=1}\neq0)}{\bbP(Y_k^{a=0}=1|E_k^{a=0}\neq0)} \cdot
        \left( \frac{\bbP(Y_k^{a=1}=2|E_k^{a=1}\neq0)}{\bbP(Y_k^{a=0}=2|E_k^{a=0}\neq0)} \right)^{-1}
    $$
    with $\bbP(Y_k^{a=0}=1|E_k^{a=0}\neq0)>0$ and $\bbP(Y_k^{a=1}=2|E_k^{a=1}\neq0)>0$.
\end{definition}
The $CCE_k$ compares the effect of the vaccine on variant 1 and variant 2 respectively, at time $k$ among the individuals who were exposed. It is a causal estimand in the sense that the comparison is made within the same subpopulations. The following proposition ensures identification:

\begin{proposition}[Time-to-event CCE]\label{PROP: CCE TTE ident}
Under Assumptions \ref{ASM: no multiple exp}, \ref{ASM: TTE Standard RCT} and \ref{ASM: TTE original exp nec}
    \begin{equation*}
        CCE_k=
        \frac{\mu_k^1(1)}{\mu_k^1(0)}\left(\frac{\mu_k^2(1)}{\mu_k^2(0)}\right)^{-1}
    \end{equation*}
    where 
    \begin{equation*}
        \mu^j_k(a)= \sum_{i=1}^k h^j_i(a) \prod_{l=0}^{i-1} h^0_l(a)
    \end{equation*}
    and
    \begin{equation*}
        h^j_k(a)=\frac{\bbP(Y_k=j\cap Y_{k-1}=0|A=a)}{\bbP(Y_{k-1}=0|A=a)} 
    \end{equation*}
    for all $a\in \{0,1\}, \, j \in \{0,1,2\}, \, k \in \{1, \dots, K\}$

\end{proposition}
The proof is provided in Appendix \ref{APP: Proof CCE TTE}.
The time-to-event $CCE$ is expressed as a ratio of ratios using the cumulative incidences of variant 1 and 2 respectively. Both of them can be calculated as a function of observed variables, without the measurement of common causes either between the exposure variables $E_k$ or the outcome variables $Y_k$.

\subsection{Comments on estimation}\label{APP: Estimation}

\subsubsection{Estimation of the $CSE_k$}\label{APP sub: CSE comment} 
In the data example in Section \ref{SEC: Data}, two regression models were fitted to estimate the hazards, for variants 1 and 2 respectively. For each of these models, outcomes corresponding to the competing variants are censoring events, as they are making the future counterfactual outcome of interest under treatment $a$ unknown \citep{young_causal_2020}. Hence, for model 1 of variant 1, individuals who developed outcomes due to variant 2 before the end of the follow-up period, are censored at the date of the observation of this event, and correspondingly for model 2 of variant 2, individuals with observed values of $Y=1$ are censored.
Then applying the semi-parametric Cox regression, the $CSE_k$ can be expressed as the ratio of the estimated coefficients using models 1 and 2 for the two variants. The estimated coefficients are independent of time, hence the $CSE_k$ will be constant over time.

\subsubsection{Testing Assumption \ref{ASM: TTE Scaled new inf}}\label{APP sub: comment on testing scaled inf}
 Assumption \ref{ASM: TTE Scaled new inf} is restrictive. However, we can construct tests to falsify this assumption even if $U$ and $Z$ are not (entirely) observed: suppose that there are some observed baseline variables $L$, such that $L \in \{U, Z\}$. Then by Assumption \ref{ASM: TTE Scaled new inf}, and by marginalizing over the variables $\{U,Z\}\setminus L$, it follows that 

\begin{equation}\label{EQ: ASsumption scaled inf placholder}
    \bbP(Y_k^{a, e_k=1}=1|Y_{k-1}^{a}=0, L=l)=\gamma_{a, k}\bbP(Y_k^{a, e_k=2}=2|Y_{k-1}^{a}=0, L=l) \text{ for all } l \in \mathcal{L}.
\end{equation}

By marginalizing over the same set of variables $\{U,Z\}\setminus L$ in Assumption \ref{ASM: TTE exposure rat of exposed} and following an argument identical to the second half of the proof of Proposition \ref{PROP: CSE ident given U} (see Appendix \ref{APP: CSE ident}), except the modification of the conditioning set from $U=u, Z=z$ to $L=l$, Equation \eqref{EQ: ASsumption scaled inf placholder} can be rewritten to the form $$\bbP(Y_k=1|Y_{k-1}=0, A=a, L=l)=\frac{\gamma_{a, k}}{\alpha_k}\bbP(Y_k=2|Y_{k-1}=0, A=a, L=l).$$
The ratio of $\gamma_{a,k}/\alpha_{k}$ is constant across the values $l \in \mathcal{L}$. Therefore, for $l_1, l_2 \in \mathcal{L}$
\begin{equation}\label{EQ: scaled inf test}
    \frac{\bbP(Y_k=1|Y_{k-1}=0, A=a, L=l_1)}{\bbP(Y_k=1|Y_{k-1}=0, A=a, L=l_2)}=
\frac{\bbP(Y_k=2|Y_{k-1}=0, A=a, L=l_1)}{\bbP(Y_k=2|Y_{k-1}=0, A=a, L=l_2)}.
\end{equation}
For example, by Cox regression, Equation \eqref{EQ: scaled inf test} can be tested by the equality between the coefficients corresponding to $L$ in the two models of the two variants.

Assumption \ref{ASM: TTE Scaled new inf} corresponds to the null
\begin{align*}
    &H_0^{strata}: &&\frac{\bbP(Y_k^{a, e_k=1}=1|Y_{k-1}^{a}=0, U=u, Z=z)}{\bbP(Y_k^{a, e_k=2}=2|Y_{k-1}^{a}=0, U=u, Z=z)}=\gamma_{a, k}.
\end{align*}
 However, $L$ is only a subset of the unobserved variables $U$ and $Z$, thus the constant proportionality assumption may hold for $L$, but it can fail to be satisfied in a smaller sub-population defined by $U$ and $Z$, thus testing based on the subset $L$ may not be consistent.

\subsubsection{Methods for rare events}\label{APP sub: rare CI}

Compared to the Cox regression model, an alternative approach for estimating the probabilities $\bbP(Y_k=j|Y_{k-1}=0, A=a)$ for all $j \in \{1,2\}$ and $k \in \{1, \dots, 6\}$ are non-parametric sample means, and then we can calculate the $CSE_k$ based on this estimates.
However, of the 15,955 observations, the outcome that is matched to the vaccine is developed in  86 (0.54\%) cases, while the mismatched outcome is developed in 24 cases (0.15\%), making them rare events. Due to the rarity of these events, the normal approximation of the estimated hazard is flawed, and consequently, the Wald-type confidence intervals of the hazards at the 6-monthly follow-ups are considerably small and invalid.

To leverage more of the observations in our estimation, the cumulative hazards were estimated instead of the hazards at times $k$, using the conventional Nelson-Aalen \citep{aalen_nonparametric_1978} estimator for the two models when the outcome is due to variants 1 and 2 respectively. Then the corresponding estimators are not for estimating the value of the $CSE_k$ at times $k$, but the cumulative hazards up until the final time point. The cumulative hazard estimates were used to test a modified null hypothesis $H_0^w$,  defined as a composite between the $CSE_k$ being equal to one, and proportionality assumption on the hazards of the competing variants. Formally,
\begin{align}\label{EQ: H0 with no waning}
    &H_0^w: \, CSE_k=1 \text{ and proportionality assumption on the hazards.}
\end{align}

The proportionality assumption on the hazards is defined as an equal proportional change of the hazard ratio between the vaccine and the placebo-treated, with respect to the two strains, over time, that is
\begin{equation}
    \begin{aligned}\label{EQ: equal vaccine waning}
        &\frac{\mathbb{P}(Y_k=1|Y_{k-1}=0, A=1)/\mathbb{P}(Y_{k+1}=1|Y_{k}=0, A=1)}{\mathbb{P}(Y_k=1|Y_{k-1}=0, A=0)/\mathbb{P}(Y_{k+1}=1|Y_{k}=0, A=0)}\\
        &=\frac{\mathbb{P}(Y_k=2|Y_{k-1}=0, A=1)/\mathbb{P}(Y_{k+1}=2|Y_{k}=0, A=1)}{\mathbb{P}(Y_k=2|Y_{k-1}=0, A=0)/\mathbb{P}(Y_{k+1}=2|Y_{k}=0, A=0)}  
    \end{aligned}    
\end{equation}
for all $k$.
Under $H_0^w$, the ratio of the cumulative hazards is equal to one.
To increase the number of events at each time step, we combined observations from the first three visits and the observations from the last three. Then the ratios for the two periods, with  Bonferroni adjusted confidence intervals are  0.23 (95\% CI: 0.000, 0.613), and 0.427 (95\% CI: 0.000, 0.876) respectively. Hence, under Assumptions \ref{ASM: TTE Standard RCT}-\ref{ASM: TTE Scaled new inf} the null hypothesis $H_0^w$ (see Equation \ref{EQ: H0 with no waning}) can be rejected.
However, the null $H_0^w$ requires that $CSE_k=1$ for all $k$, and proportionality between the hazards, defined in Equation \ref{EQ: equal vaccine waning}. Thus, even if $H_0^w$ is rejected, we cannot conclude that sieve effect is implied.
Regardless, under the assumption of no sieve effect ($CSE_k=1 \, \forall k$), the ratio of the cumulative hazards must be equal to 1 in the first period, as then the ratio of cumulative hazards is equivalent to the hazard ratio. However, the same problem arises when Wald-type confidence intervals are used for rare events.

As shown in the previous paragraph, both the semi-parametric and the non-parametric approach face problems in approximation of the variance, due to the low number of uncensored events. However, when events are rare, the time-fixed and time-varying estimators will be similar. Many vaccine trials that concern infectious diseases, such as HIV, include a small proportion of events, and this implies that the vaccine effect on the competing variants can be approximated by the estimands introduced in Section \ref{SEC: fixed time estimands}.
In the context of rare diseases, we recommend using the introduced time-fixed estimands, namely the $CCS$, $CCE$, $SEIE$ and $SEITE$. Although Assumption \ref{ASM: no multiple exp} might be considered restrictive, the probability of multiple infectious contacts occurring over the course of the trial is negligible for diseases with low prevalence.
As an illustrative example, let us denote the number of contacts that can potentially lead to infection with $m$, for example in the case of HIV, needle sharing, or sexual contact, and let us denote with $r$ the relative prevalence of infectious individuals in the population. Then the probability of making more than $1$ infectious contact is $1-(1-r)^m+(1-r)^{m-1}r$. The relative prevalence of HIV in the US was approximately 0.36\% \citep{noauthor_volume_2023}, of which a considerable amount of individuals were untransmittable, making this an over-approximation of the relative prevalence of the potentially infectious. Considering $m=5$, is also an overestimation, as having more than 2 sexual partners was considered a high-risk behaviour by \citet{rolland_increased_2012}. Then the multi-exposure probability is $0.00013$ which can be regarded as negligible, making the time-fixed estimates suitable tools for assessing the differential effect of the vaccine in the two variants.

\subsection{Derivation of the variance and confidence intervals}\label{APP: Var calc}
If the exposure status is observed, then the $CCS$ can be identified from the observed data as
$$
\frac{\mathbb{P}(Y=1|A=1, E=1)}{\mathbb{P}(Y=1|A=0, E=1)}\cdot \frac{\mathbb{P}(Y=2|A=1, E=2)}{\mathbb{P}(Y=2|A=0, E=2)}.
$$
As the ratio of the probabilities are conditional on $E=1$ and $E=2$ respectively, they are independent. Hence the log-transformed version of this estimand can be approximated as the sum of independent normally distributed random variables. Then the "True" variance can be derived according to \citet{katz_obtaining_1978}.

However, when $E$ is concealed, independence no longer holds. Regardless of this dependence between 
\begin{align*}
    &\frac{\mathbb{P}(Y=1|A=1)}{\mathbb{P}(Y=1|A=0)}
    && \text{and}
    &&& \frac{\mathbb{P}(Y=2|A=1)}{\mathbb{P}(Y=2|A=0)}
\end{align*}
the respective variances of the two relative risks can be derived under the log approximation and the variance of the log-transformed estimator be defined as the sum of the two. Due to the possible dependence between the two relative risks, this is an overestimation of the variance, leading to conservative confidence intervals. However, the simulation study in Appendix \ref{APP: Simulation} shows that the loss in power endured by ignoring the correlation is negligible.

Finally, in $\mathbb{P}(Y=2|A=1)/\mathbb{P}(Y=2|A=0)$ the conditioning set could be extended $(A=a, Y \neq 1)$, therefore the formula for $CCS$ is defined as
$$
CCS= \frac{\mathbb{P}(Y=1|A=1)}{\mathbb{P}(Y=1|A=0)} \cdot \frac{\mathbb{P}(Y=2|A=0, Y \neq 1)}{\mathbb{P}(Y=2|A=1, Y \neq 1)} \cdot \frac{\mathbb{P}(Y \neq 1 | A=0)}{\mathbb{P}(Y \neq 1 | A=1)}
$$
What can be noted now, is that the middle term is jointly independent of the other two, while $\mathbb{P}(Y=1|A=1)/\mathbb{P}(Y=1|A=0)$ and $\mathbb{P}(Y\neq1|A=1)/\mathbb{P}(Y\neq1|A=0)$ have a known correlation of $\rho = -1$. Thus the variance for
$$log(\mathbb{P}(Y=2|A=0, Y \neq 1)/\mathbb{P}(Y=2|A=1, Y \neq 1))$$ 
can be derived using Method C from \citet{katz_obtaining_1978}, while the variance for
$$log(\mathbb{P}(Y=1|A=1)/\mathbb{P}(Y=1|A=0) \cdot \mathbb{P}(Y\neq1|A=1)/\mathbb{P}(Y\neq1|A=0))$$ follows from \citet{katz_obtaining_1978} and the formula $Var(X+Y)=Var(X)+Var(Y)+2\rho \sqrt{Var(X)}\sqrt{Var(Y)}$. Lastly, as these log-transformed expressions are independent, the variance can be defined as the sum of them.

The $EET$ is identified using the baseline variable $L$ and data of the treated individuals only. Hence the number of cases can no longer be modeled as two independent Binomial distributions, therefore the previous method described by \citet{katz_obtaining_1978} cannot be applied. Instead, we model the estimator of the $EET$ as a ratio of the proportions sampled from a trinomial distribution, as by Assumption \ref{ASM: no multiple exp}, $\mathbb{P}(E=\mathbf{B})=0$. Then we can use the conservative asymptotic confidence intervals for this ratio, following the approach of \citet{nelson_statistical_1972}.

Let us denote the observed number of individuals who developed variants 1 and 2 with $y_1$ and $y_2$ respectively.
Then, the two-sided $100\alpha \%$ confidence interval for the estimator of the $EET$ is defined as
$$
\left(\frac{y_2+1}{y_1}F(1-(1-\alpha)/2, 2\cdot(y_2+1), 2\cdot y_1)   \right)^{-1}
$$
$$
\frac{y_1+1}{y_2}F(1-(1-\alpha)/2, 2\cdot(y_1+1), 2\cdot y_2)
$$
for the upper and the lower limits respectively. $F(\alpha, a, b)$ denotes $\alpha$ quantile of the F-distribution with $a$ and $b$ degrees of freedom.

\subsection{Ratio of the absolute $CECE$s}\label{APP: Absolute measures}
\citet{stensrud_identification_2023} derived sharp bounds for the absolute causal effect conditional on exposure in the single variant setting. Those bounds can be equivalently used when the two competing variants are present, using Assumptions \ref{ASM: no multiple exp}-\ref{ASM: No cross}. Then the absolute $CECE$ ratio ($aCECEr$) is partially identified by the bounds, when\\
$\mathbb{P}(Y=1|A=0)>\mathbb{P}(Y=1|A=1)$ and $\mathbb{P}(Y=2|A=0)>\mathbb{P}(Y=2|A=1)$
\begin{align*}
    \frac{\mathbb{P}(Y=1|A=0)-\mathbb{P}(Y=1|A=1)}{1 - \frac{\mathbb{P}(Y=2|A=1)}{\mathbb{P}(Y=2|A=0)}}&\leq \frac{\mathbb{P}(Y^{a=1}=1|E=1)-\mathbb{P}(Y^{a=0}=1|E=1)}{\mathbb{P}(Y^{a=1}=2|E=2)-\mathbb{P}(Y^{a=0}=2|E=2)}\\
    &\leq \frac{1 - \frac{\mathbb{P}(Y=1|A=1)}{\mathbb{P}(Y=1|A=0)}}{\mathbb{P}(Y=2|A=0)-\mathbb{P}(Y=2|A=1)}
\end{align*}
equivalently 
\begin{align*}
   &\frac{\mathbb{P}(Y=1|A=0)-\mathbb{P}(Y=1|A=1)}{\mathbb{P}(Y=2|A=0)-\mathbb{P}(Y=2|A=1)} \cdot \mathbb{P}(Y=2|A=0)\leq aCECEr\\ &\leq \frac{\mathbb{P}(Y=1|A=0)-\mathbb{P}(Y=1|A=1)}{\mathbb{P}(Y=2|A=0)-\mathbb{P}(Y=2|A=1)} \cdot \frac{1}{\mathbb{P}(Y=1|A=0)} 
\end{align*}
The implication of these bounds is that point identification of the absolute $CECE$ ratio is only possible when $$\mathbb{P}(Y=1|A=0)=\frac{1}{\mathbb{P}(Y=2|A=0)}$$
that can only hold, given that probabilities are bounded between $0$ and $1$, when\\
$\mathbb{P}(Y=1|A=0)=\mathbb{P}(Y=2|A=0)=1$

Derivation of the vaccine efficacy ratio
according to the definition, $VE=1-RR$ follows similarly, after rewriting as 
$$\frac{VE_1}{VE_2}=\frac{\mathbb{P}(Y^{a=1}=1|E=1)-\mathbb{P}(Y^{a=0}=1|E=1)}{\mathbb{P}(Y^{a=1}=2|E=2)-\mathbb{P}(Y^{a=0}=2|E=2)}\cdot \frac{\mathbb{P}(Y^{a=0}=2|E=2)}{\mathbb{P}(Y^{a=0}=1|E=1)}.$$

Arguments for the other estimands defined in the manuscript follow analogously.

\end{document}